\documentclass[useAMS,usenatbib]{mn2e}
\usepackage{graphicx}
\usepackage{rotating,times,pictex,graphicx,latexsym}
\usepackage{color}
\usepackage{longtable,amsmath}
\usepackage{lscape}

\title{NTT follow-up observations of star cluster candidates from the FSR
catalogue}  

\author[Froebrich, Meusinger \& Scholz]{D.~Froebrich$^{1}$\thanks{Based on
observations collected at ESO, Chile; ESO 077.B-0074(A)}\thanks{E-mail:
df@star.kent.ac.uk}, H.~Meusinger$^{2}$, and A.~Scholz$^{3}$\\ $^1$ Centre for
Astrophysics and Planetary Science, University of Kent, Canterbury, CT2 7NH, UK
\\ $^2$ Th\"uringer Landessternwarte Tautenburg, Sternwarte 5, 07778 Tautenburg,
Germany \\ $^3$ SUPA, University of St. Andrews, Department of Physics \& Astronomy,
North Haugh, St. Andrews, Fife, KY16 9SS, Scotland, United Kingdom } 

\begin{document}

\date{Received sooner; accepted later}
\pagerange{\pageref{firstpage}--\pageref{lastpage}} \pubyear{2007}
\maketitle

\label{firstpage}

\begin{abstract}

We are conducting a large program to classify newly discovered Milky Way star
cluster candidates from the list of Froebrich, Scholz \& Raftery
\cite{2007MNRAS.374..399F}. Here we present deep NIR follow-up observations from
ESO/NTT of 14 star cluster candidates. We show that the combined analysis of
star density maps and colour-colour/magnitude diagrams derived from deep
near-infrared imaging is a viable tool to reliably classify new stellar
clusters. This allowed us to identify two young clusters with massive stars,
three intermediate age open clusters, and two globular cluster candidates among
our targets. The remaining seven objects are unlikely to be stellar clusters.
Among them is the object FSR\,1767 which has previously been identified as a
globular cluster using 2MASS data by Bonatto et al. \cite{2007MNRAS.381L..45B}.
Our new analysis shows that FSR\,1767 is not a star cluster. We also summarise
the currently available follow-up analysis of the FSR candidates and conclude
that this catalogue may contain a large number of new stellar clusters, probably
dominated by old open clusters.

\end{abstract}

\begin{keywords}
Galaxy: globular clusters: individual; Galaxy: open clusters, individual
\end{keywords}

\section{Introduction}
 
Star clusters are the building blocks of the stellar component of galaxies.
Identifying and characterising clusters is thus a crucial step for our
understanding of structure formation and mass assembly in the Milky Way. The
benefits of wide-field studies of stellar clusters are twofold: 

i) Open clusters are currently the most important sites of star formation and 
early stellar evolution. In particular, the formation of massive stars is
intrinsically linked to the formation of star clusters. For example, the cluster
mass is thought to correlate with the mass of the most massive star in the
cluster (Weidner \& Kroupa \cite{2006MNRAS.365.1333W}). Investigating age
spread, morphology, and mass segregation in a large sample of young open
clusters has the potential to put limits on models for massive star formation
(see review by Beuther et al. \cite{2007prpl.conf..165B}). Furthermore, by
probing the distribution of masses in a diverse sample of clusters allows us to
constrain the impact of environment on the outcome of star formation and to
probe the diversity and the origin of the Initial Mass Function -- fundamental
problems in current star formation theory (see review by Bonnell et al.
\cite{2007prpl.conf..149B}).

As open clusters dissolve, the stars migrate into the field. Therefore, the
study of old open clusters can shed light on the timescales for cluster
disruption and the underlying physical processes such as tidal interactions with
giant molecular clouds, evaporation of low-mass objects, mass loss due to
stellar evolution, as well as mass segregation. The currently known sample of
old open clusters is very incomplete (e.g. Bonatto \& Bica
\cite{2007A&A...473..445B}). It is hence an important task to enlarge the sample
of known and well classified galactic open clusters. 

ii) Cluster surveys have the potential to discover new Globular Clusters (GC), a
particular interesting avenue given the outstanding importance of this type of
cluster. As emphasised by Harris \cite{1996AJ....112.1487H}, Milky Way GCs have
proven throughout the last century to be ``irreplaceable objects in an amazingly
wide range of astrophysical studies''. The most important issue is the
continuing debate about the key processes in galaxy formation and evolution. GCs
have been considered for many years to be the most valuable tracers of the
oldest stellar population in our galaxy.

Recent advances point to a complex picture of the genesis of our Galaxy, driven
by a mixture of processes including rapid protogalactic collapse, accretion,
cannibalism, galaxy collisions, and star bursts. The rich source of historical
details provided by the (inhomogeneous) Milky Way GC system is among the most
promising approaches to disentangle the many processes (West et al.
\cite{2004Natur.427...31W}). A complete census of the Milky Way GC system is
therefore very important. More recently, a few examples of new GC classes were
detected in nearby galaxies: the so-called FF (`faint fuzzies') clusters in two
lenticular galaxies (Brodie \& Larsen \cite{2002AJ....124.1410B}), even more
extended GCs in the halo of M31 (Huxor et al. \cite{2005MNRAS.360.1007H}), very
massive GCs in NGC\,5128 (Martini \& Ho \cite{2004ApJ...610..233M}) and
ultracompact objects (perhaps bridging the gap in parameter space to dwarf
galaxies) in the Fornax cluster (Mieske et al.
\cite{2002A&A...383..823M,2004A&A...418..445M}). There are no known analogues
for such unusual GCs in our galaxy where they should have been discovered unless
they are hidden by dust in the classical `Zone of Avoidance', the least complete
area for the Milky Way GC sample. 

Recent studies suggest that the currently known sample of Milky Way GCs is
incomplete (see Bonatto et al. \cite{2007MNRAS.381L..45B} or Bica et al.
\cite{2007A&A...472..483B}), particularly at the low-luminosity end (the
Palomar-type GCs). The number of missing GCs close to the Galactic Plane ($|z| <
0.5$\,kpc) and within 3\,kpc from the Galactic Centre has been estimated to
$\sim 10\pm3$ (Ivanov et al. \cite{2005A&A...442..195I}).

Driven by the aforementioned science goals, we have recently carried out a
systematic large-scale cluster survey based on star density maps derived from
the 2MASS database (Froebrich, Scholz \& Raftery \cite{2007MNRAS.374..399F},
hereafter FSR). The FSR survey covers the entire Galactic Plane ($|b| <
20^\circ$) and detected a total number of 1788 potential star clusters, from
which 767 have been known beforehand and 1021 are unknown cluster candidates.
The contamination of those candidates has been estimated to be about 50\,\%,
indicating that the catalogue may contain up to 500 new star clusters. 

In particular, the FSR survey revealed several promising GC candidates in the 
Zone of Avoidance. Four of them have been discussed already in detail elsewhere 
(Froebrich et al \cite{2007MNRAS.377L..54F}, \cite{2008MNRAS.383L..45F}; Bica et
al. \cite{2007A&A...472..483B}, Bonatto et al. \cite{2007MNRAS.381L..45B}).
Given the small number ($\sim 150$) of known galactic GCs on the one hand and
the diversity of the GC species on the other hand (see above), every new GC is
of value. 

In this paper, we present detailed follow-up analysis for 14 cluster candidates
from the FSR survey, selected to be among the best candidates for new GCs. The
paper is structured as follows. In Sect.\,\ref{dataanalysis} we present our new
observations and the reduction of the data. The detailed analysis and results
for each individual cluster, including the appearance of the cluster, the
contamination with field stars and the isochrone fitting to determine the
cluster properties are presented in Sect.\,\ref{results}. Finally in
Sect.\,\ref{discussion} we discuss and conclude our findings.

\section{Data Analysis}

\label{dataanalysis}

\subsection{Cluster Selection}

We have selected a number of cluster candidates from the FSR list for further
follow up investigation. Originally 15 cluster candidates were selected, 14 of
which have been observed and the results are presented in this paper. About half
of the selected candidates were possible globular clusters according to the
analysis in Froebrich et al. \cite{2007MNRAS.374..399F}. The remainder of the 
objects where selected because of their interesting appearance in the 2MASS
images. Since the analysis of the cluster properties was refined after the
target selection for the observations, only four of the targets are still
considered to be potential globular cluster candidates in the FSR list. These
are FSR\,0002, 0089, 1716, and 1767. Analysis of 2MASS data for three of our
targets has been published (FSR\,0089 - Bonatto \& Bica
\cite{2007A&A...473..445B}; FSR\,1754 - Bica et al. \cite{2008MNRAS.385..349B};
FSR\,1767 - Bonatto et al. \cite{2007MNRAS.381L..45B}). FSR\,0089 was classified
as a 1\,Gyr old open cluster, FSR\,1754 as an uncertain case with two apparent
main sequences, and FSR\,1767 as a nearby Palomar type globular cluster. Another
selected cluster candidate (FSR\,1570 or Teutsch\,143a) has been published by
Pasquali et al. \cite{2006A&A...448..589P} just before the final version of the
FSR list was compiled, and is hence not in the FSR list. The cluster is very
young (slightly older than 4\,Myr) and contains the luminous blue variable star
WRA\,751. The remaining cluster candidates (FSR\,0088, 0094, 1527, 1530, 1659,
1712, 1716) are investigated here for the first time in detail. Our data for
FSR\,1735 has already been published in Froebrich et al.
\cite{2007MNRAS.377L..54F}. The object is most likely a globular cluster in the
inner Milky Way. For completeness reasons we have added a short analysis of
FSR\,1735 in this paper as well.

\subsection{Data}\label{data}

Observations have been performed in service mode using SofI at the NTT for the
project ESO\,077.B-0074(A). We observed each cluster candidate in J, H, and K
with a pixel scale of 0.288". An eight point mosaic in the cluster candidate
area was observed to cover a large enough control field in the vicinity. The
candidate area was observed at the beginning and end of each mosaic, ensuring
twice the per pixel integration time in the cluster candidate area. The per
pixel integration time in each filter was 225\,s. In general the observing
conditions were clear or thin cirrus was present. In a few cases the cirrus was
more thick and hence the per pixel integration time was doubled. Only for object
FSR\,0002 variable conditions occurred during the observations in the K band
(see Sect.\,\ref{res_0002}). The mosaics cover in total an area of about
11.7'x11.7', with the corners and the centre missing.

Standard NIR data reduction procedures were followed when creating the mosaics.
Sky flats were used for flat-fielding and the xdimsum task in IRAF\footnote{IRAF
is distributed by the National Optical Astronomy Observatories, which are
operated by the Association of Universities for Research in Astronomy, Inc.,
under cooperative agreement with the National Science Foundation.} was used for
sky-subtraction and mosaicing. The average seeing FWHM in the final co-added JHK
frames is between 2.5 and 3.0 pixels, corresponding to 0.7" to 0.85". This
results in many regions in a severe crowding due to the high star density.
Therefore, the completeness limit of the photometry is highly variable from
cluster candidate to cluster candidate. The completeness limits and photometric
uncertainties for each cluster candidate will be shown and discussed
individually in Sect.\ref{results}. 

\subsection{Photometry}

For source detection and photometry we used the SExtractor software (Bertin \&
Arnouts \cite{1996A&AS..117..393B}). Due to the high star density in most
fields, the limiting factor for the photometry is the confusion limit.
Calibration of the images was performed using the wealth of 2MASS sources
available in each field. The {\it rms} scatter occurring when comparing the
2MASS photometry with our measurements is rather large, in the order of 0.1\,mag.
This is mostly caused by our much better spatial resolution and the high star
density in the images. Hence, our photometric uncertainties are also at least
0.1\,mag. 

Furthermore, bright stars in the images are saturated and hence their photometry
is unreliable. The saturation starts for stars with a brightness between 10 and
11\,mag, depending on the weather conditions and/or the seeing. At brighter
magnitudes we therefore apply a separate fit of the 2MASS colours to our
measured brightnesses for calibration. Still, the magnitudes become increasingly
unreliable for brighter stars.

\subsection{Star Density Maps}

To assess the amount of stellar overdensity in the areas of the cluster
candidates we created for each of the observed mosaics a star density map (SDM).
For this we have used only the stars in the field with reliable photometry
(quality flag of less than 4 from the SExtractor software) in all three bands.
The SDM maps have a pixel size of 30" and the pixel values indicate the star
density. It is determined by measuring the distance to the 50th nearest
neighboring star and converting this to the star density. The combination of
pixelsize and 50th nearest neighbour was chosen because of the typical star
densities (10..20 stars/arcmin$^2$ with reliable photometry) in our images and
the sizes of the cluster candidates. A typical cluster candidate should show up
as a stellar overdensity in the SDMs with an extend of up to 3x3 pixels,
corresponding to a radius of about 45".

The SDMs presented here show the star density in gray-scale. White corresponds
to the lowest density and dark to the highest. Areas that are not covered by the
mosaic or where the distance to the 50th nearest neighbour is above a set
threshold are shown in black. The scaling of the pixel values from black to
white is linear but different in each of the maps for the individual cluster
candidates. This has been done to enhance as much as possible the contrast
between the cluster candidate and control area. Since we are only interested in
the relative change of the star density within an image, we will not note the
maximum and minimum values for each image.

\subsection{Relative Extinction Maps}

Based on the SDMs, we have also created relative extinction maps (REM) for each
of the mosaics. We used the same pixel size for these maps as in the SDMs. The
relative extinction value is determined using the 50 nearest stars to the centre
of each pixel. Two maps of the median J$-$H and H$-$K colour excess of these
stars with respect to the field without cluster stars are determined. This
colour excess is converted into optical extinction and both maps are averaged to
obtain the final REM (see e.g. Froebrich et al. \cite{2007MNRAS.378.1447F} for
the conversion factors). 

The presentation of the extinction values is done in gray-scale, with high
extinction values in white and low extinction in black. Again, areas that are
not covered by the mosaic, or were the distance to the 50th nearest neighbour is
above a certain threshold, are shown in black. The scaling of extinction values
is linear and, as for the SDMs, different in each of the maps to enhance as much
as possible the contrast in each case. Since we are only interested in the
relative extinction values within an image, we will not note the maximum and
minimum values in each case. 

Together, the SDMs and REMs are used during the analysis of the cluster
candidates to decide which area of the mosaic is to be chosen as control field.

\subsection{Decontamination of Foreground and Background Stars}

The cluster candidates observed in this project are all situated close to the
Galactic Plane and/or the Galactic Centre. Hence, to analyse the potential
clusters we need to decontaminate the cluster field from foreground and
background stars. We used the technique described in Bonatto \& Bica
\cite{2007MNRAS.377.1301B}. It counts the stars per unit area in cells of J-band
magnitude and J$-$H and J$-$K colour in the cluster and the control field. Stars
are then randomly removed from the cluster field according to the difference in
the stellar densities in these cells. The combination of J-band magnitude and
J$-$H and J$-$K colour is the optimum choice to decontaminate cluster main
sequences in crowded fields (Bonatto \& Bica \cite{2007MNRAS.377.1301B}). The
size of the cells was varied depending on the number of stars in the field.
Typical values for the cell sizes are $\Delta$J\,=\,0.5\,mag,
$\Delta$(J$-$H)\,=\,0.2\,mag, and $\Delta$(J$-$K)\,=\,0.2\,mag. 

The correct choice of control field is obviously important for an as accurate as
possible decontamination. For each cluster candidate we carefully examined the
SDMs, REMs, and colour images created for each field. Control field regions were
chosen so that they: i) are as close as possible to the cluster; ii) are as
large as possible; iii) have no apparently higher or lower extinction values
than regions close to the cluster candidate. This ensures that the control field
has a foreground and background population of stars which is as similar to the
cluster field as possible and contains a large enough number of stars for a
sufficiently accurate statistics. 

In regions where the choice of control field was difficult due to variable
extinction, we compared the decontamination procedure using different control
fields to ensure our choice does not influence the decontamination. In the
detailed analysis of our cluster candidates in Sect.\,\ref{results}, we indicate
the choice and reasons for the control field selection in each case. We also
repeated the random decontamination process several times for each particular
control field, to ensure that remaining features are real and not just due to
the random nature of the process.

\subsection{Colour Magnitude Diagrams}

To analyse our cluster candidates in detail, we used the decontaminated J$-$K
vs. K colour magnitude diagrams (CMD). The plotted datapoints remained in the
area of the cluster after one particular realisation of the decontamination
process. The solid red line in the diagrams indicates the completeness limit of
the photometry. It is calculated as the peak in the luminosity function of the
stars in the cluster area in each band. The completeness limit changes
significantly depending on the crowding in the field and the observing
conditions (as indicated above in Sect.\,\ref{data}). In some diagrams we plot
two completeness limits. In these cases the upper one corresponds to the 2MASS
limit in that area, the lower one represents our new data. This shows the
improvement of our data compared to 2MASS and helps comparing our analysis to
already published work on some of the cluster candidates. The solid black line
is the best fitting isochrone to the cluster candidate. Isochrones are taken
from Girardi et al. \cite{2002A&A...391..195G}. 

\subsection{Colour Colour Diagrams}

As a second tool to analyse the cluster candidates we use decontaminated H$-$K
vs J$-$H colour colour diagrams (CCD). They show the same stars (in the same
colouring and symbols) as the CMDs. Overplotted are the best fitting isochrone
(solid black) as well as the same isochrone without extinction (dashed black) to
indicate where un-reddened main sequence and giant stars are located. The
reddening path for stars is also shown, enclosed by the two straight solid black
lines. The slope of the reddening path is determined using $A_\lambda \propto
\lambda^\beta$ and the effective wavelength of the 2MASS
filters\footnote{$\lambda_J = 1.235\mu$m, $\lambda_H = 1.662\mu$m, $\lambda_K =
2.159\mu$m from\\
http://www.ipac.caltech.edu/2mass/releases/allsky/doc/explsup.html}, the system
our brightnesses are calibrated in. If not mentioned otherwise, we use a
standard value of $\beta$\,=\,1.6.

\subsection{Isochrone Fitting}

To confirm the nature of the cluster candidates as real star clusters, and if
so, to determine their main parameters, such as distance, age, reddening, and
metallicity we use model isochrones from Girardi et al.
\cite{2002A&A...391..195G}. They are overplotted into the CMDs and CCDs of the
individual decontaminated cluster fields. We varied the isochrone parameters
(age, metallicity) as well as the 'environmental' parameters (distance,
reddening, $\beta$) until a simultaneous fit of both diagrams was found.
Parameters of the best 'by-eye' fit are then taken as the cluster properties. In
some cases a range of parameters can lead to a satisfying fit, and hence the
cluster parameters cannot be determined accurately. These uncertainties are
discussed separately for the individual objects.

\subsection{Radial Star Density Profiles}

For all objects that we classified as stellar cluster, we determined radial star
density profiles (RDP) in order to determine their size. We used the
decontaminated photometry to determine the RDPs, because this will limit
statistical noise from foreground and background stars, which is significant in
most cases due to the position of the clusters near the Galactic Plane. To
determine the radius of the cluster, we need to obtain the RDP out to a large
enough radius. Hence the decontamination has to be done for a large field around
the cluster centre. In contrast, the decontamination used for the classification
of the cluster candidate via CMDs and CCDs only used the central part of the
candidate, to minimise the number of remaining foreground and background stars. 
The determined RDPs are fit using a King like profile of the form $\sigma(r) =
\sigma_b + \sigma_0 / (1 + (r/r_c)^2)$, where $\sigma_b$ is the density of
foreground and background stars remaining after the decontamination of the large
field, $\sigma_0$ the central star density of the cluster, and $r_c$ the core
radius. See Sect.\,\ref{discussion} for a discussion of the uncertainties in the
determined parameters.

\section{Results}\label{results}

In the following section we will discuss in detail our results for each of the
cluster candidates using the above described SDMs, REMs, CMDs, and CCDs.
Together with the isochrone fitting and additional information, the plots will
be used to classify the objects as a star cluster or not, and to determine its
parameters. A summary of the main classification results and the parameters can
be found in Table\,\ref{properties}.

\begin{figure}
\centering

\beginpicture
\setcoordinatesystem units <1mm,1mm> point at 0 0
\setplotarea x from 0 to 80 , y from 0 to 70
\put {\includegraphics[height=4.0cm]{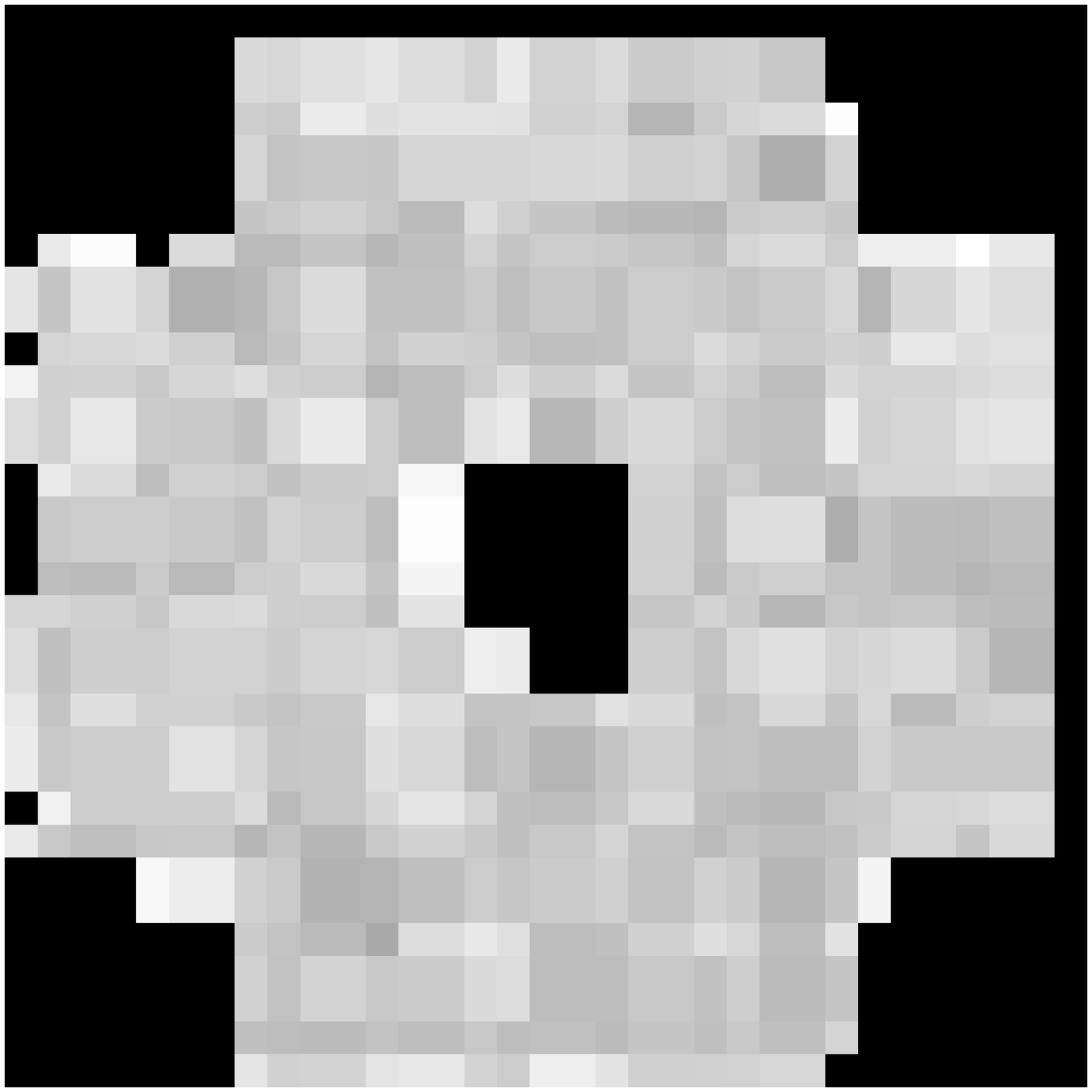}} at -20 50
\put {\includegraphics[height=4.0cm]{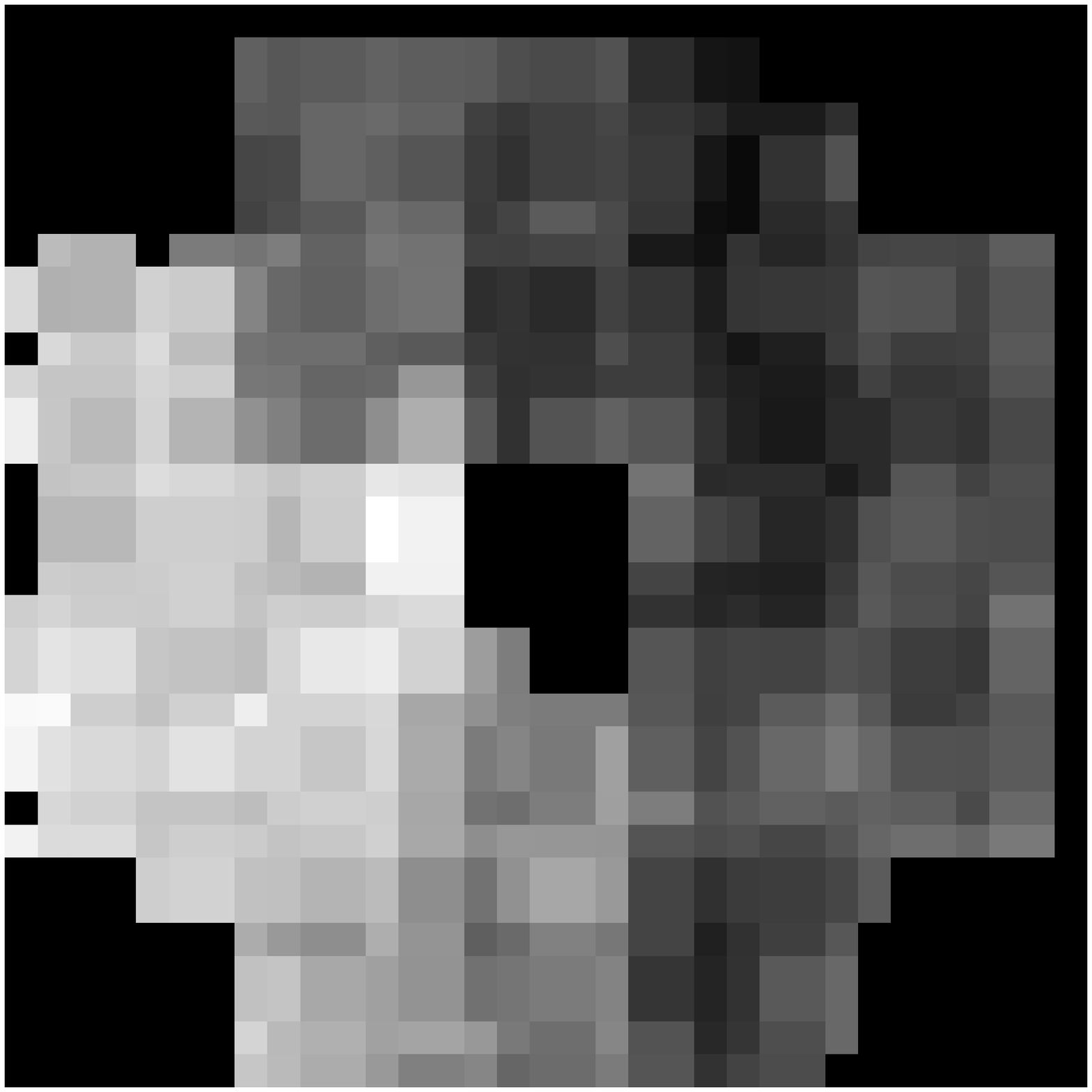}} at 25 50
\put {\includegraphics[height=8cm, angle=-90]{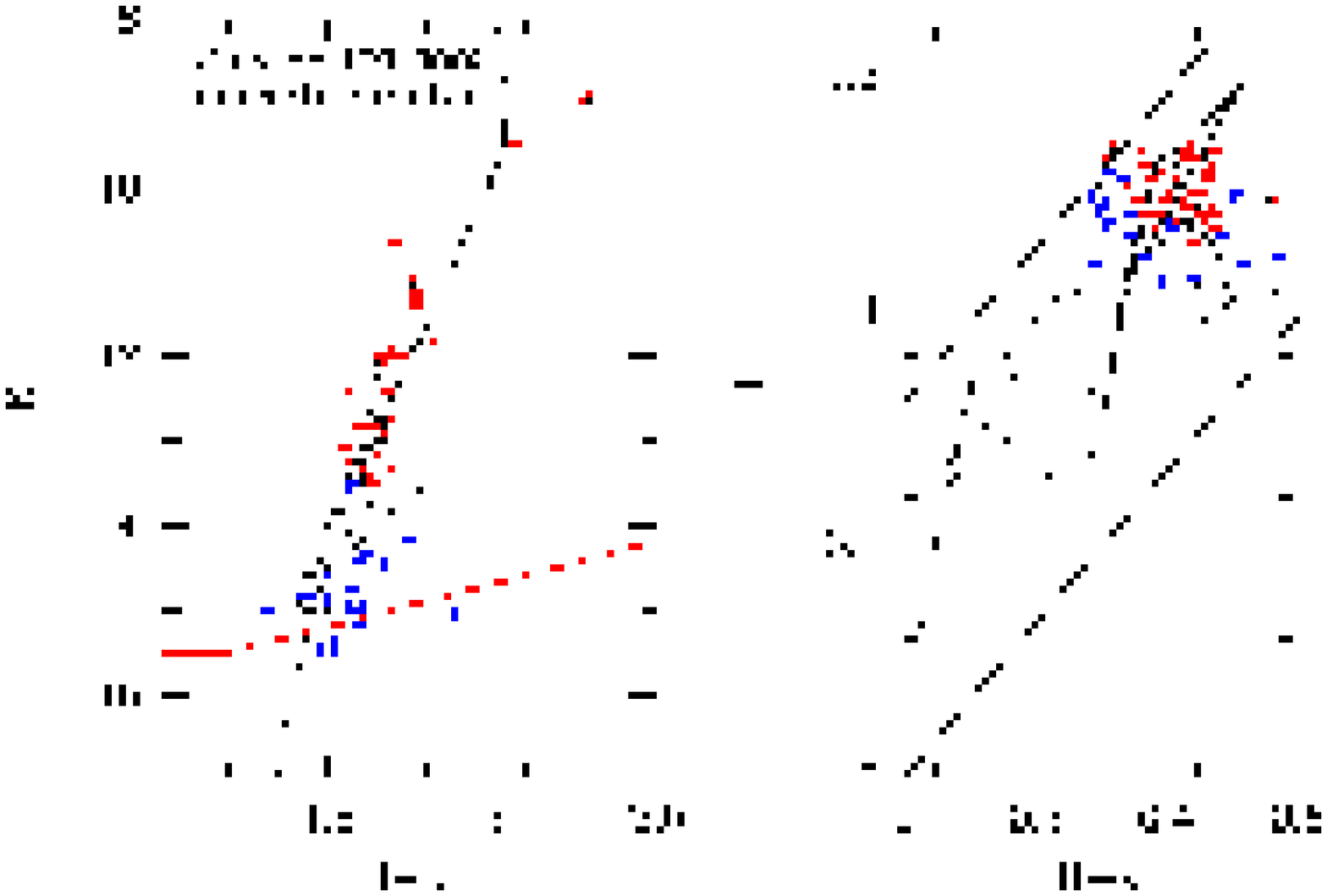}} at 0 0
\put {{\color{blue}\oval(30,30)}} at -25 58
\put {{\color{blue}\oval(30,30)}} at 20 58
\endpicture

\caption{\label{cl0002_jkk} {\bf Top Left:} SDM in the area of FSR\,0002,
obtained for all stars with reliable photometry in all three bands. Bright
colours correspond to low star density and dark colours to high star density.
The image size is 12'x12'. The circle indicates the cluster area. {\bf Top
Right:} Same scale image of the REM of the area around the cluster candidate 
FSR\,0002 obtained from colour excess calculations. {\bf Bottom Left:} One
realisation of the decontaminated CMD. The black solid line indicates the best
fitting isochrone (log(age)\,=\,9.7, for other parameters see text) and the red
line shows the completeness limit of the data. {\bf Bottom Right:}
Decontaminated CCD for FSR\,0002. The best fitting isochrone is shown as solid
line and its unreddened position as dashed line. The reddening path is indicated
by the two straight solid lines. the colouring of the symbols is the same as in
the CMD.}  

\end{figure}

\subsection{FSR\,0002}\label{res_0002}

The visual inspection of the cluster candidate image gives the impression of a
homogeneous field of stars. This is confirmed by the SDM (top left panel in
Fig.\,\ref{cl0002_jkk}), which shows only small variations in the star density.
The REM, however, shows large systematic differences in the colours of the stars
from one side of the mosaic to the other. These differences show indications of
the positions of the individual frames in the mosaic. It turns out that in this
case some of the images in the K-band mosaic have been taken during cloudy
conditions. Since the entire mosaic is calibrated against 2MASS sources in the
field, the colours of all stars observed under cloudy conditions are
systematically wrong. The calibration of the photometry with 2MASS was done
using only stars in the western half of the mosaic to ensure an accurate
calibration for the largest part of the field. 

Given the problems with the K-band data, we have selected a control field in the
western part of the mosaic. The decontamination leaves a number of stars,  which
in principle could be fit as a RGB/AGB of a cluster of stars. Depending on the
exact position of the control field, however, a highly variable number of stars
remains. There are about 1000 stars in the cluster region and depending on the
choice of the control field, between 25 and 140 remain - a very small and
variable fraction. Moreover, the CMDs of the cluster and any control field are
extremely similar and resemble a RGB/AGB of a cluster as well. 

In Fig.\,\ref{cl0002_jkk} we show one realisation of the decontamination in the
CMD and CCD. Stars are represented by two different symbols/colours to
facilitate the identification of groups of stars in both diagrams (i.e.
distinguish between bright and faint main sequence/giant stars in the CCD). The
overplotted isochrone has the parameters: Z\,=\,0.019, log(age)\,=\,9.7,
d\,=\,15\,kpc, A$_K$\,=\,0.65\,mag. The only possibility to fit both diagrams
simultaneously is to use a high metallicity and old age. Lower metallicity
isochrones require a higher extinction to fit the CMD and hence the CCD will not
be fit, except assuming very unreasonable dust properties with $\beta$ much less
than 1.6. Given our data analysis discussed above, we conclude that it is highly
unlikely that the cluster candidate FSR\,0002 is a star cluster. 

\begin{figure}
\centering

\beginpicture
\setcoordinatesystem units <1mm,1mm> point at 0 0
\setplotarea x from 0 to 80 , y from 0 to 70
\put {\includegraphics[height=4.0cm]{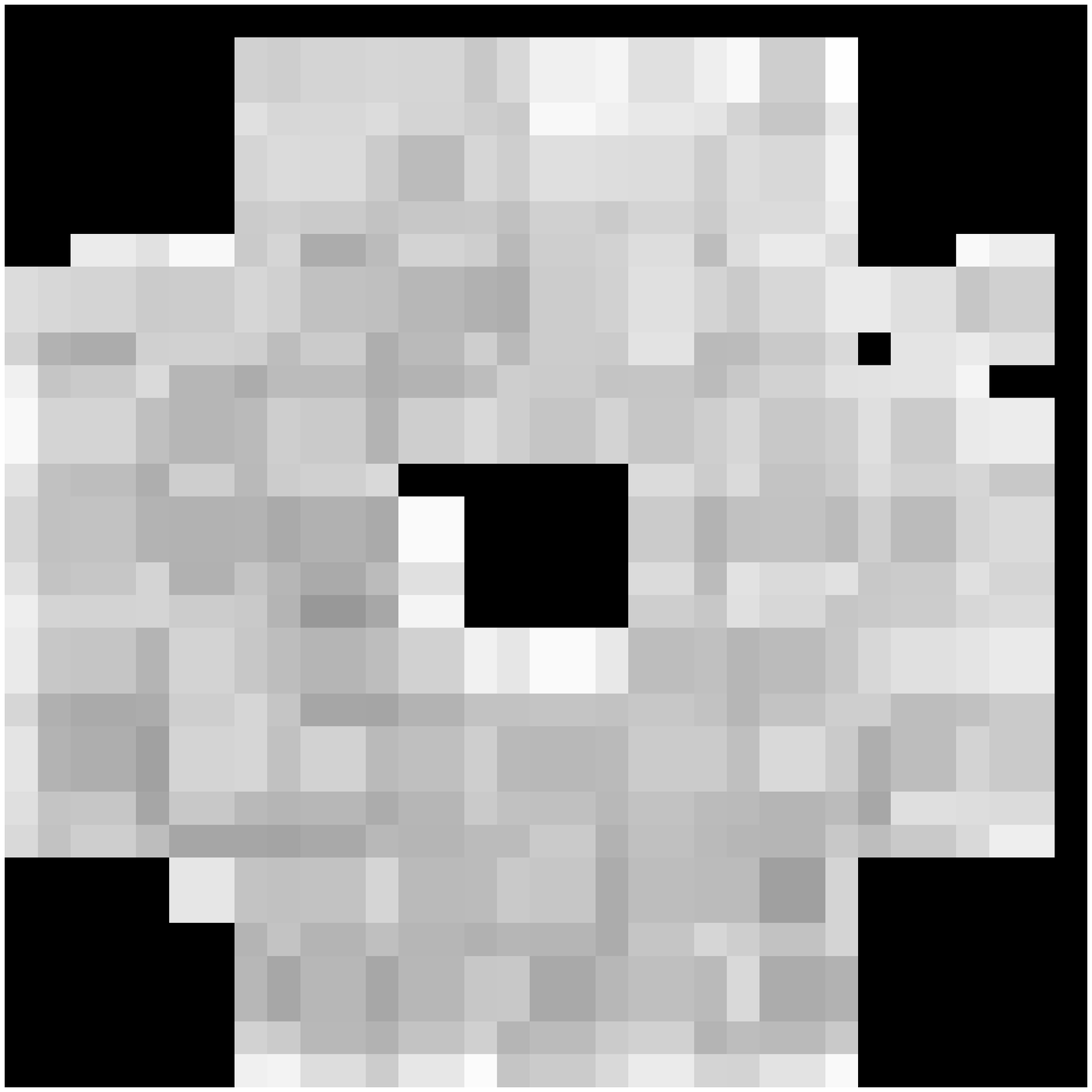}} at -20 50
\put {\includegraphics[height=4.0cm]{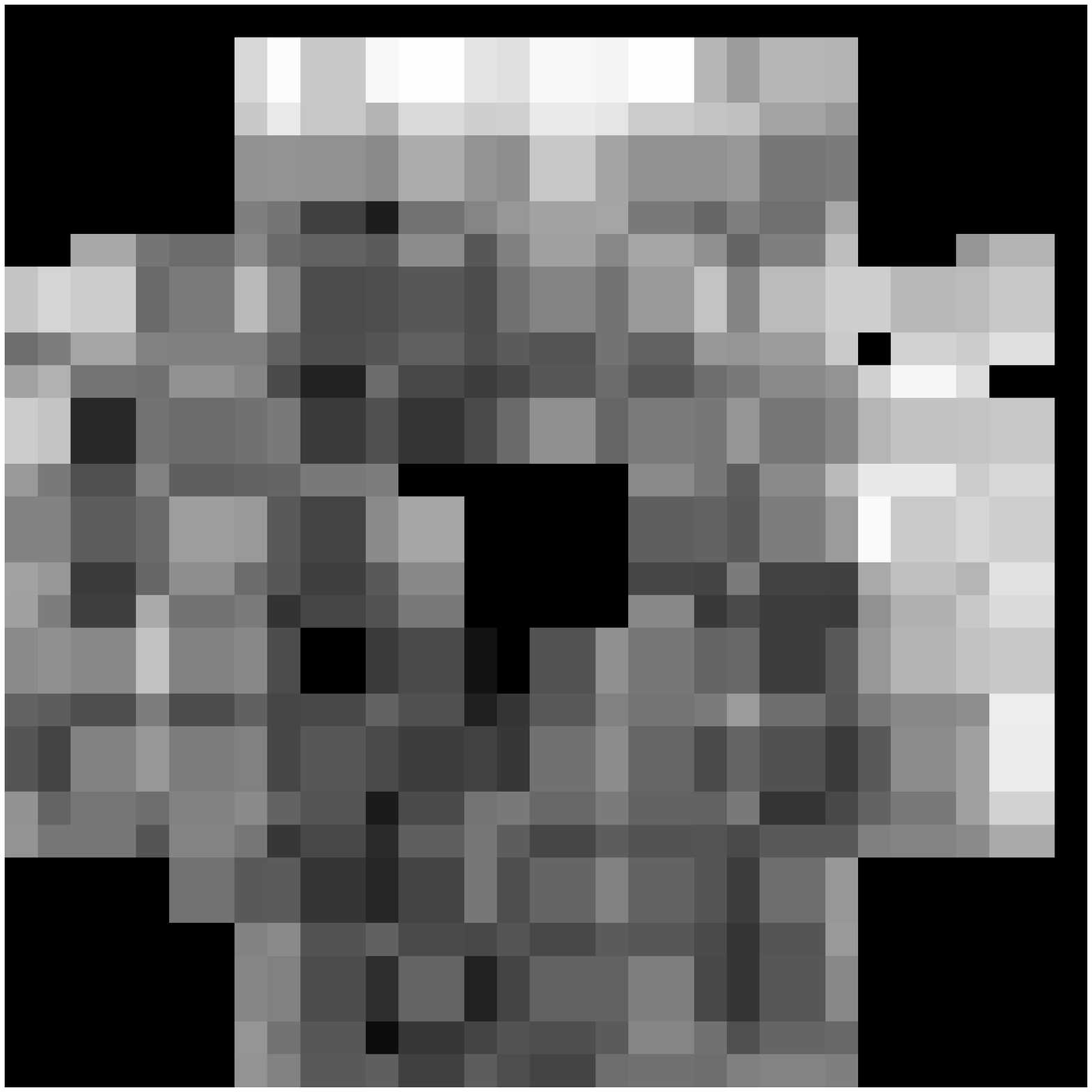}} at 25 50
\put {\includegraphics[height=8cm, angle=-90]{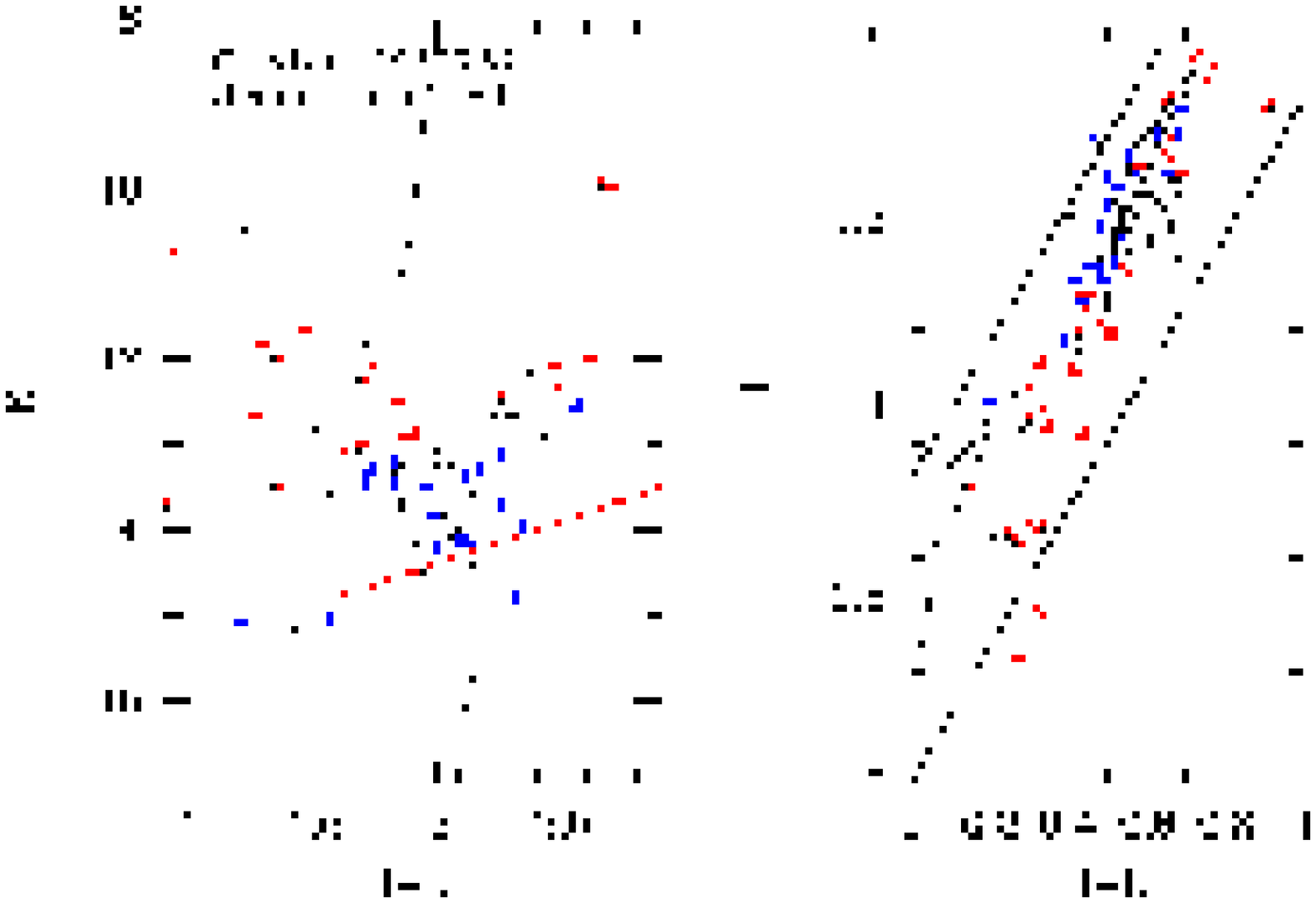}} at 0 0
\put {{\color{blue}\oval(30,30)}} at -25 58
\put {{\color{blue}\oval(30,30)}} at 20 58
\endpicture

\caption{\label{cl0023_jkk} As Fig.\,\ref{cl0002_jkk} but for the cluster
candidate FSR\,0023. The plotted isochrone has log(age)\,=\,10. See text for
other parameters. Clearly no isochrone can fit all the data.}   

\end{figure}

\subsection{FSR\,0023}

The image of the region around the cluster candidate FSR\,0023 shows no apparent
cluster, as well as no significant changes of the star density. This is
confirmed by the SDM. Only a slight underdensity towards the northern and
western edges of the mosaic is seen. Similarly, the REM shows higher extinction
values in the same regions. The lower star density and higher extinction hint
that these areas are influenced by a cloud of gas and dust. Nevertheless, these
regions cover only a small fraction of the entire mosaic, and we have therefore
chosen as control field the entire area outside the cluster field.  Choosing a
smaller control field outside the cloud area in the east or south  of the mosaic
does not change the results.

Only a small fraction of about 10\,\% of the stars remain after the
decontamination procedure. In the CCD and CMD it is not possible to reliably fit
the remaining objects by a single isochrone with plausible parameters. In
Fig.\,\ref{cl0023_jkk} we show one realisation of the decontamination with an
isochrone overplotted. The parameters of the isochrone are: Z\,=\,0.019,
log(age)\,=\,10, d\,=\,450\,pc, A$_K$\,=\,0.9\,mag. Again, as for FSR\,0002 one
needs a high metallicity old isochrone to fit the data. Given these implausible
parameters, the bad agreement of any isochrone with the data, and the small
number of stars remaining after the decontamination, we conclude that FSR\,0023
is not a star cluster.

\begin{figure}
\centering

\beginpicture
\setcoordinatesystem units <1mm,1mm> point at 0 0
\setplotarea x from 0 to 80 , y from 0 to 70
\put {\includegraphics[height=4.0cm]{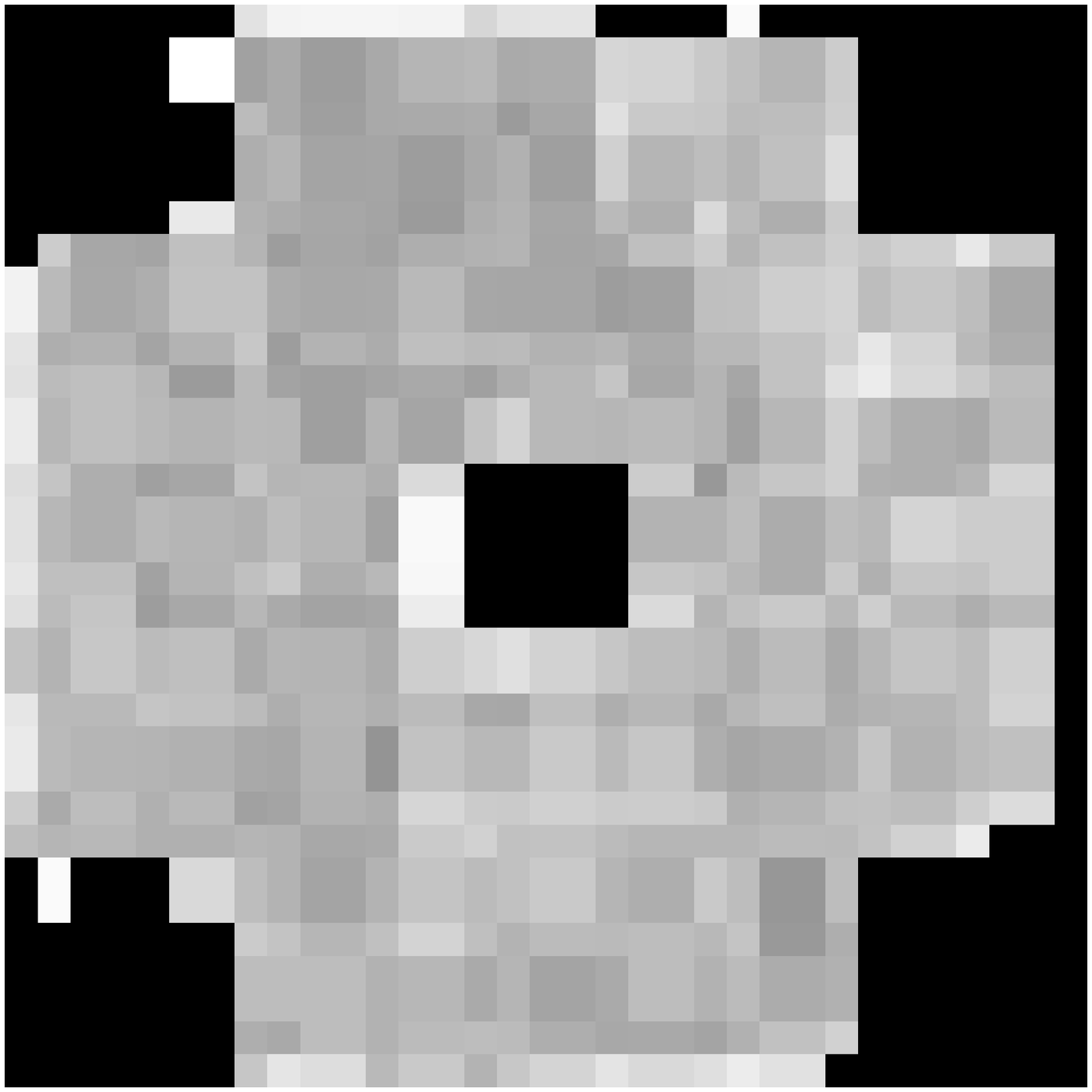}} at -20 50
\put {\includegraphics[height=4.0cm]{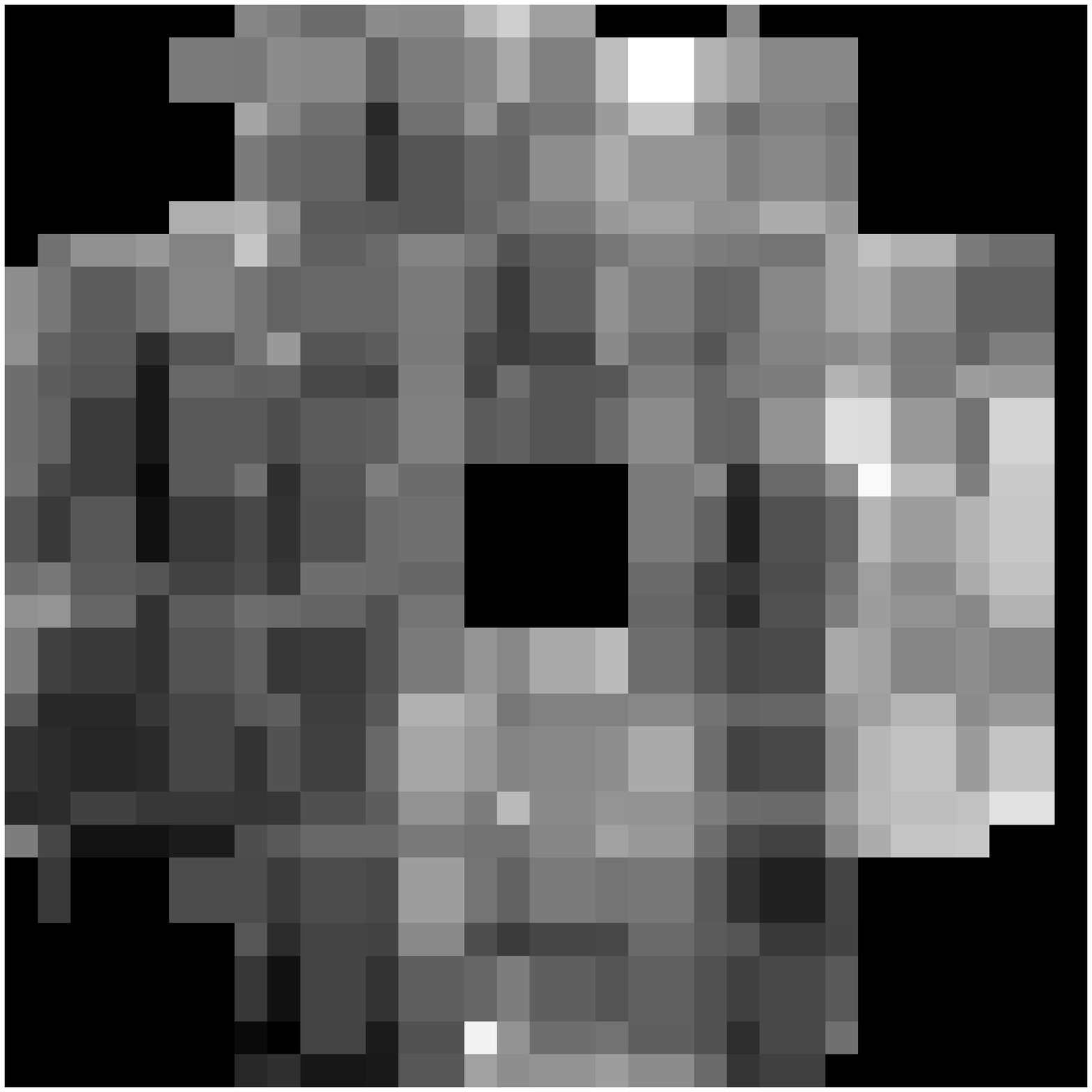}} at 25 50
\put {\includegraphics[height=8cm, angle=-90]{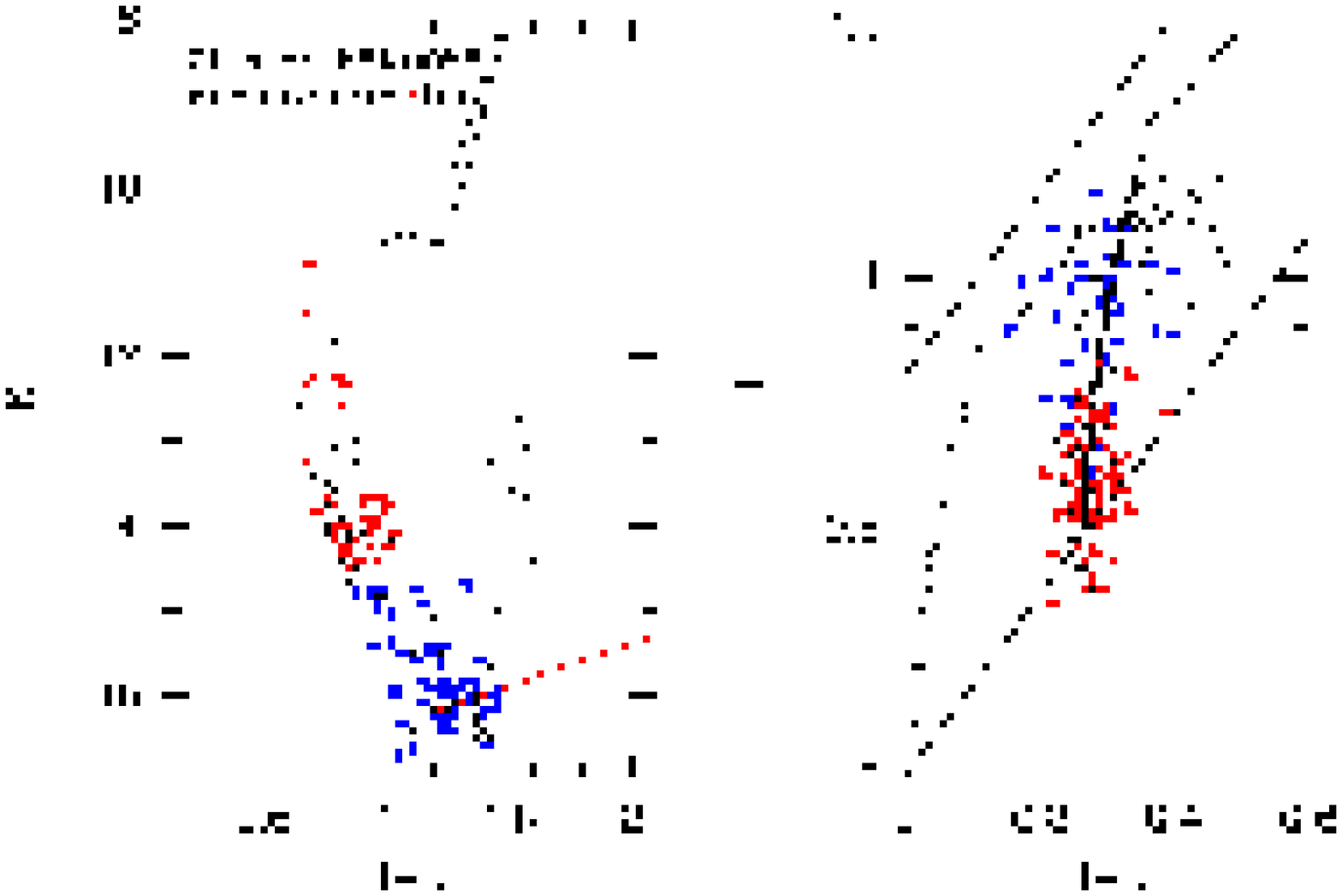}} at 0 0
\put {{\color{blue}\oval(30,30)}} at -25 58
\put {{\color{blue}\oval(30,30)}} at 20 58
\endpicture

\caption{\label{cl0088_jkk} As Fig.\,\ref{cl0002_jkk} but for the cluster
candidate FSR\,0088. The plotted isochrone has log(age)\,=\,8.7. See text for
other parameters.}  
\end{figure}

\subsection{FSR\,0088}

The image of FSR\,0088 shows nothing that looks like a cluster of stars. There
seems to be a very slight overdensity of stars in the region around the cluster
candidate, which is tentatively seen in the SDM (see Fig.\,\ref{cl0088_jkk}).
The REM of the entire field also shows no significant systematic differences.
There are fluctuation in the map with a standard deviation of about 0.6\,mag of
optical extinction. These are, however, distributed in the entire mosaic and we
thus use the entire area outside the cluster as the control field.

There is a large number of stars remaining after the decontamination procedure
(about 25\,\% of the stars in the region around the cluster candidate). They are
aligned along a main sequence in the CMD and the CCD. The brighter objects are
located near the bottom edge of the reddening path in the CCD, indicating that
those objects are stars of spectral type A. There are no or only individual
giant branch stars remaining after the decontamination. Hence, there is no
possibility to deduce the metallicity. The position of the cluster inside the
solar circle and its apparent young/intermediate age, however, justifies the
assumption of at least solar metallicity. Assuming slightly higher or lower
metallicities does not change any of the following results. 

We have fit an isochrone to the stars (see Fig.\,\ref{cl0088_jkk}), using the
upper end of the main sequence and its shape in the CMD to constrain the cluster
parameters. Furthermore, the distribution of the stars in the CCD was used to
determine the dust properties. We find that the best fit can be achieved with
$\beta =$\,1.6, an age of 500\,Myr, a K-band extinction of 0.5\,mag, and a
distance of 2.0\,kpc. {\bf The uncertainty for the age is about 60\,\%. The
distance can be estimated within 300\,pc, and the reddening within 0.05\,mag
K-band extinction. Outside these ranges the isochrones will clearly not fit the
CMD and CCD simultaneously.} The age of 0.5\,Gyr means, that stars with masses
of about 2.7 solar masses have just left the main sequence. These are stars of
spectral type A-F, in agreement with the low J$-$H colours seen of those objects
in the CCD.

The fact that we see no clear cluster in the image, nor a significant star
density enhancement in the nearest neighbour map, could be due to the fact that
we see a cluster in the process of dissolving into the field star population,
which is certainly in agreement with the determined age. The RDP is very noisy
and strongly depends on which realisation of the decontamination is used (see
Fig.\,\ref{krad_0088} for one example). It is hence not possible to measure the
cluster size accurately enough.

\begin{figure}
\centering

\beginpicture
\setcoordinatesystem units <1mm,1mm> point at 0 0
\setplotarea x from 0 to 80 , y from 0 to 70
\put {\includegraphics[height=4.0cm]{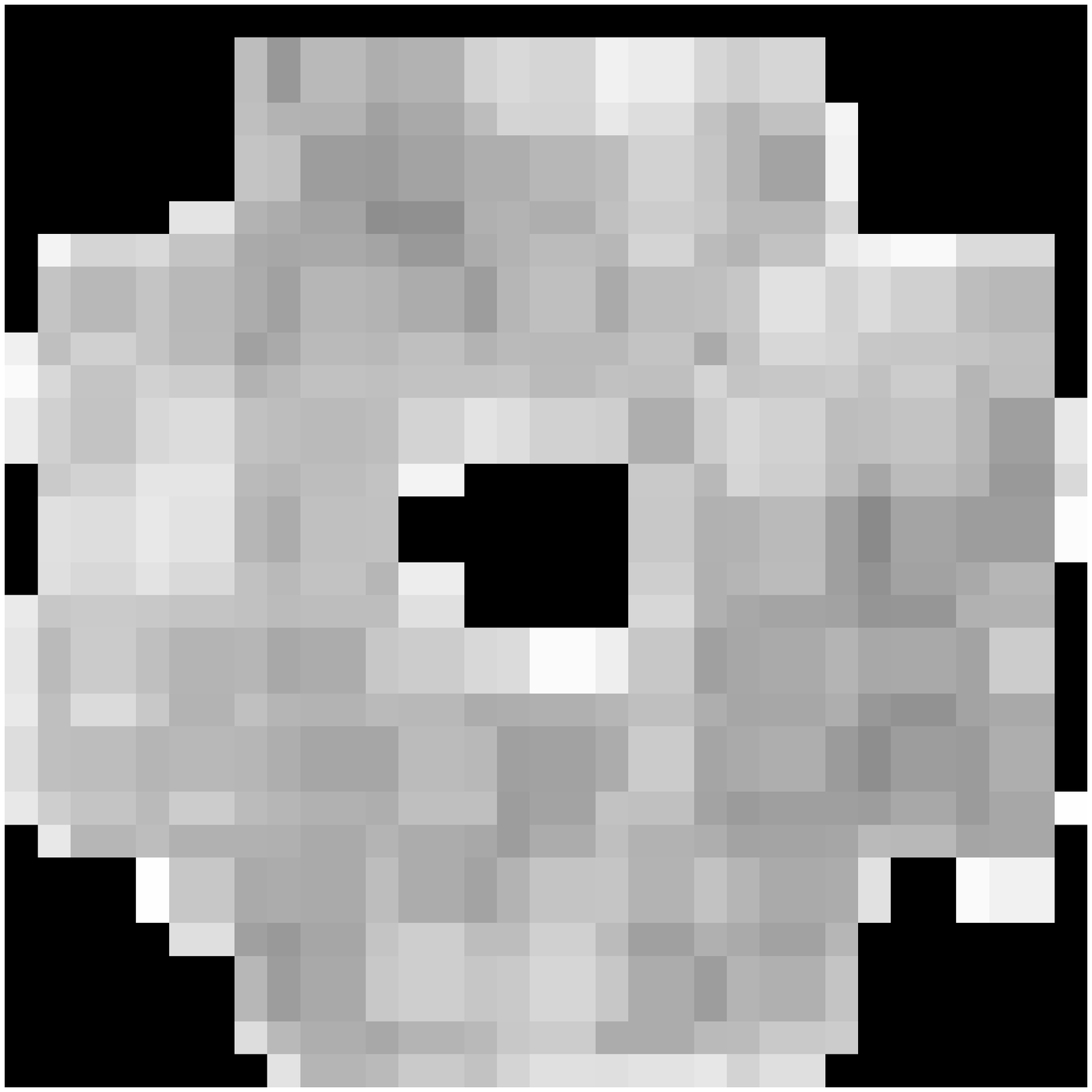}} at -20 50
\put {\includegraphics[height=4.0cm]{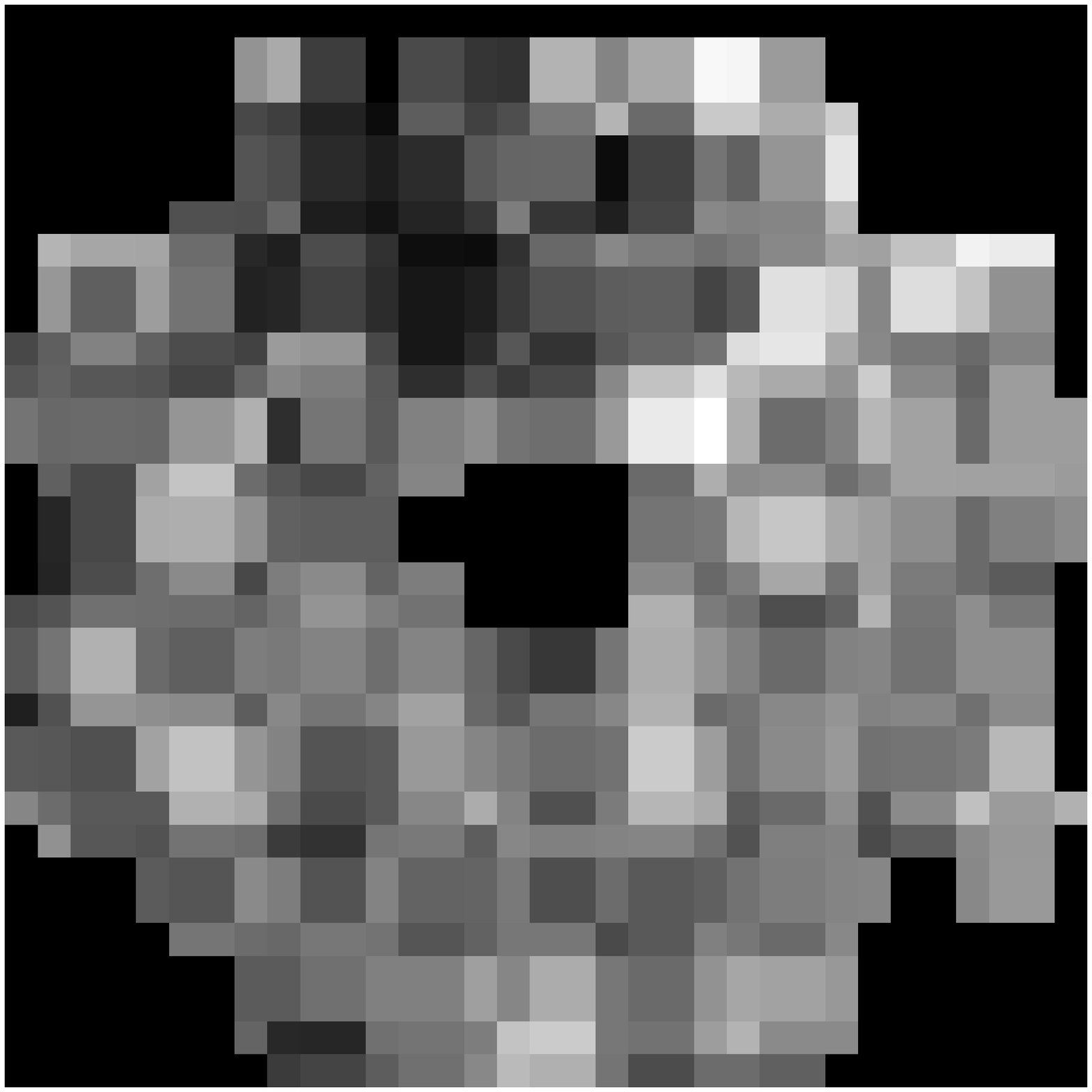}} at 25 50
\put {\includegraphics[height=8cm, angle=-90]{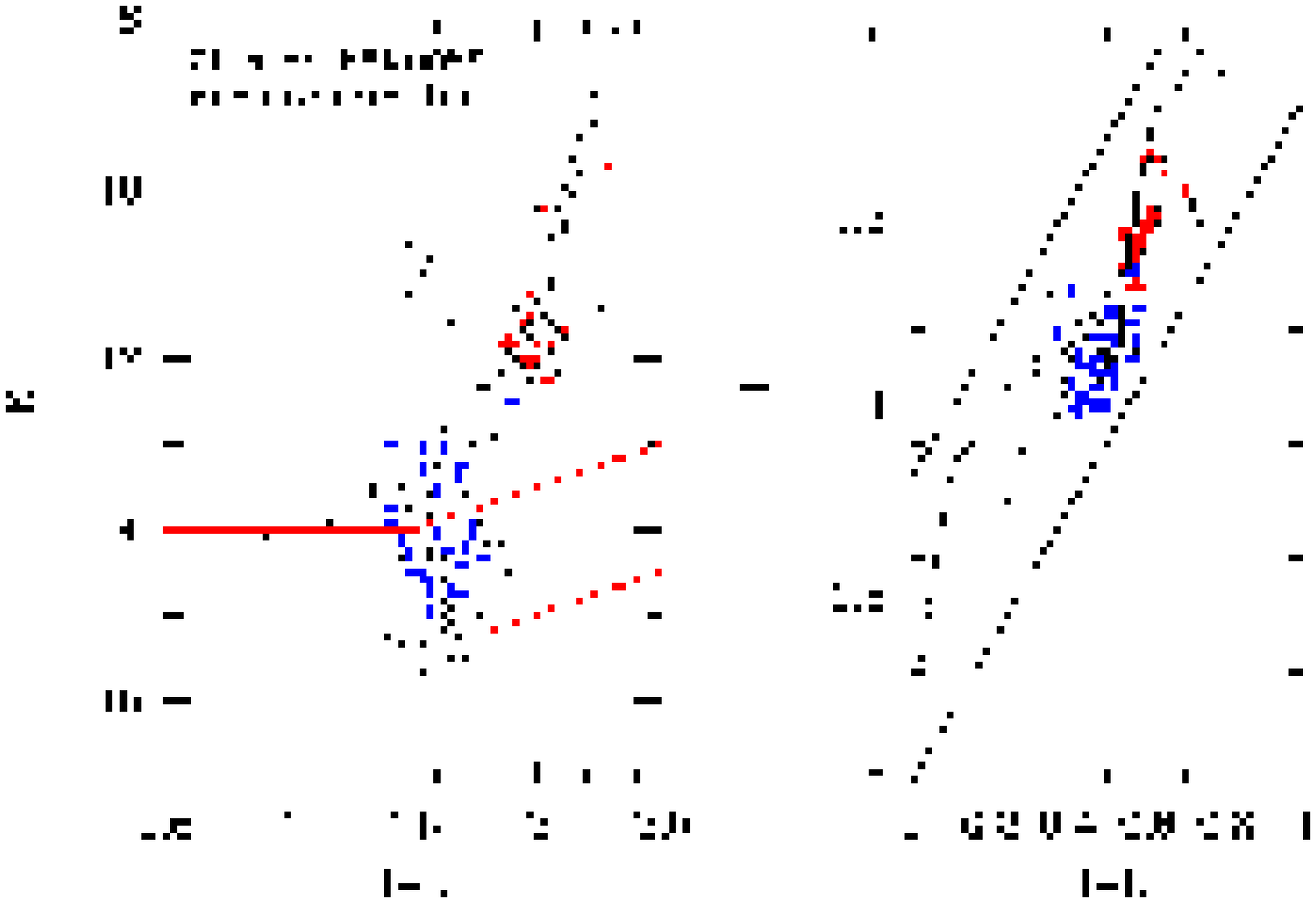}} at 0 0
\put {{\color{blue}\oval(30,30)}} at -25 58
\put {{\color{blue}\oval(30,30)}} at 20 58
\endpicture

\caption{\label{cl0089_jkk} As Fig.\,\ref{cl0002_jkk} but for the cluster
candidate FSR\,0088. The plotted isochrone has log(age)\,=\,9. See text for
other parameters. The upper completeness limit in the CMD indicates the 2MASS
limit.}   
\end{figure}

\subsection{FSR\,0089}

The image of the cluster candidate shows a slight overdensity of stars,
especially the number of brighter stars seems to be slightly higher. This is
confirmed in the SDM (see Fig.\,\ref{cl0089_jkk}), which shows a higher
concentration of stars in the cluster area. The REM of the mosaic shows a dip in
the extinction at the position of the cluster. The A$_V$ values in the cluster
candidate area are about 2\,mag lower than in the surrounding field. This either
means that the stars in the cluster area are on average closer to Earth and
hence bluer, the increased number of stars is caused by lower extinction at the
position of the cluster candidates, or the cluster stars are intrinsically bluer
than the surrounding field stars. 

The 2MASS photometry of this object has been investigated in Bonatto \& Bica
\cite{2007A&A...473..445B}. They classified it as a stellar cluster with an age
of 1\,Gyr, a distance of 2.2\,kpc, and a reddening corresponding to
$A_V$\,=\,9.1\,mag. The data presented here for the object is of much better
spatial resolution than the 2MASS photometry, and about 1\,mag deeper (see e.g.
the difference in the completeness limits shown in Fig.\,\ref{cl0089_jkk}). If
the interpretation of Bonatto \& Bica \cite{2007A&A...473..445B} is correct, an
isochrone with their parameters should fit the new data as well. 

In the decontaminated CMD of our data we have two groups of stars (about 30\,\%
of the stars in the cluster area remain). One can be identified as giant stars
and the other one as main sequence objects in a star cluster. We have
overplotted our and the 2MASS observational limits in Fig.\,\ref{cl0089_jkk} to
show the differences in the data. We have overplotted in the CMD an isochrone
corresponding to the Bonatto \& Bica \cite{2007A&A...473..445B} parameters. As
can be seen the deeper data still agrees with the parameters found using the
2MASS photometry. We detect more of the main sequence stars, which are too faint
for 2MASS. Our best fit to both, the CMD and CCD requires Z\,=\,0.019,
$\beta$\,=\,1.6, log(age)\,=\,9, d\,=\,2.2\,kpc, and A$_K$\,=\,1.0\,mag. {\bf
Due to the presence of main sequence and giant stars we can estimate the age
within 30\,\%. The distance is accurate within 300\,pc and the reddening within
0.05\,mag K-band extinction.} Similar to FSR\,0088, the RDP is noisy and
variable. Hence no radius can be determined. Note that Bonatto \& Bica
\cite{2007A&A...473..445B} find a core radius of about 0.4\,pc for this object.

There is, however, a minor detail in the data that the isochrone cannot fit: In
the CCD the scatter in colours of the main sequence stars seems to point towards
smaller colours, while the scatter in colour of the red giants points towards
larger colours. A similar behavior, fainter stars in a cluster candidate seem to
show lower extinction than brighter stars in the same candidate, can also be
seen in some other examples (see below: FSR\,1712, 1735). As a possible reason
we have identified the calibration of our photometry with 2MASS data.
Additionally to the large scatter of about 0.1\,mag of the residuals in the
calibration, there are some cases with even larger discrepancies for fainter
magnitudes. In the case of FSR\,0089, about 1/3 of the stars fainter than
12\,mag in the K-band seem to be shifted by about 0.05\,mag towards fainter
magnitudes compared to 2MASS. This will not influence the decontamination
procedure, since the colour-magnitude cells are chosen much larger than this,
but it will show up in the CCDs.

\begin{figure}
\centering

\beginpicture
\setcoordinatesystem units <1mm,1mm> point at 0 0
\setplotarea x from 0 to 80 , y from 0 to 70
\put {\includegraphics[height=4.0cm]{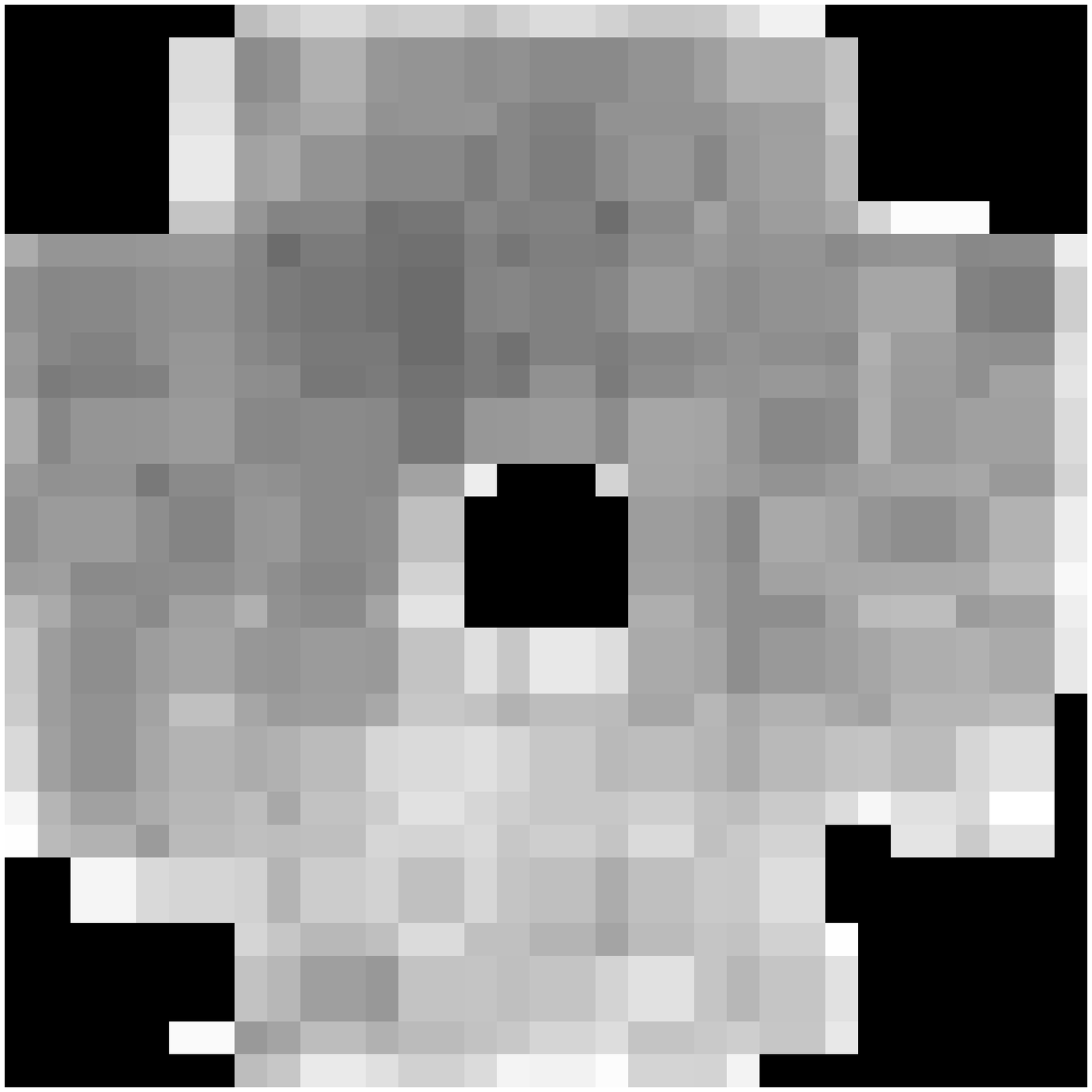}} at -20 50
\put {\includegraphics[height=4.0cm]{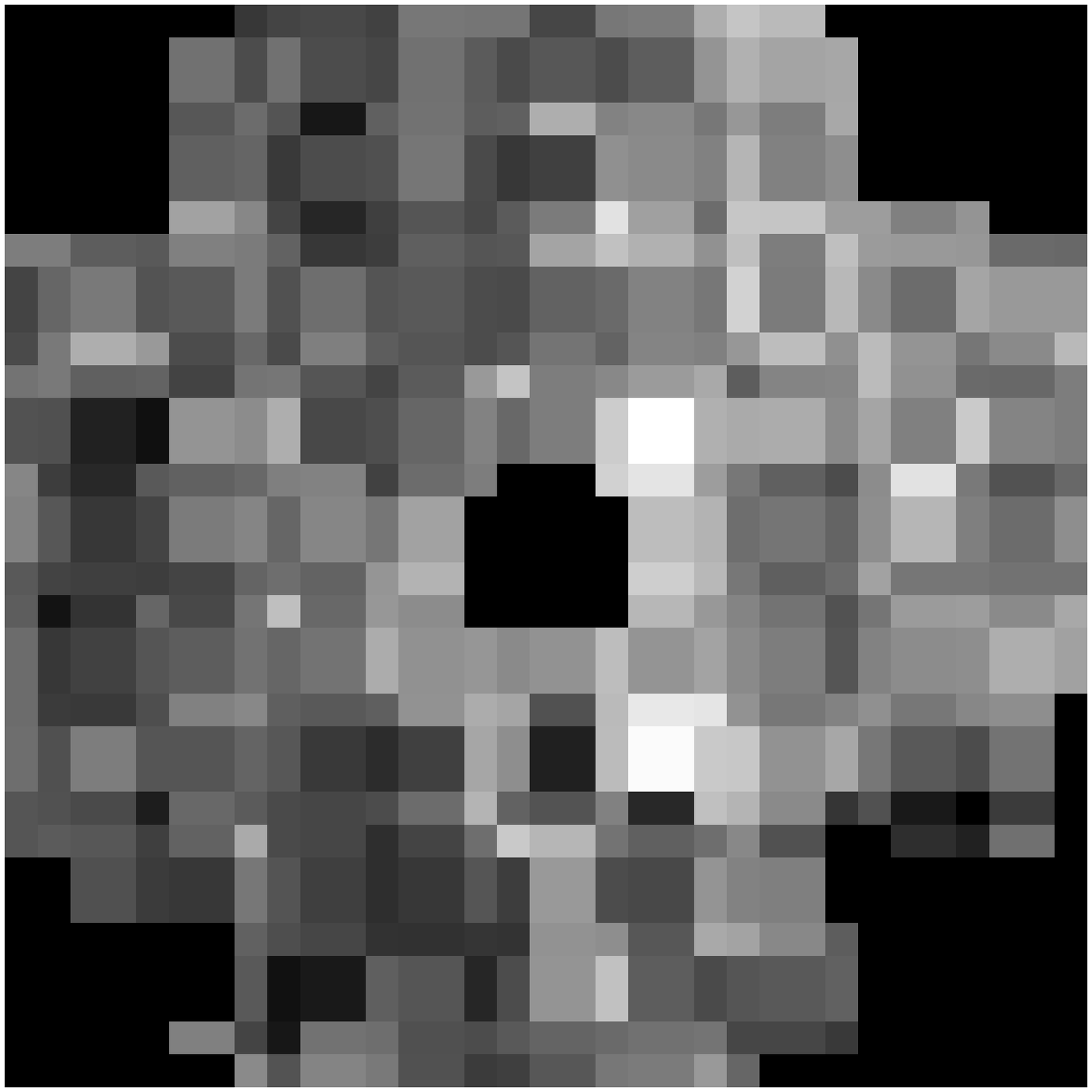}} at 25 50
\put {\includegraphics[height=8cm, angle=-90]{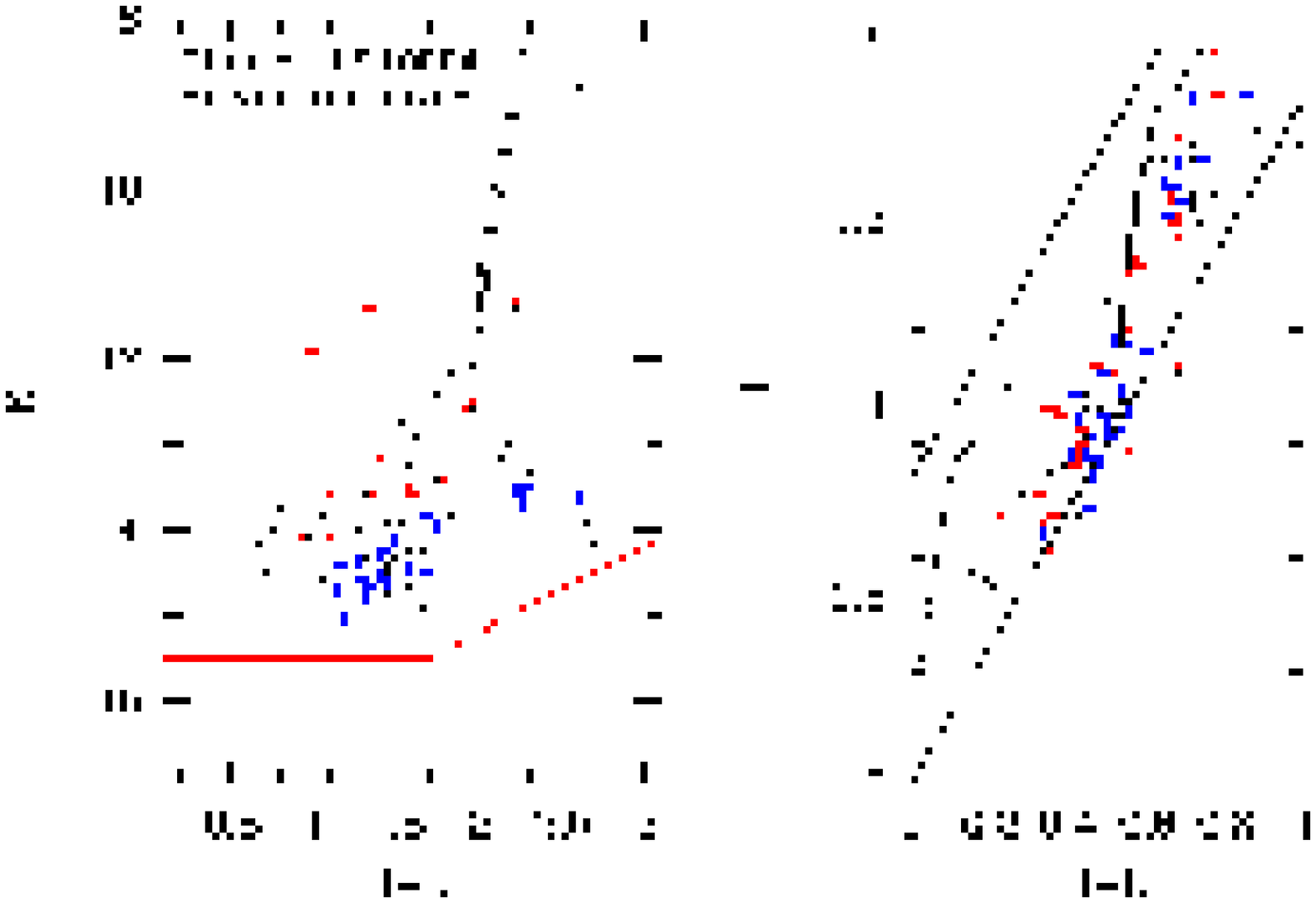}} at 0 0
\put {{\color{blue}\oval(30,30)}} at -25 58
\put {{\color{blue}\oval(30,30)}} at 20 58
\endpicture

\caption{\label{cl0094_jkk} As Fig.\,\ref{cl0002_jkk} but for the cluster
candidate FSR\,0094. The plotted isochrone has log(age)\,=\,9. See text for
other parameters. Clearly no isochrone can fit all the data.}  
\end{figure}

\subsection{FSR\,0094}

The image of FSR\,0094 seems to show a clear indication of an enhanced star
density in the area of the cluster candidate. This is also apparent in the SDM
(see Fig.\,\ref{cl0094_jkk}). The REM shows that the western and southern part
of the mosaic suffer from an increased amount of extinction compared to the area
of the cluster. We have thus chosen the eastern part of the mosaic as the
control area.

The stars remaining (up to 20\,\%) after the decontamination in the CMD and CCD
clearly follow a distribution that cannot be fit by a single isochrone. Rather a
range of extinction values and distances is needed. Hence, we conclude that
FSR\,0094 is not a cluster of stars. The isochrone in Fig.\,\ref{cl0094_jkk} is
just plotted to clarify our argument that it cannot fit the data, and has the
following parameters: Z\,=\,0.019, $\beta$\,=\,1.6, log(age)\,=\,9,
d\,=\,2.0\,kpc, and A$_K$\,=\,1.0\,mag. 

\begin{figure}
\centering

\beginpicture
\setcoordinatesystem units <1mm,1mm> point at 0 0
\setplotarea x from 0 to 80 , y from 0 to 70
\put {\includegraphics[height=4.0cm]{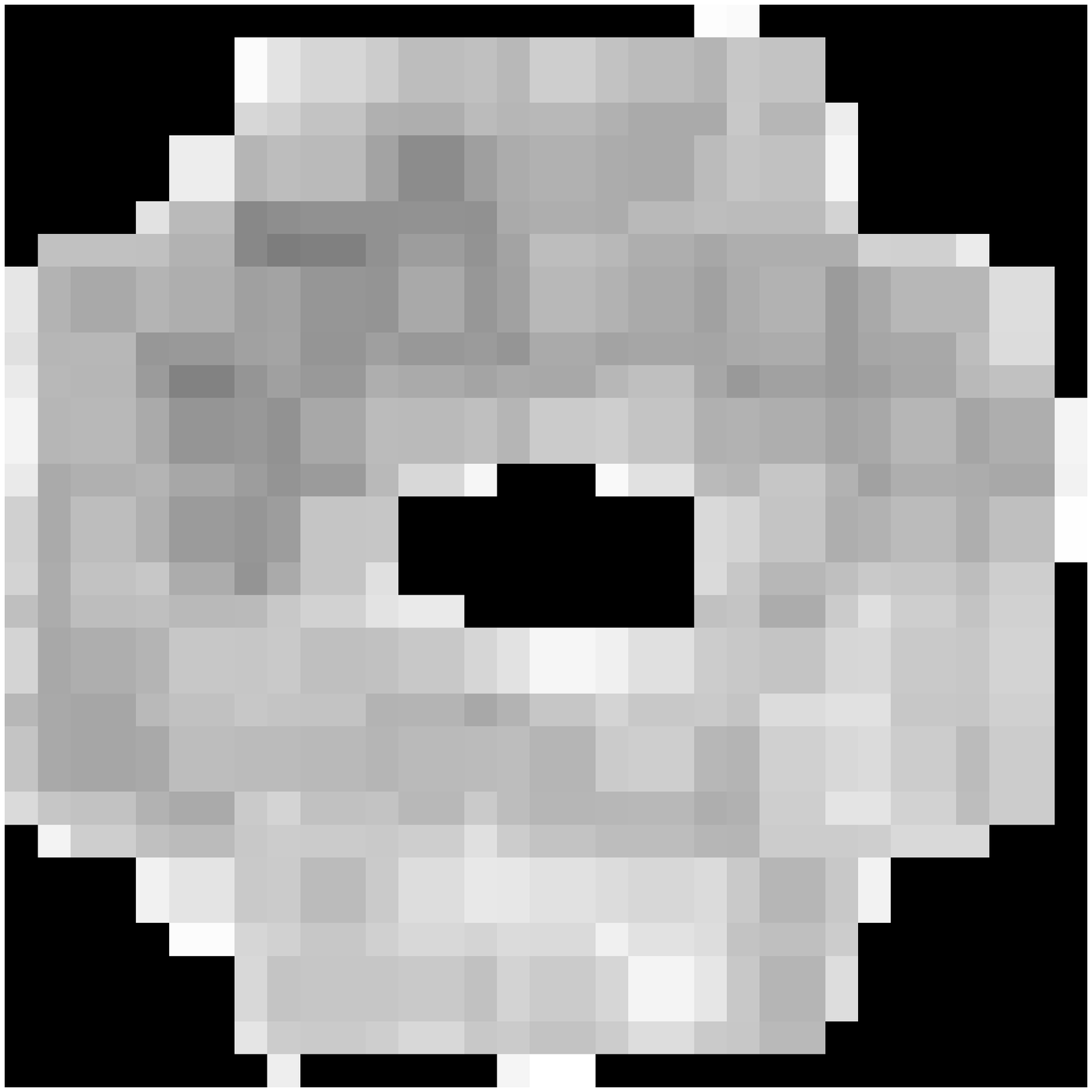}} at -20 50
\put {\includegraphics[height=4.0cm]{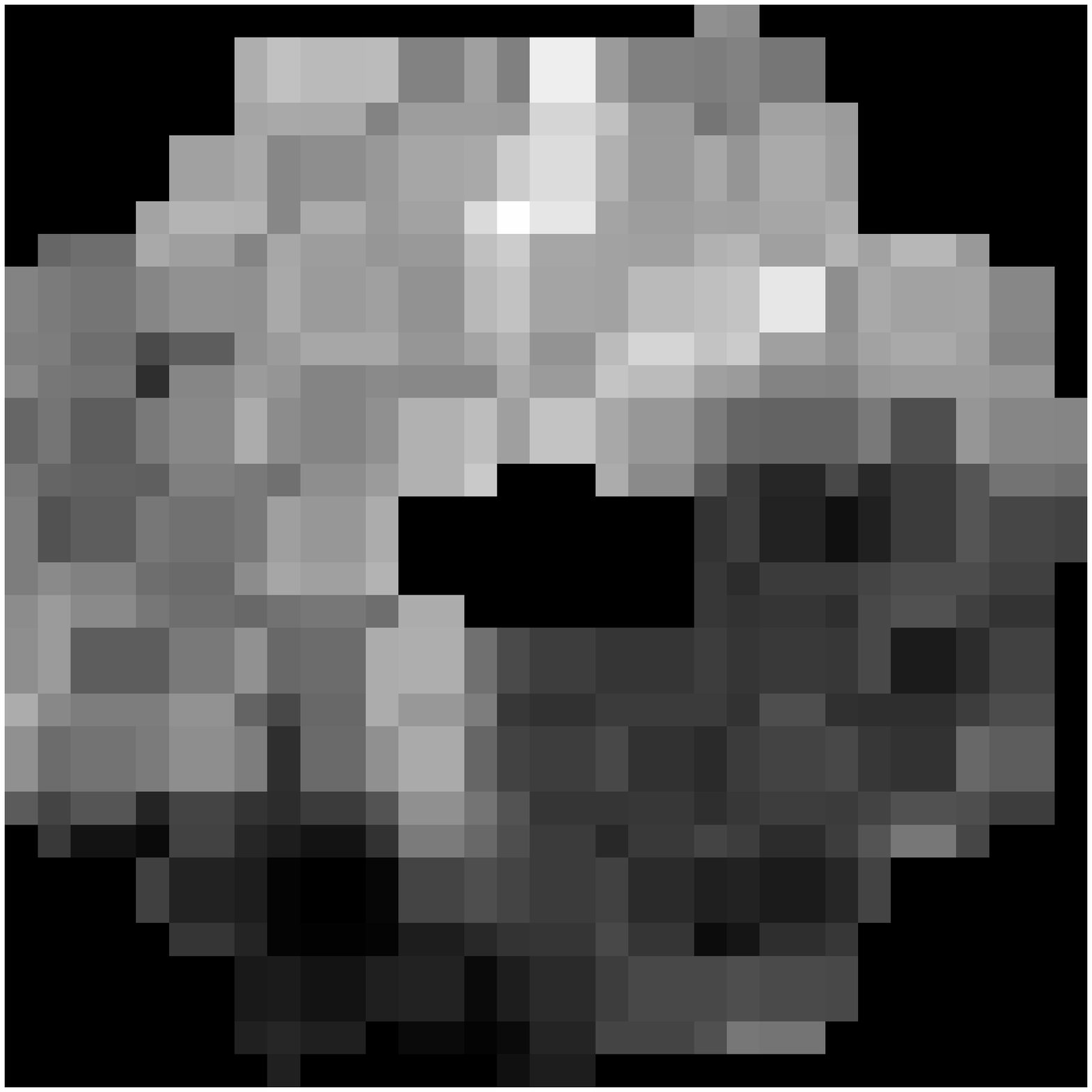}} at 25 50
\put {\includegraphics[height=8cm, angle=-90]{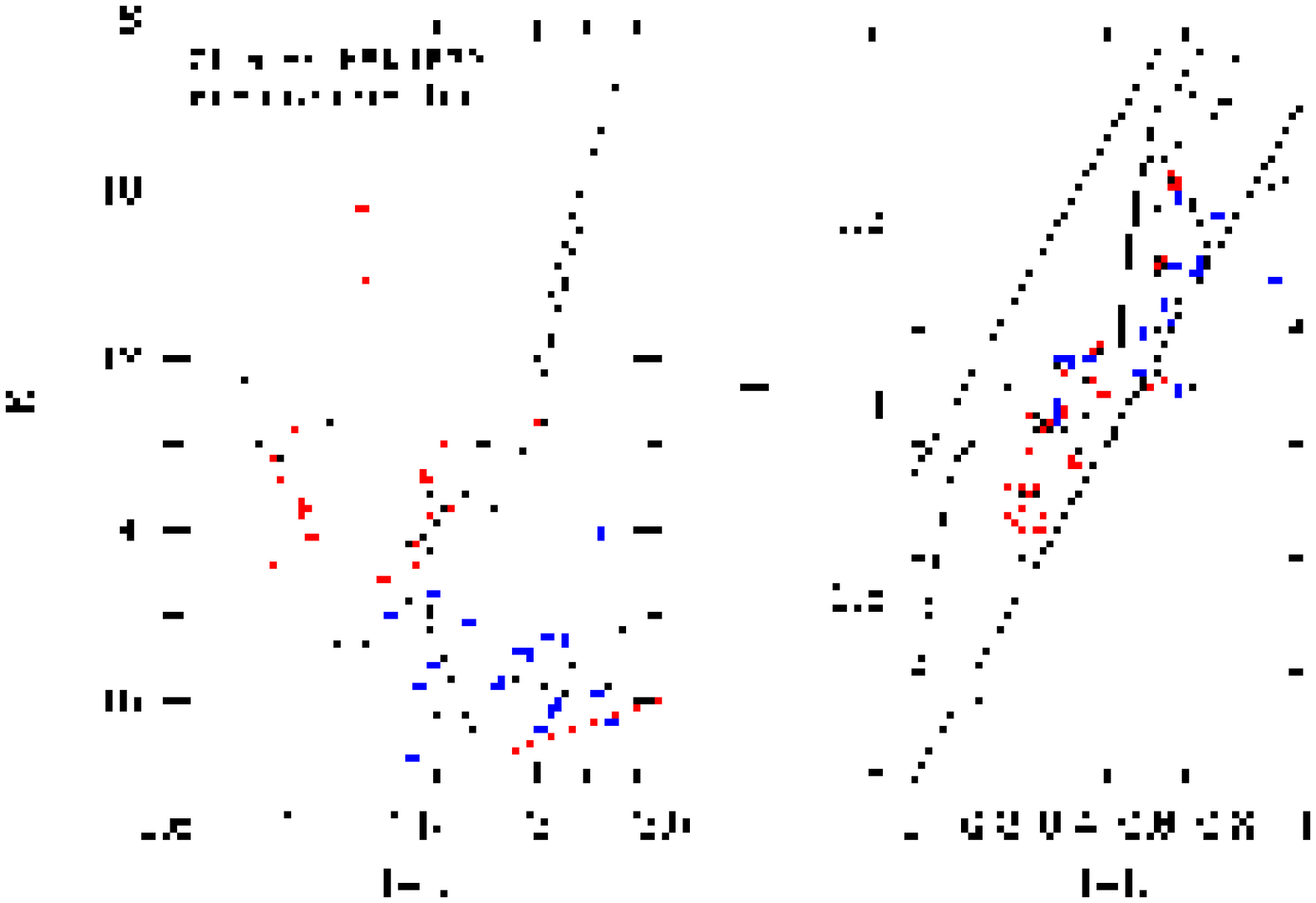}} at 0 0
\put {{\color{blue}\oval(30,30)}} at -25 58
\put {{\color{blue}\oval(30,30)}} at 20 58
\endpicture

\caption{\label{cl1527_jkk} As Fig.\,\ref{cl0002_jkk} but for the cluster
candidate FSR\,1527. The plotted isochrone has log(age)\,=\,9. See text for
other parameters. Clearly no isochrone can fit all the data.}  
\end{figure}

\subsection{FSR\,1527}

There is an indication of a cluster in the image of this region, which seems to
come from a slightly larger number of brighter stars. The SDM also shows an
increase in the stellar density at the position of the cluster candidate. In the
REM we can identify that the southern and western side of the mosaic are subject
to a lower amount of extinction than the rest of the field (see
Fig.\,\ref{cl1527_jkk}). We therefore choose the north-eastern part (outside the
cluster candidate area) of the mosaic as control field.

The remaining stars after the decontamination (about 25\,\%) could to some
extend be explained by an isochrone in the CMD. However, the scatter of the
stars along the reddening path in the CCD is inconsistent with this proposal,
i.e. with a constant reddening for all cluster stars. We conclude, that
FSR\,1527 is not a cluster. The isochrone in Fig.\,\ref{cl1527_jkk} is plotted
to clarify our argument that it cannot fit the data, and has the following
parameters: Z\,=\,0.019, $\beta$\,=\,1.6, log(age)\,=\,9, d\,=\,3.0\,kpc, and
A$_K$\,=\,1.0\,mag. 

\begin{figure}
\centering

\beginpicture
\setcoordinatesystem units <1mm,1mm> point at 0 0
\setplotarea x from 0 to 80 , y from 0 to 70
\put {\includegraphics[height=4.0cm]{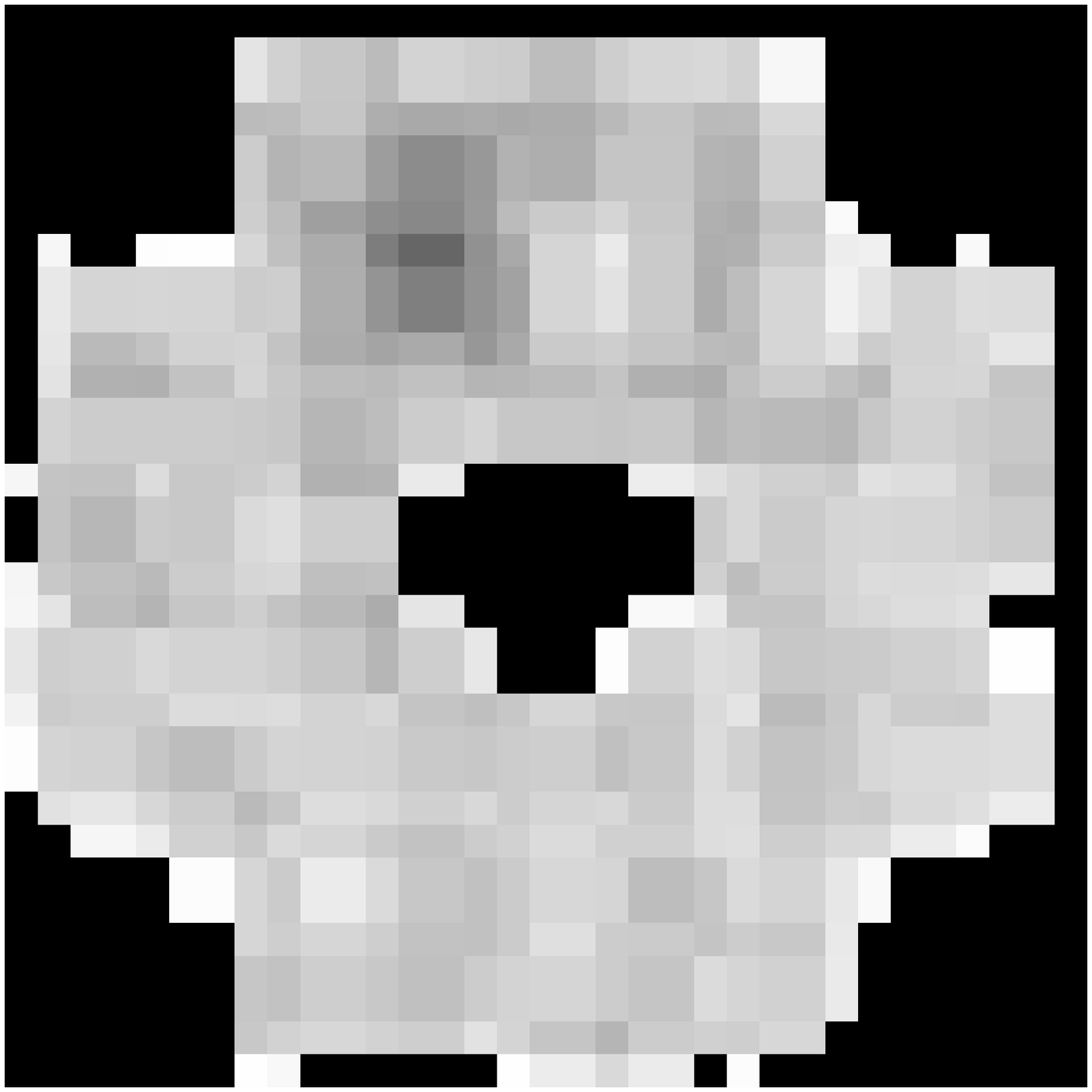}} at -20 50
\put {\includegraphics[height=4.0cm]{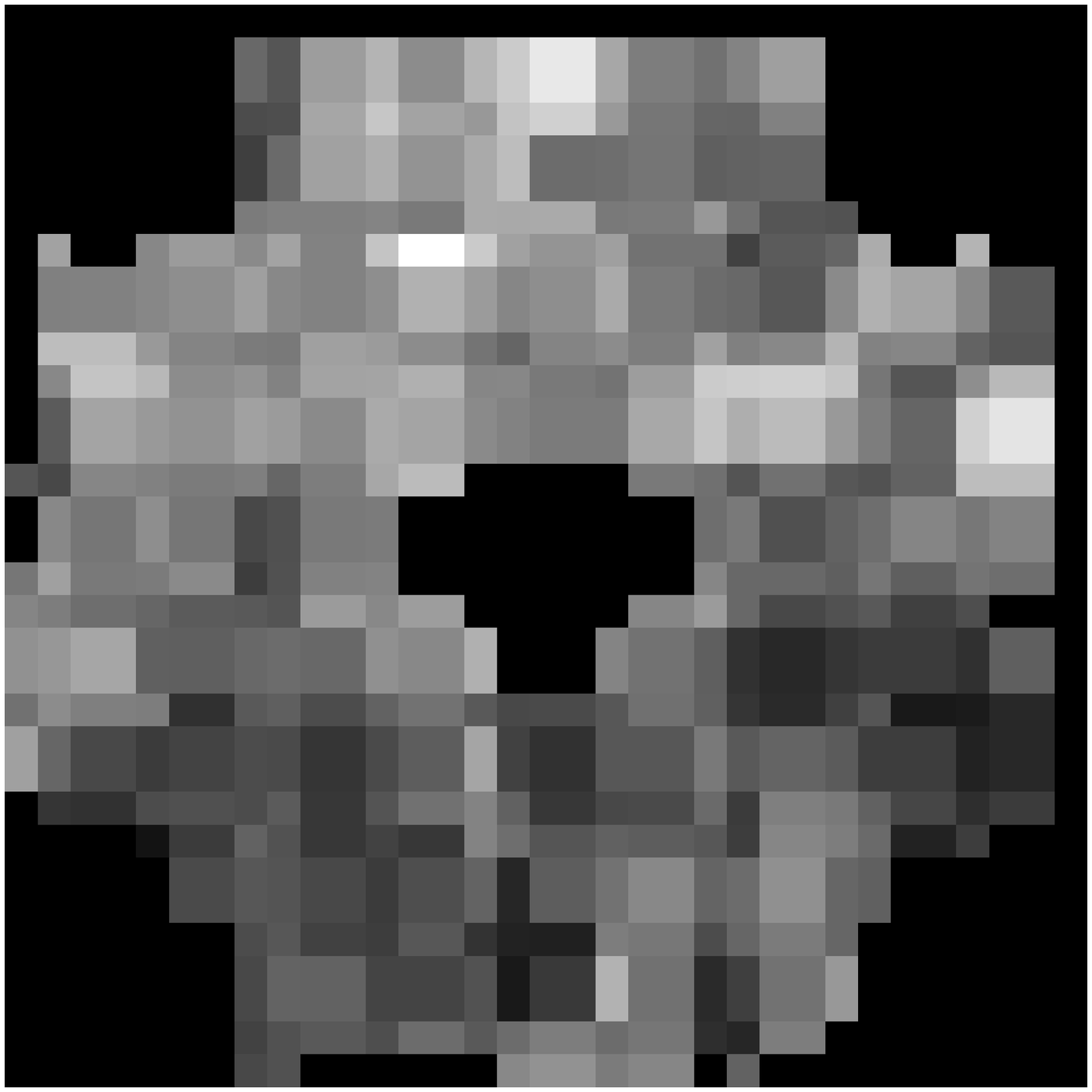}} at 25 50
\put {\includegraphics[height=8cm, angle=-90]{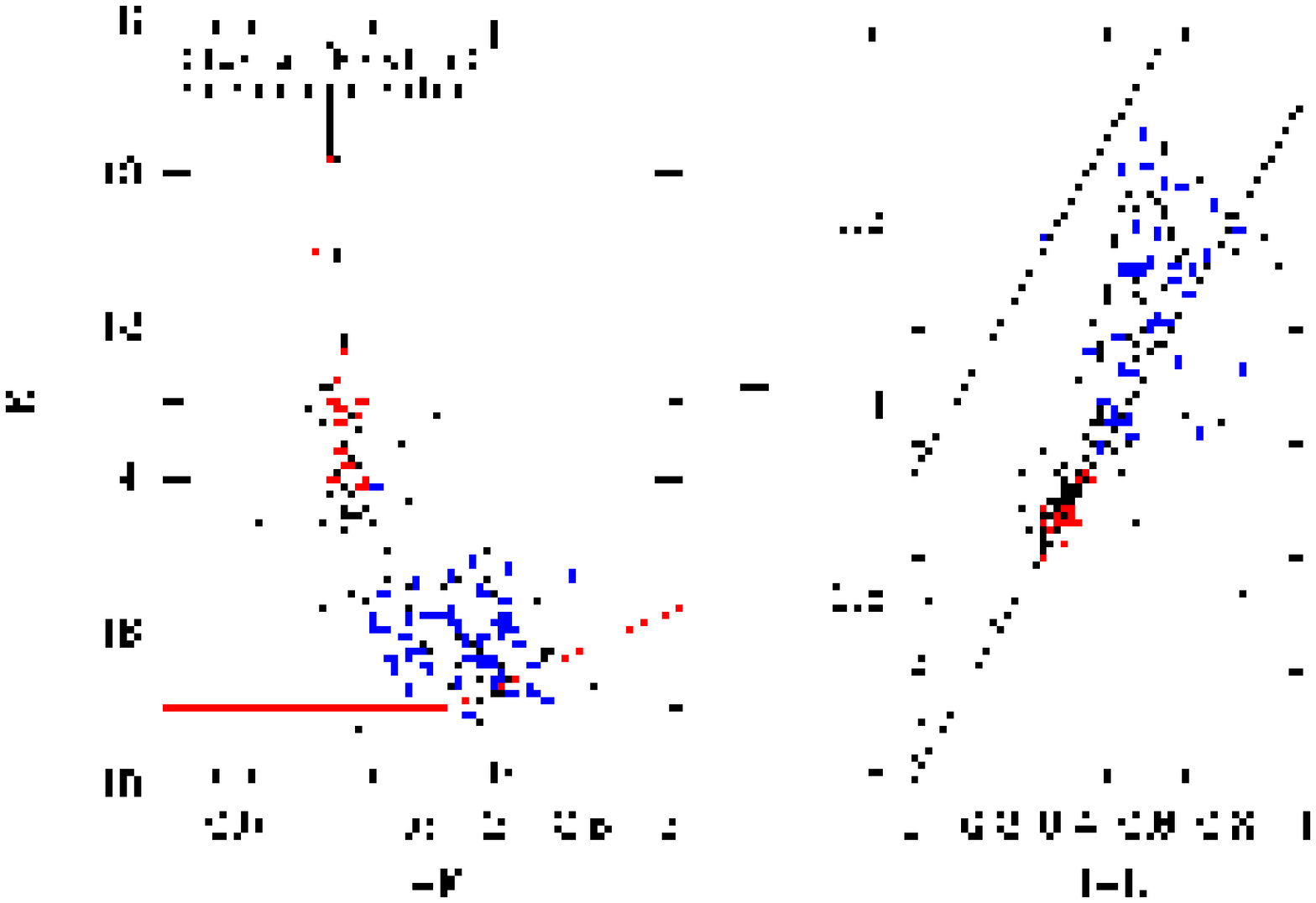}} at 0 0
\put {{\color{blue}\oval(30,30)}} at -25 58
\put {{\color{blue}\oval(30,30)}} at 20 58
\endpicture

\caption{\label{cl1530_jkk} As Fig.\,\ref{cl0002_jkk} but for the cluster
candidate FSR\,1530. The plotted isochrone has log(age)\,=\,6.6. See text for
other parameters.}  
\end{figure}

\subsection{FSR\,1530}

The image of the cluster candidate area shows a clear enhancement of the density
of stars, centred on a very bright object. This is confirmed by the SDM which
has a clear indication of an overdensity of stars at the cluster candidate
position. The REM shows that the southern half of the field is slightly less
influenced by extinction. Therefore, we positioned the control field just west
of the cluster candidate area. Furthermore, at the position of the cluster, the
median colour of the stars is much redder than in the field, indicating a
population of young stars, i.e. the presence of a young cluster.

The stars remaining after the decontamination (about 50\,\%) can be fit with an
isochrone of a young cluster (see Fig.\,\ref{cl1530_jkk}). The upper main
sequence stars form a very compact group in the CCD with colours lying on the
bottom of the reddening path, indicating spectral types earlier than A0. The
other stars can be interpreted as lower mass, (pre-main sequence) objects. The
best isochrone fit can be achieved using solar metallicity, an age of 4\,Myr, a
distance of 2.5\,kpc, and $A_K$\,=\,0.9\,mag. It is not fully possible to fit
the lower mass main sequence stars in the CCD perfectly using this isochrone.
The reason might be that a lower age is required, but Girardi et al. 
\cite{2002A&A...391..195G} do not provide isochrones for younger populations. 
{\bf As the age is small and possibly an upper limit, the determined distance is
also rather uncertain. A range from 2.0 to 3.5\,kpc would fit the data. The
K-band extinction is uncertain by 0.05\,mag.} The RDP is slightly hampered by
the presence of the bright star in the cluster centre. However, the profile
(Fig.\,\ref{krad_1530}) shows a small concentrated cluster with a core radius of
about 0.15\,pc.

We have remeasured the central coordinates of the cluster in our deeper images.
The central coordinates are RA\,=\,10:08:58.3, DEC\,=\,-57:17:11 (J2000), about
half an arcminute away from the values given in Froebrich et al.
\cite{2007MNRAS.374..399F}. The brightest  star in the cluster area,
[M81]\,I-296, projected close to the central coordinates is identified as an
H$_\alpha$ emission line star. Its 2MASS colours and brightness are
K\,=\,7.31\,mag and J$-$K\,=\,1.22\,mag, possibly making it a high mass cluster
member.

\begin{figure}
\centering

\beginpicture
\setcoordinatesystem units <1mm,1mm> point at 0 0
\setplotarea x from 0 to 80 , y from 0 to 70
\put {\includegraphics[height=4.0cm]{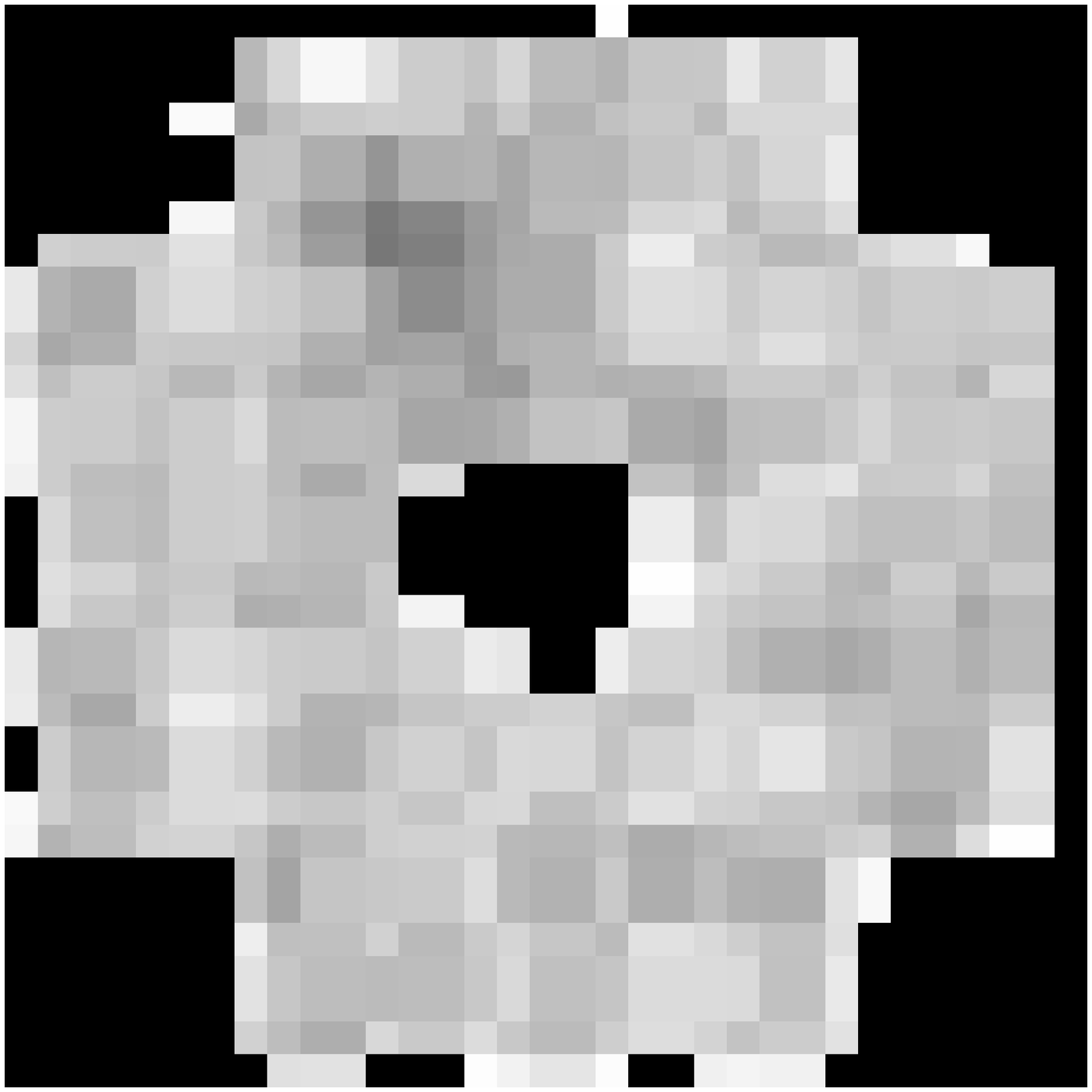}} at -20 50
\put {\includegraphics[height=4.0cm]{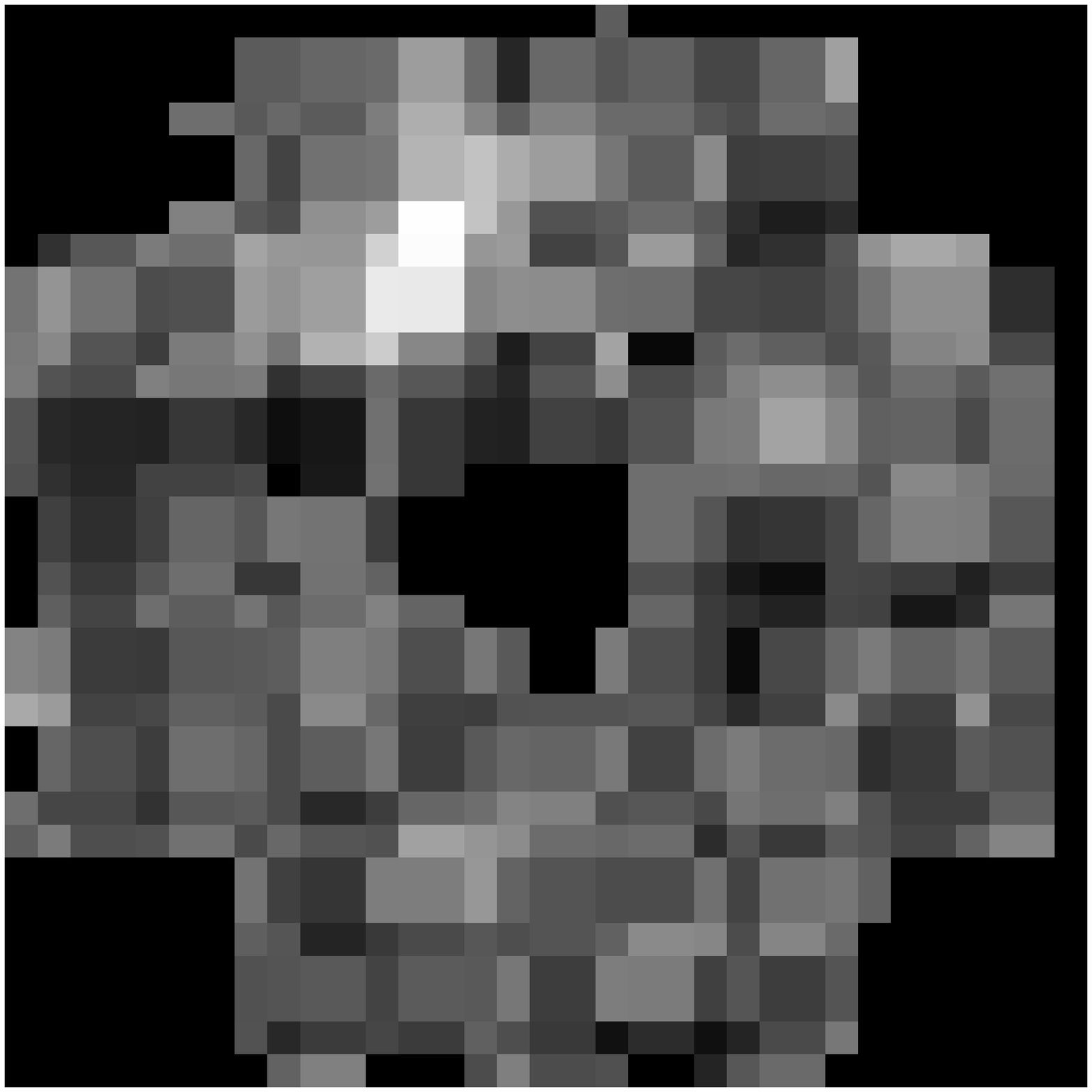}} at 25 50
\put {\includegraphics[height=8cm, angle=-90]{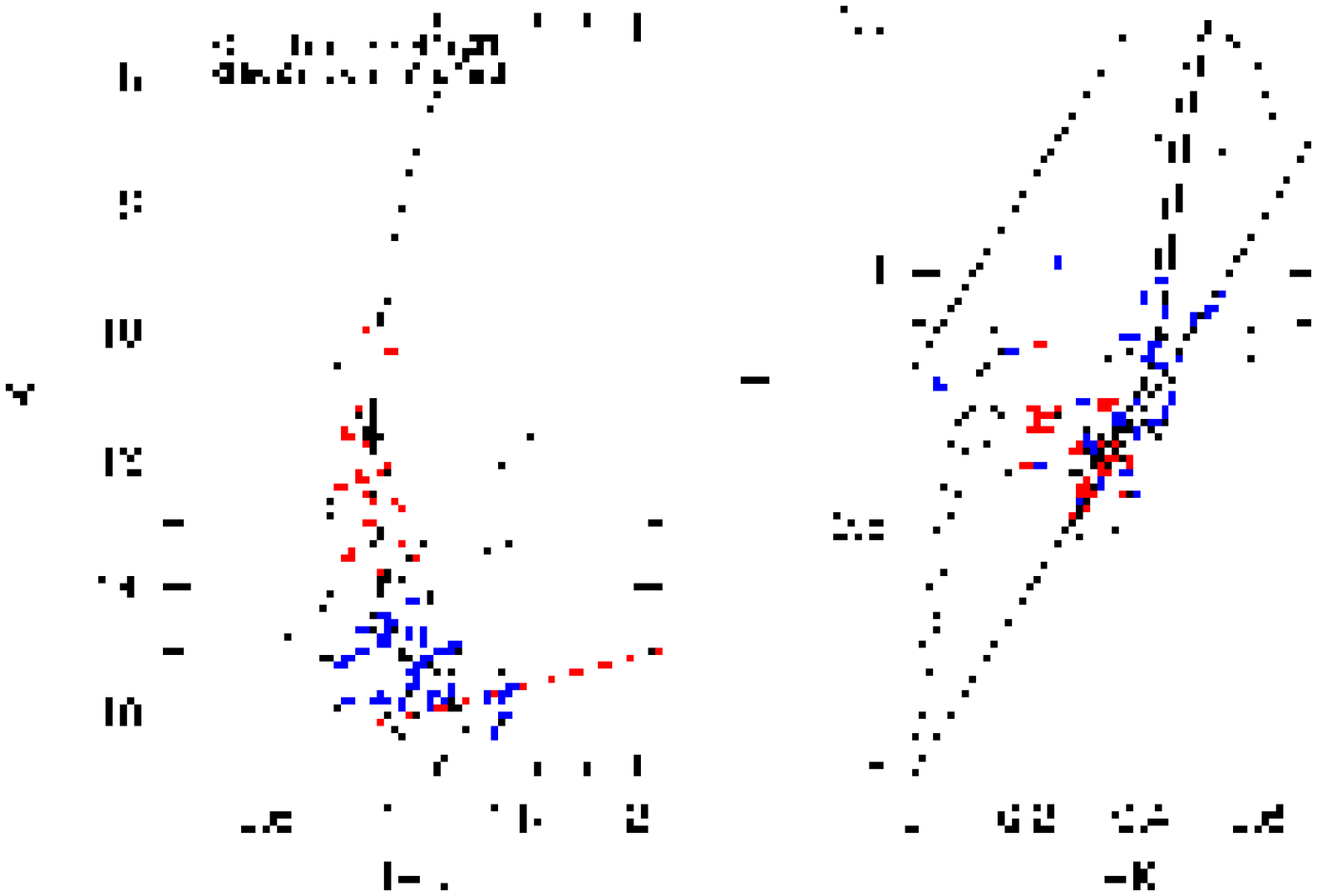}} at 0 0
\put {{\color{blue}\oval(30,30)}} at -25 58
\put {{\color{blue}\oval(30,30)}} at 20 58
\endpicture

\caption{\label{cl1570_jkk} As Fig.\,\ref{cl0002_jkk} but for the cluster
candidate FSR\,1570. The plotted isochrone has log(age)\,=\,6.9. See text for
other parameters. The $+$ in the CMD indicates the WRA\,751 as measured by
2MASS.}

\end{figure}

\subsection{FSR\,1570}

The image shows a clear concentrated cluster of stars around a very bright star.
This is also evident in the SDM. The REM shows that the colours of the stars in
the cluster are much redder than in the surrounding field. This indicates a
young cluster. The surrounding field shows no significant fluctuations (less
than 0.35\,mag A$_V$) in the stellar colours, hence the entire area outside the
cluster is used as a control field.

The cluster is known as Teutsch\,143a and has been investigated with optical
photometry by Pasquali et al. \cite{2006A&A...448..589P}, who identified it as
the birth cluster of the galactic luminous blue variable WRA\,751. The authors
determine a distance of 6\,kpc, an extinction of $A_V$\,=\,6.1\,mag, and an age
of above 4\,Myr.

After the decontamination about 40\,\% of the stars remain in the cluster field.
Using $\beta$\,=\,1.6 and an extinction of $A_K$\,=\,0.8\,mag, we can fit the
remaining stars with an 8\,Myr isochrone at a distance of 6\,kpc (see
Fig.\,\ref{cl1570_jkk}; in the CMD we have marked with a $+$ the position of
WRA\,751 obtained from 2MASS photometry since the star is saturated in our
images). The bright main sequence stars are all located at the bottom of the
reddening path, and are thus of spectral type earlier than A0, in agreement with
Pasquali et al. \cite{2006A&A...448..589P} who identified 24 stars with spectral
types earlier than B3 in the cluster region. {\bf As the cluster is very young,
the age estimate from the Girardi et al. \cite{2002A&A...391..195G} isochrones
is uncertain by a factor of two. The distance can be constrained within 1\,kpc,
and the K-band extinction within 0.05\,mag.} We have measured a core radius of
about 0.35\,pc for the cluster using the RDP in Fig.\,\ref{krad_1570}. However,
the profile is influenced by the presence of WRA\,751 near the centre,
preventing the detection of stars nearby.

\begin{figure}
\centering

\beginpicture
\setcoordinatesystem units <1mm,1mm> point at 0 0
\setplotarea x from 0 to 80 , y from 0 to 70
\put {\includegraphics[height=4.0cm]{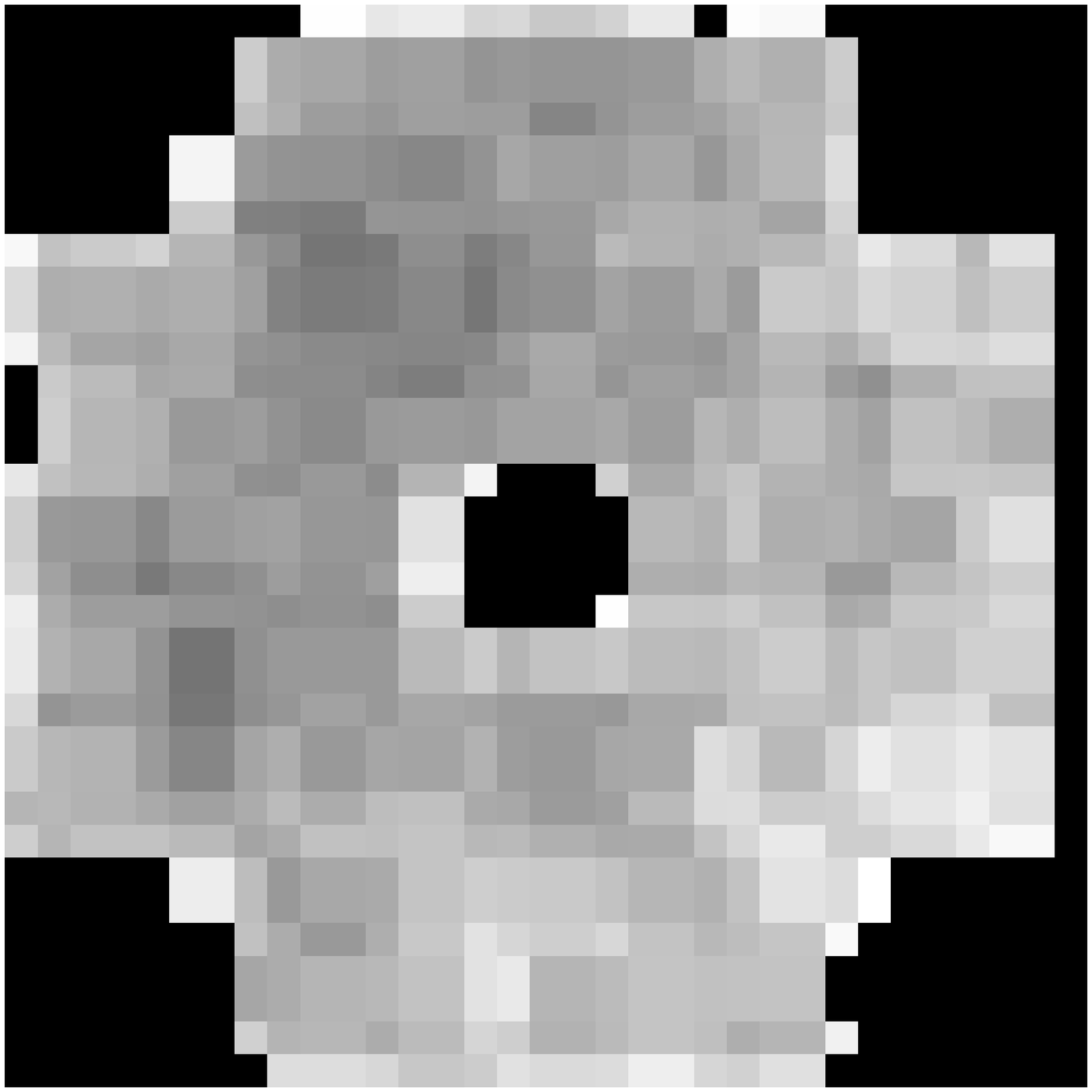}} at -20 50
\put {\includegraphics[height=4.0cm]{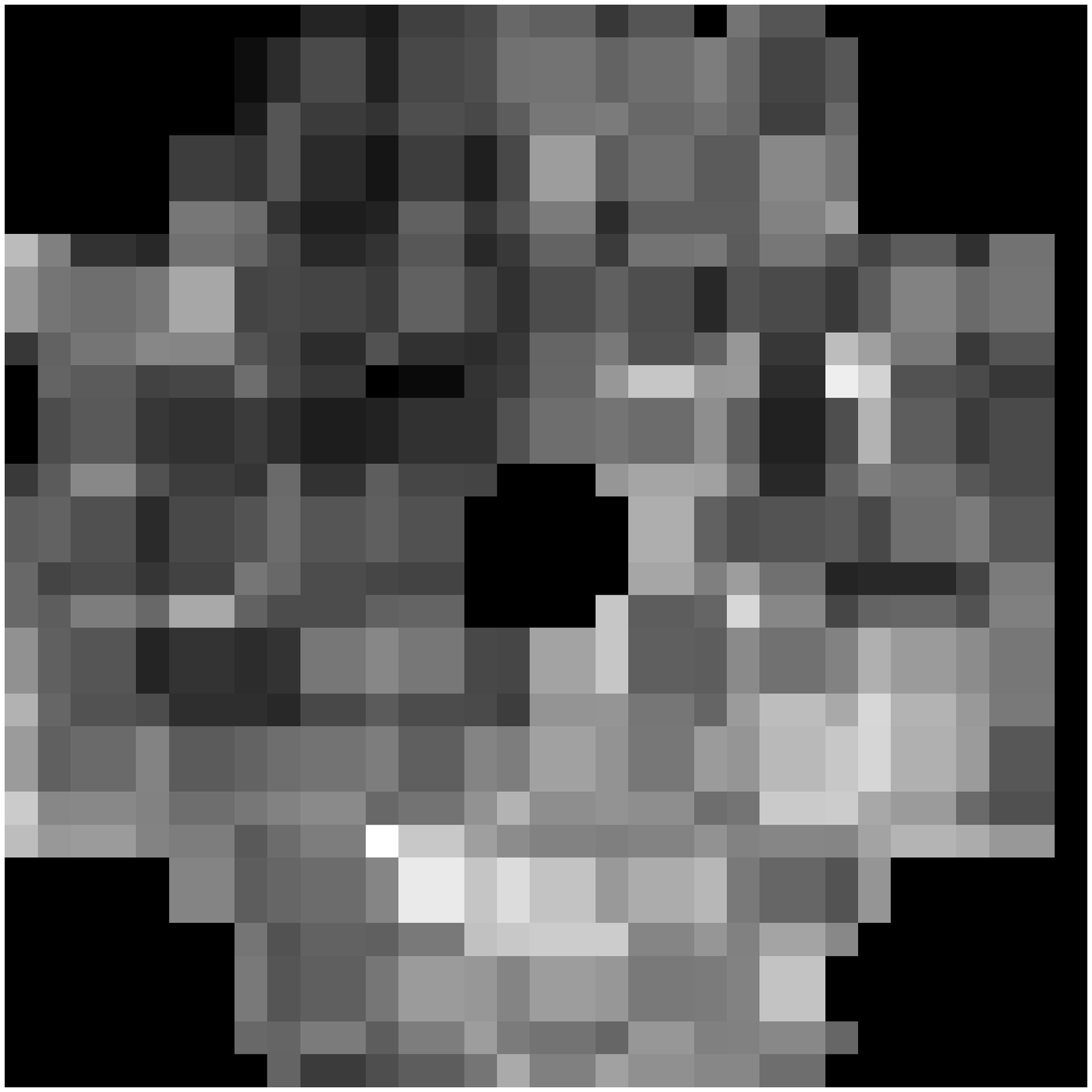}} at 25 50
\put {\includegraphics[height=8cm, angle=-90]{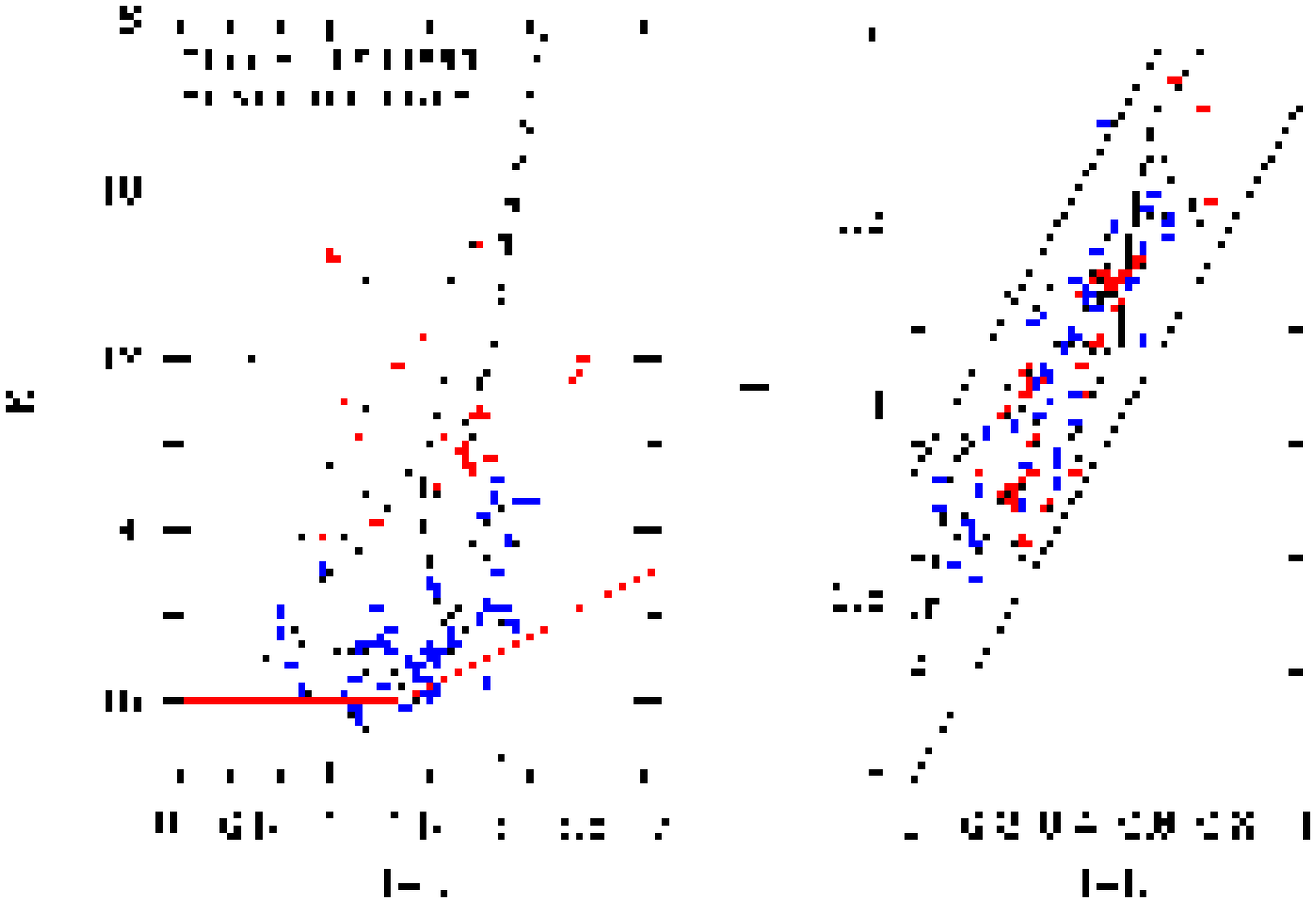}} at 0 0
\put {{\color{blue}\oval(30,30)}} at -25 58
\put {{\color{blue}\oval(30,30)}} at 20 58
\endpicture

\caption{\label{cl1659_jkk} As Fig.\,\ref{cl0002_jkk} but for the cluster
candidate FSR\,1659. The plotted isochrone has log(age)\,=\,9. See text for
other parameters. Clearly no isochrone can fit all the data.}  
\end{figure}

\subsection{FSR\,1659}

The impression from the image of FSR\,1659 is that there is an overdensity of
stars at the position of the cluster candidate. This is confirmed in the SDM.
However, as can be seen from the REM of the mosaic, the western and southern
part of the area are influenced by increased amounts of extinction. Hence, the
overdensity might just be caused by this effect. Therefore we have chosen the
eastern part of the mosaic as the control field.

After the decontamination about 30\,\% of the stars remain in the cluster
candidate area. However, their distribution in the CMD and CCD (see
Fig.\,\ref{cl1659_jkk}) cannot be fit by a single isochrone. The isochrone in
Fig.\,\ref{cl1659_jkk} is plotted to clarify this argument and has the following
parameters: Z\,=\,0.019, $\beta$\,=\,1.6, log(age)\,=\,9, d\,=\,2.0\,kpc, and
A$_K$\,=\,1.0\,mag. We conclude that FSR\,1659 is not a star cluster.

\begin{figure}
\centering

\beginpicture
\setcoordinatesystem units <1mm,1mm> point at 0 0
\setplotarea x from 0 to 80 , y from 0 to 70
\put {\includegraphics[height=4.0cm]{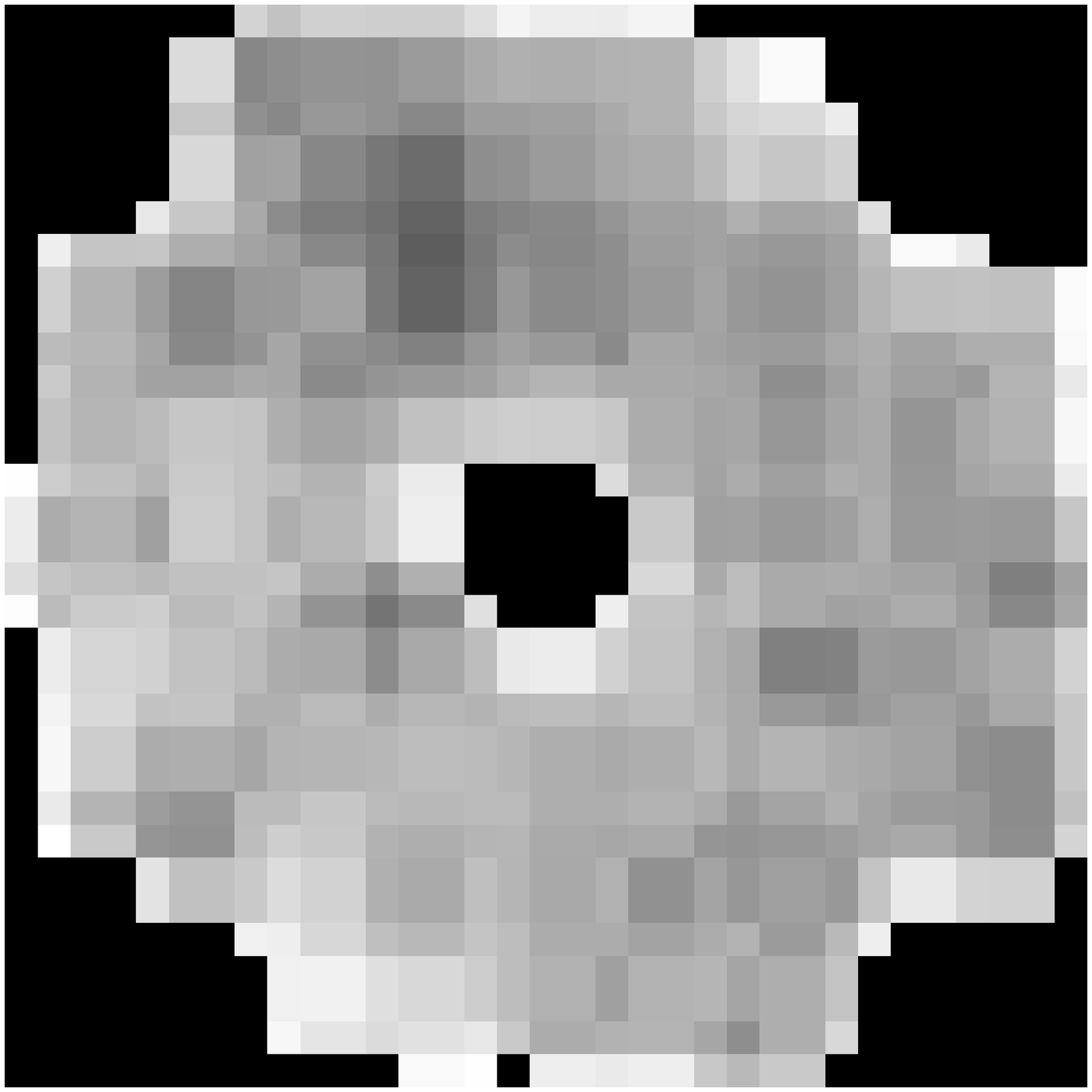}} at -20 50
\put {\includegraphics[height=4.0cm]{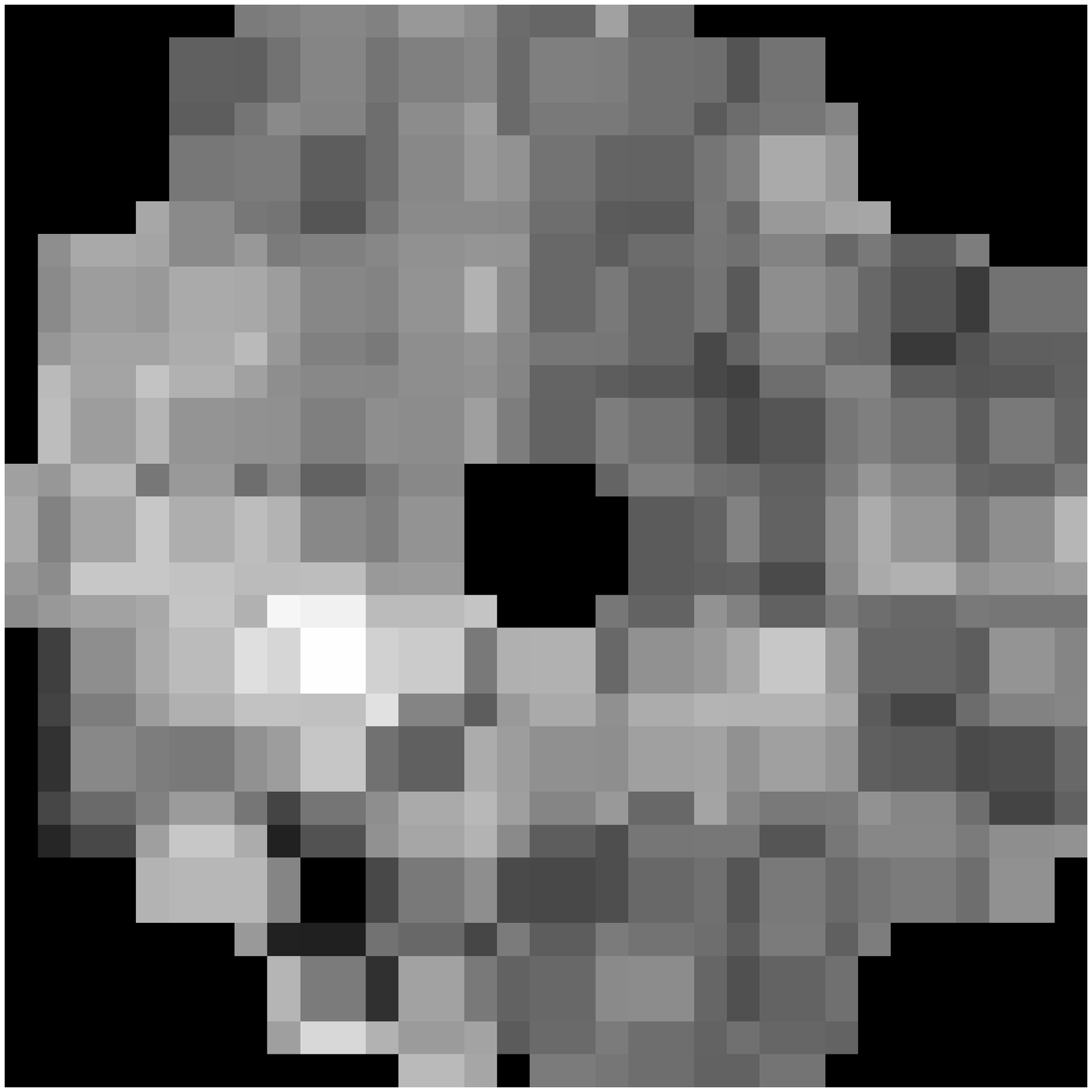}} at 25 50
\put {\includegraphics[height=8cm, angle=-90]{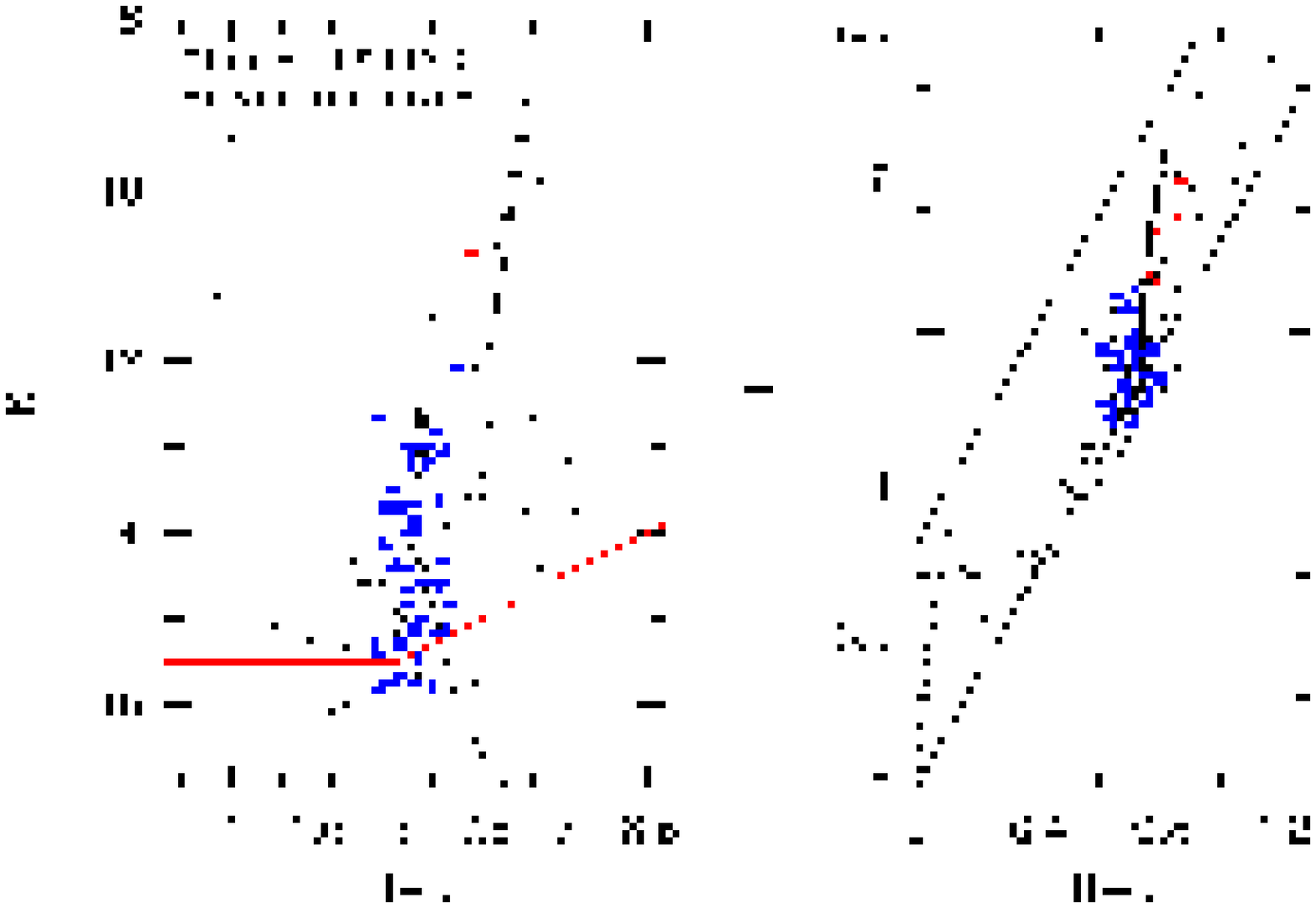}} at 0 0
\put {{\color{blue}\oval(30,30)}} at -25 58
\put {{\color{blue}\oval(30,30)}} at 20 58
\endpicture

\caption{\label{cl1712_jkk} As Fig.\,\ref{cl0002_jkk} but for the cluster
candidate FSR\,1712. The plotted isochrone has log(age)\,=\,8.9. See text for
other parameters.}  
\end{figure}

\subsection{FSR\,1712}

The image of FSR\,1712 shows clearly a concentrated cluster of stars. This is
confirmed in the SDM. The REM (see Fig.\,\ref{cl1712_jkk}) of the region shows
that the south-western part of the mosaic suffers from an increased value of
extinction. The rest of the field shows no significant fluctuations. We have
chosen the western part of the mosaic as the control area.

The decontamination procedure leaves about 40\,\% of the stars. They can be fit
by a main sequence in the CMD (see Fig.\,\ref{cl1712_jkk}). The brighter stars 
lie close to the bottom of the reddening path, hence should be stars of spectral
type A of F. Since this implies a relatively recent formation and there are no
or very few red giants to determine the metallicity otherwise, we use solar
metallicity isochrones.

We can fit the main sequence in the CMD and CCD using $\beta$\,=\,1.6, an age of
0.8\,Gyr, a distance of 1.8\,kpc, and an extinction of $A_K$\,=1.4\,mag towards
the cluster. There seem to be no or only a very small number of possible red
giants in the cluster. Like for FSR\,0088, they seem to be shifted towards
slightly redder colours in the CCD. The reason in this case might be that these
few objects are actually not related to the cluster and are background red
giants. {\bf This small number of giants also influences the accuracy of the age
estimate, which is not better than a factor of two. The distance is accurate
within 300\,pc and the K-band extinction within 0.1\,mag.} A core radius of
0.2\,pc is found for the cluster (Fig.\,\ref{krad_1712}), confirming the
concentrated appearance in the picture.

We have remeasured the central coordinates of the cluster in our deeper images.
The central coordinates are RA\,=\,15:54:46.3, DEC\,=\,-52:31:47 (J2000), about
one and a half arcminute south of the values given in Froebrich et al.
\cite{2007MNRAS.374..399F}. There are two ROSAT sources about 4' north and
south-east of the cluster. It is not known if they are related to the cluster,
but there are in total only 3 ROSAT sources within half a degree around the
cluster coordinates.

\begin{figure}
\centering

\beginpicture
\setcoordinatesystem units <1mm,1mm> point at 0 0
\setplotarea x from 0 to 80 , y from 0 to 70
\put {\includegraphics[height=4.0cm]{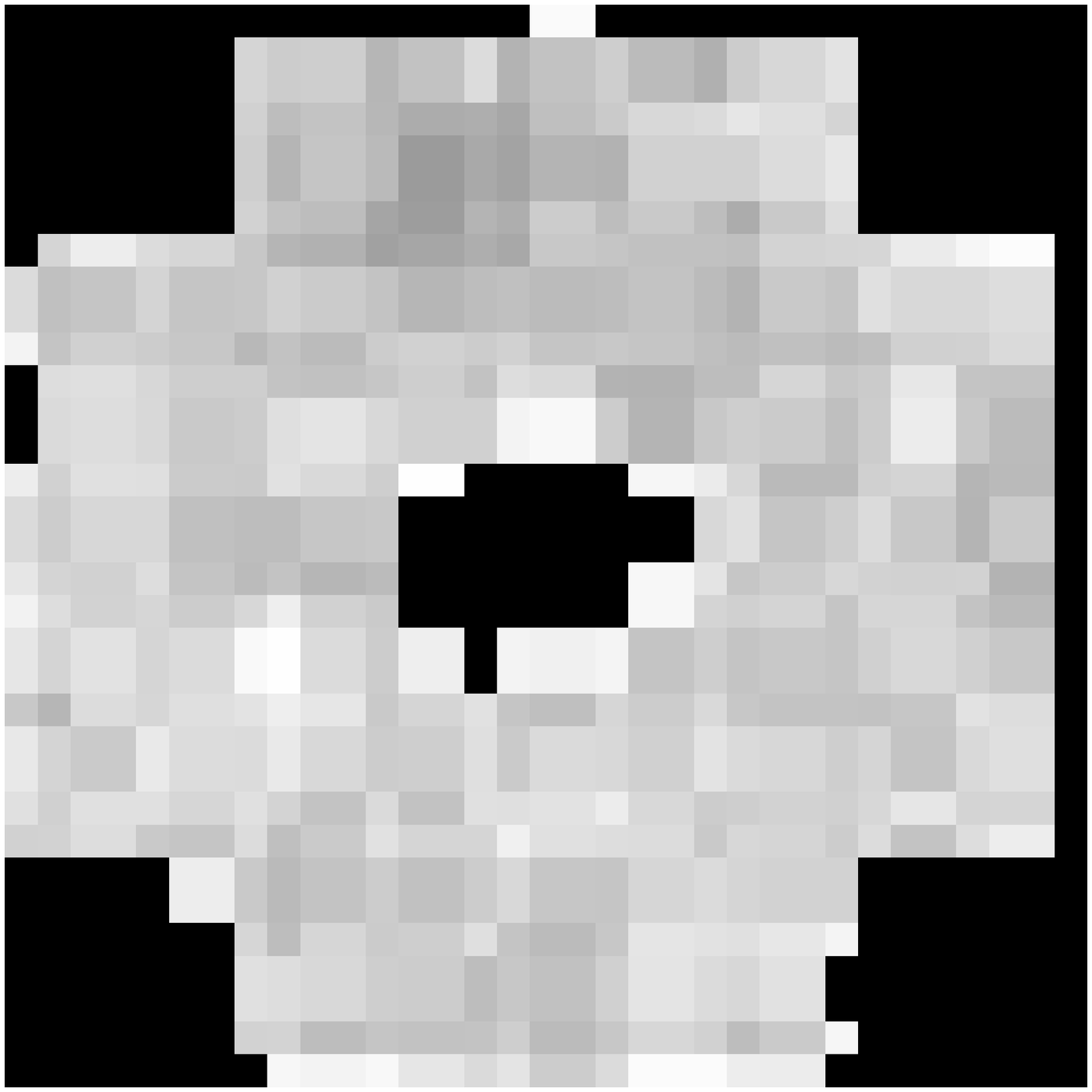}} at -20 50
\put {\includegraphics[height=4.0cm]{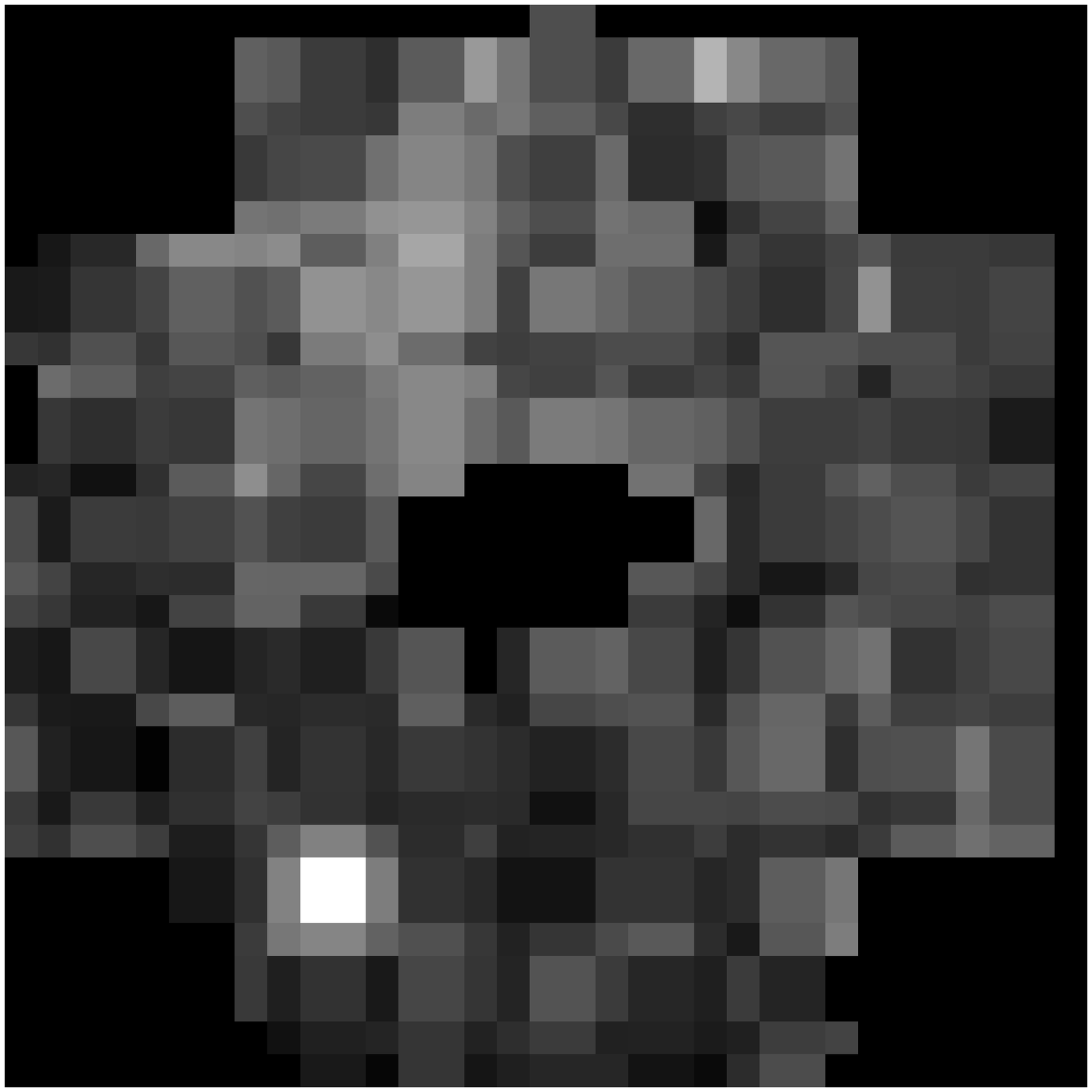}} at 25 50
\put {\includegraphics[height=8cm, angle=-90]{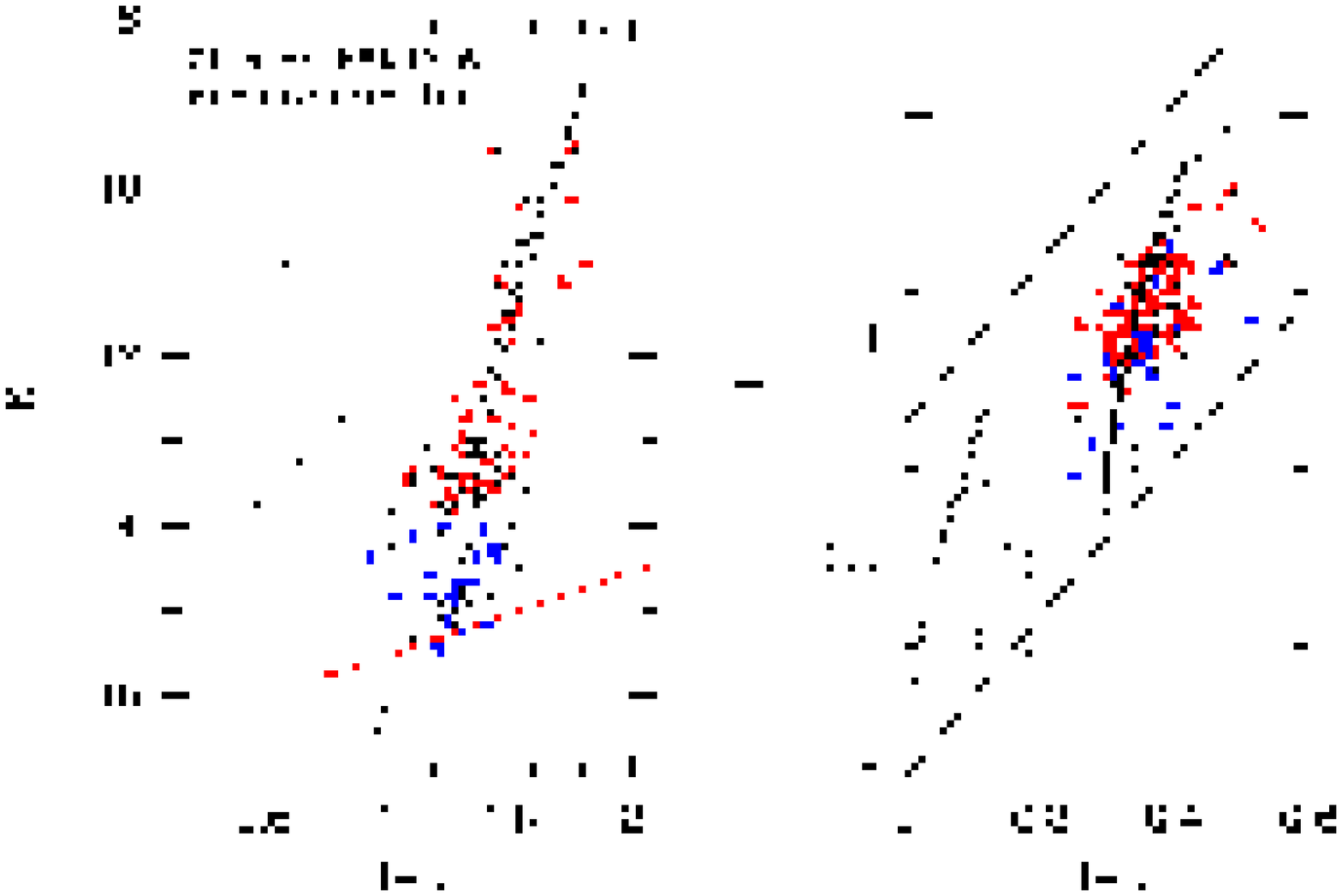}} at 0 0
\put {{\color{blue}\oval(30,30)}} at -25 58
\put {{\color{blue}\oval(30,30)}} at 20 58
\endpicture

\caption{\label{cl1716_jkk} As Fig.\,\ref{cl0002_jkk} but for the cluster
candidate FSR\,1716. The plotted isochrone has log(age)\,=\,9.3. See text for
other parameters.}  
\end{figure}

\subsection{FSR\,1716}

The image of this object shows a clear overdensity of stars, which is also
confirmed in the SDM. The REM shows no significant A$_V$ fluctuation across the
field, except in the cluster candidate region, where the average colour of the
stars seems to be redder than in the surrounding field (the high extinction peak
in the south-east is caused by a very bright star). The redder colour of stars
in this area should be intrinsic to the stars, since it comes along with an
increased star density, contradicting a locally enhanced extinction due to a
small cloud, which would decrease the star density. We hence use the entire
field except the cluster candidate region as control area.

After the decontamination about 30\,\% of the stars remain. Their CMD can be
interpreted as a well populated giant branch (see Fig.\,\ref{cl1716_jkk}). There
are two peaks in the K-band luminosity function (see Fig.\,\ref{klum_1716}). One
at about K\,=\,13.1\,mag, the other at K\,=\,13.7\,mag. We interpret the former
as the core helium burning objects. Note, that if we choose the second peak as
the core helium burning objects, the determined distance to the cluster would
increase by a factor of 1.3. The slope of the giant branch can be best fit using
a metallicity of Z\,=\,0.004. A lower metallicity results in a too steep slope
of the RGB and a higher value in a too shallow slope. {\bf However, a range of
Z\,=\,0.001 to 0.008 can in principle explain the data.} The RGB stars also form
a compact group in the CCD, where the brighter stars show a smaller scatter in
the colours than the fainter stars, in agreement with the photometric
uncertainties.

We can fit an isochrone to the RGB in the CMD and CCD using the above mentioned
metallicity, $\beta$\,=\,1.6, a distance of 7\,kpc, and an extinction of
$A_K$\,=\,0.57\,mag. Since we do not detect any main sequence stars we cannot
determine the age of the cluster. If we use our completeness limit as an
indicator for how faint the main sequence stars need to be in order that we
cannot detect them, the age of the cluster has to be at least 2\,Gyr. Virtually
no change in the quality of the fit is seen when using ages of 10\,Gyr or above.
{\bf The core helium burning objects allow to estimate the cluster distance
within 500\,pc. However, the upper limit for the age means that the cluster
could be as close as 5\,kpc (if the age is 12\,Gyr). As for the other clusters
the K-band extinction is uncertain by 0.05\,mag.}

The well populated giant branch, its low metallicity and the clusters position
close to the Galactic Center, suggest that this object might indeed be much
older. It could well be another (Palomar type?) globular cluster, similar to
FSR\,0190 (Froebrich et al. \cite{2008MNRAS.383L..45F}). The core radius of the
object is determined to 0.9\,pc using the RDP (Fig.\,\ref{krad_1716}). 

We have remeasured the central coordinates of the cluster in our deeper images.
The central coordinates are RA\,=\,16:10:29.0, DEC\,=\,-53:44:48 (J2000), about
one arcminute north-east of the values given in Froebrich et al.
\cite{2007MNRAS.374..399F}. There is an IRAS source (detected at 100\,$\mu$m
only) about 2.7' south of the cluster centre which seems, however, unrelated to
the cluster.

\begin{figure}
\centering

\beginpicture
\setcoordinatesystem units <1mm,1mm> point at 0 0
\setplotarea x from 0 to 80 , y from 0 to 70
\put {\includegraphics[height=4.0cm]{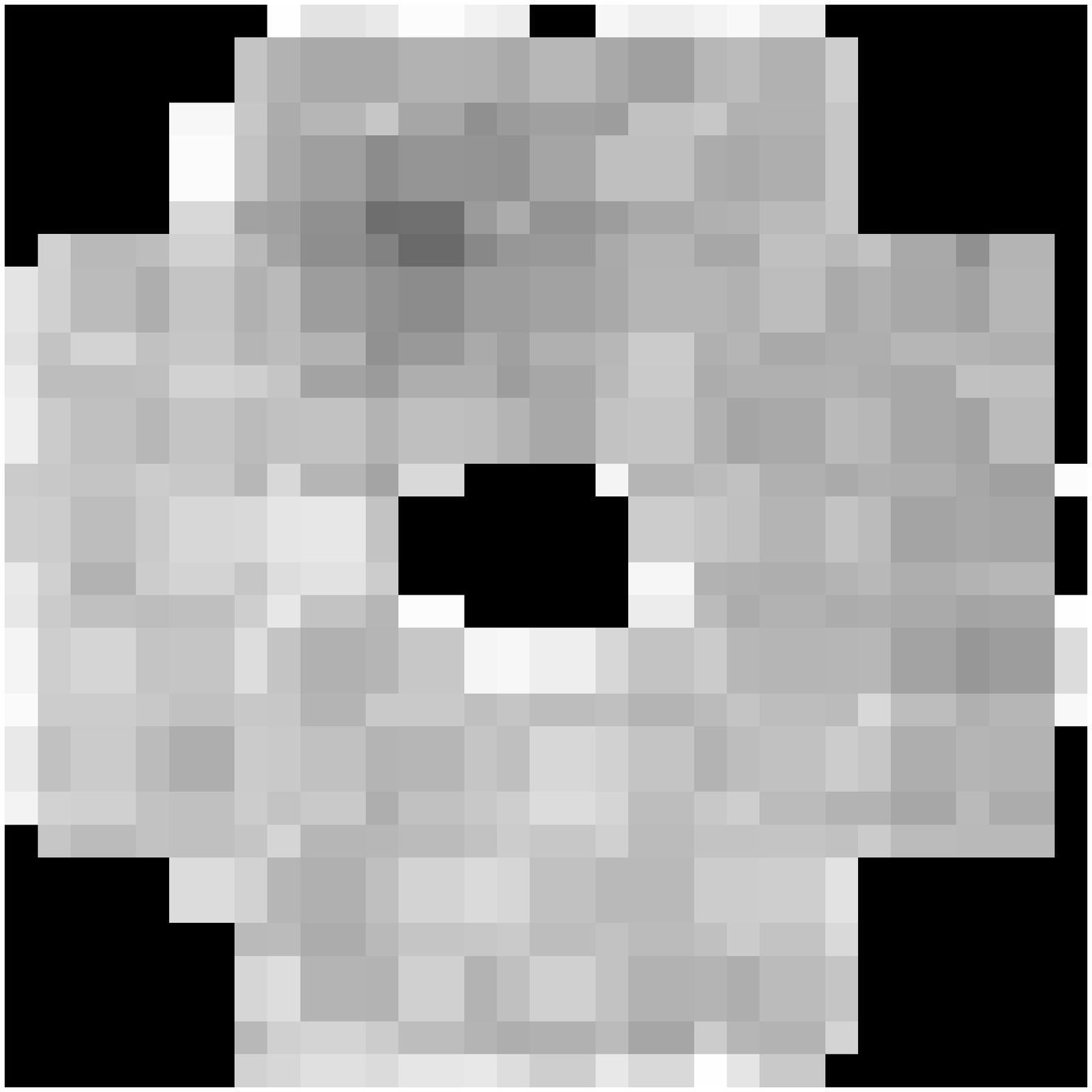}} at -20 50
\put {\includegraphics[height=4.0cm]{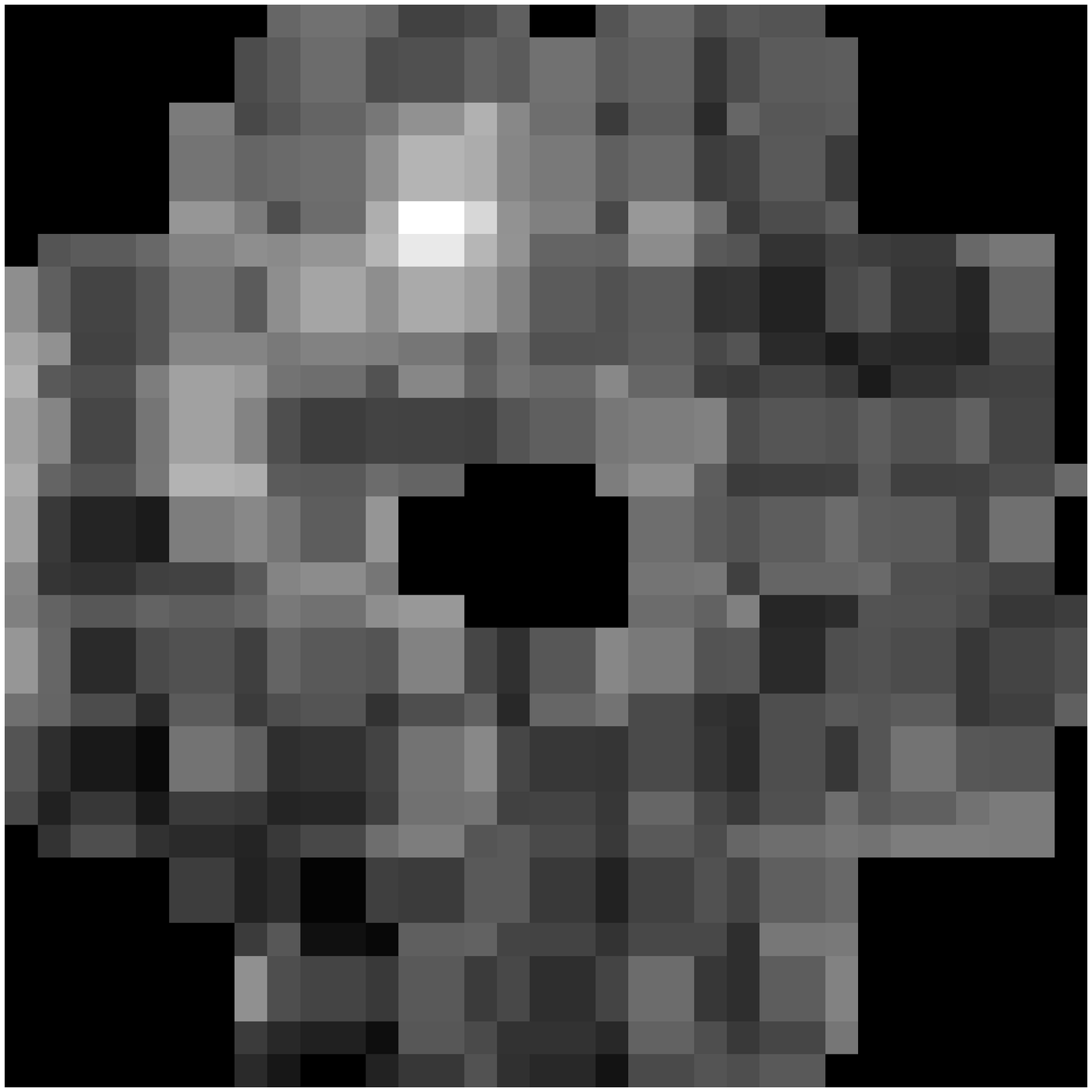}} at 25 50
\put {\includegraphics[height=8cm, angle=-90]{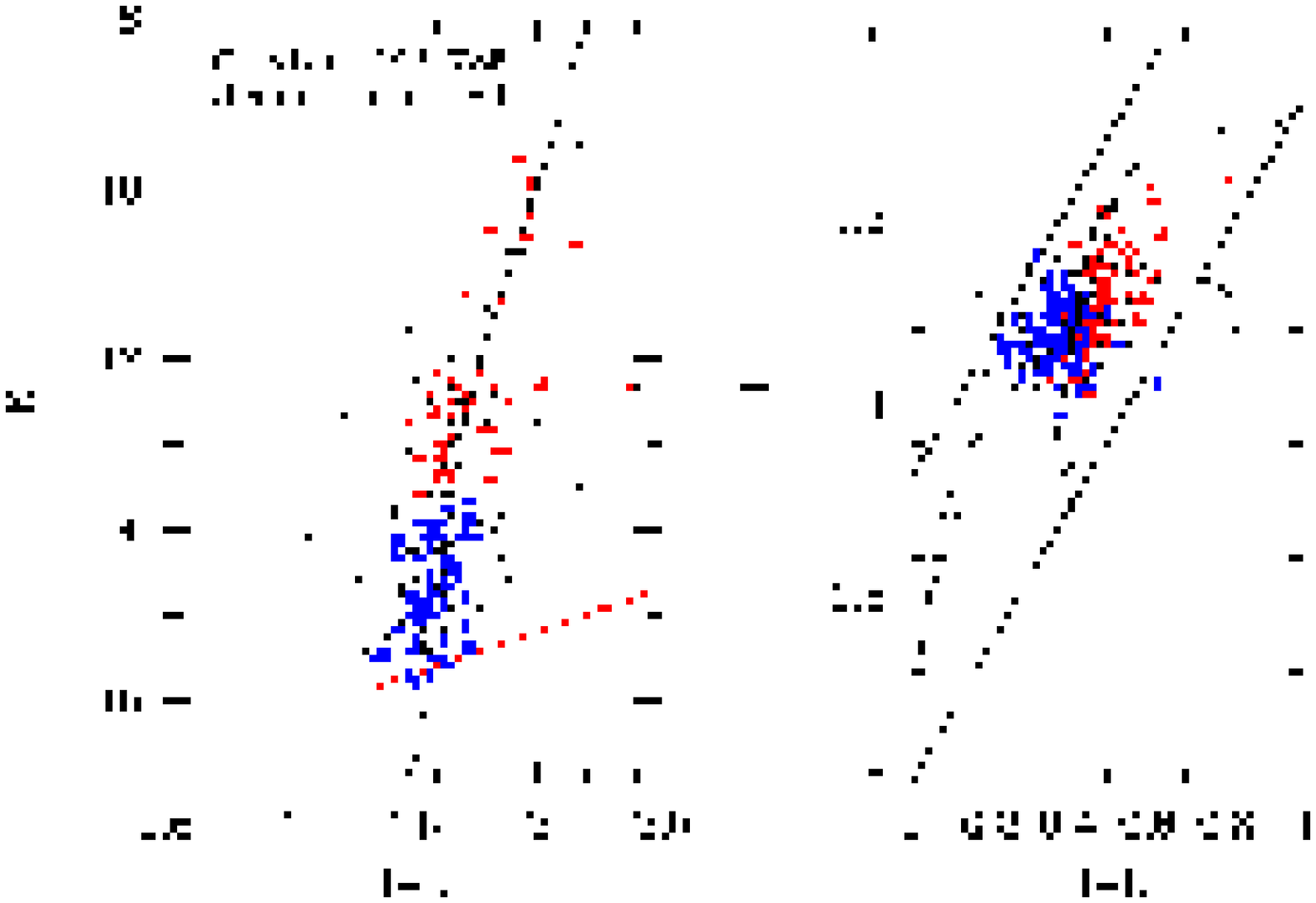}} at 0 0
\put {{\color{blue}\oval(30,30)}} at -25 58
\put {{\color{blue}\oval(30,30)}} at 20 58
\endpicture

\caption{\label{cl1735_jkk} As Fig.\,\ref{cl0002_jkk} but for the cluster
candidate FSR\,1735. The plotted isochrone has log(age)\,=\,9.9. See text for
other parameters.}  
\end{figure}

\subsection{FSR\,1735}

Here we re-analyse the data of FSR\,1735, already presented as a globular
cluster candidate in Froebrich et al. \cite{2007MNRAS.377L..54F}. This is to
ensure a homogeneous analysis and interpretation of all the cluster candidates
observed in this project. We re-analyse FSR\,1735 using the entire field of
observations, compared to just the small region around the cluster used in
Froebrich et al. \cite{2007MNRAS.377L..54F}. The cluster image clearly shows a
compact and populated cluster of stars. The SDM verifies this. In the REM we see
that the colour of the stars in the cluster area is redder than in the field,
most probably caused by the fact the we only see red giant cluster stars. This
is very similar to the above discussed maps of FSR\,1716. The south-eastern
corner of the mosaic seems to possess a slightly smaller number of stars. We
hence chose the western side of the image as the control field.

About 40\,\% to 45\,\% of the stars in the cluster area remain after the
decontamination procedure. The decontaminated CMD of the cluster shows a well
populated RGB/AGB (see Fig.\,\ref{cl1735_jkk}). There is a peak in the K-band
luminosity function at about K\,=\,14\,mag (see Fig.\,\ref{klum_1735}), which is
interpreted as the core helium burning objects. The slope of the RGB is best fit
using a metallicity of Z\,=\,0.004, {\bf but as for FSR\,1716 a range from
Z\,=\,0.001 to 0.008 can in principle explain the data.} There are no main
sequence stars detected, hence the age cannot be constrained. Using the
detection limits of our photometry, we find that the age has to be larger than
2\,Gyr. Indeed, we can increase this limit to an age above 8\,Gyr, since such an
age can better explain the CMD and CCD simultaneously. Slightly depending on the
age, the distance to the cluster is 8.5\,kpc and the reddening
$A_K$\,=\,0.7\,mag. The used dust properties are $\beta$\,=\,1.6. {\bf The
uncertainty for the distance estimate is 500\,pc, including the fact that we
only have a lower limit for the age. The K-band extinction can be estimated
within 0.05\,mag.} These results are in agreement with the parameters published
for FSR\,1735 in Froebrich et al. \cite{2007MNRAS.377L..54F}. The differences in
the determined parameters are entirely due to the assumed age of 12\,Gyr in the
earlier publication. We have also remeasured the core radius of the cluster.
Using the RDP (Fig.\,\ref{krad_1735}) it is determined as 0.95\,pc, virtually
identical to the value determined in Froebrich et al.
\cite{2007MNRAS.377L..54F}.

\begin{figure}
\centering

\beginpicture
\setcoordinatesystem units <1mm,1mm> point at 0 0
\setplotarea x from 0 to 80 , y from 0 to 70
\put {\includegraphics[height=4.0cm]{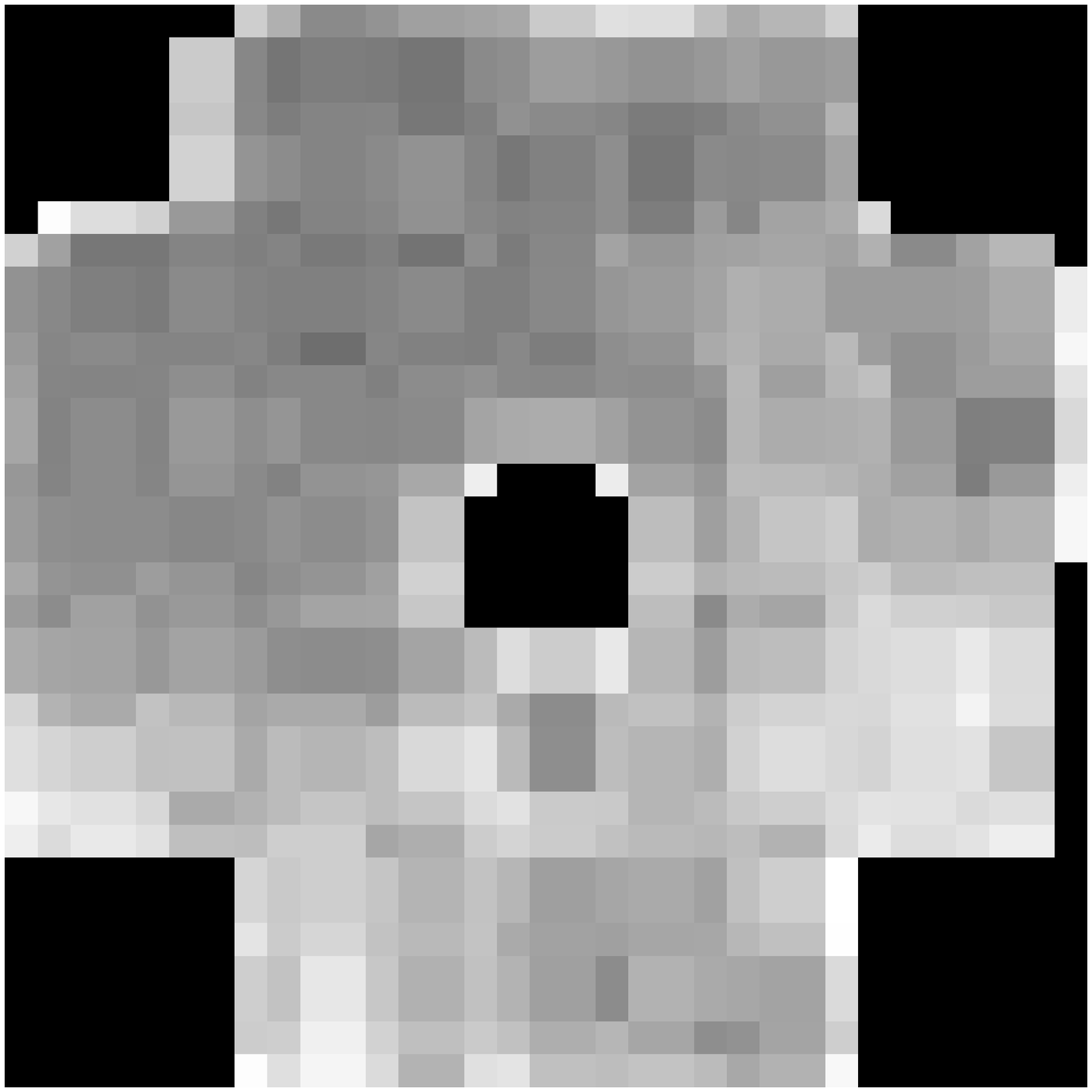}} at -20 50
\put {\includegraphics[height=4.0cm]{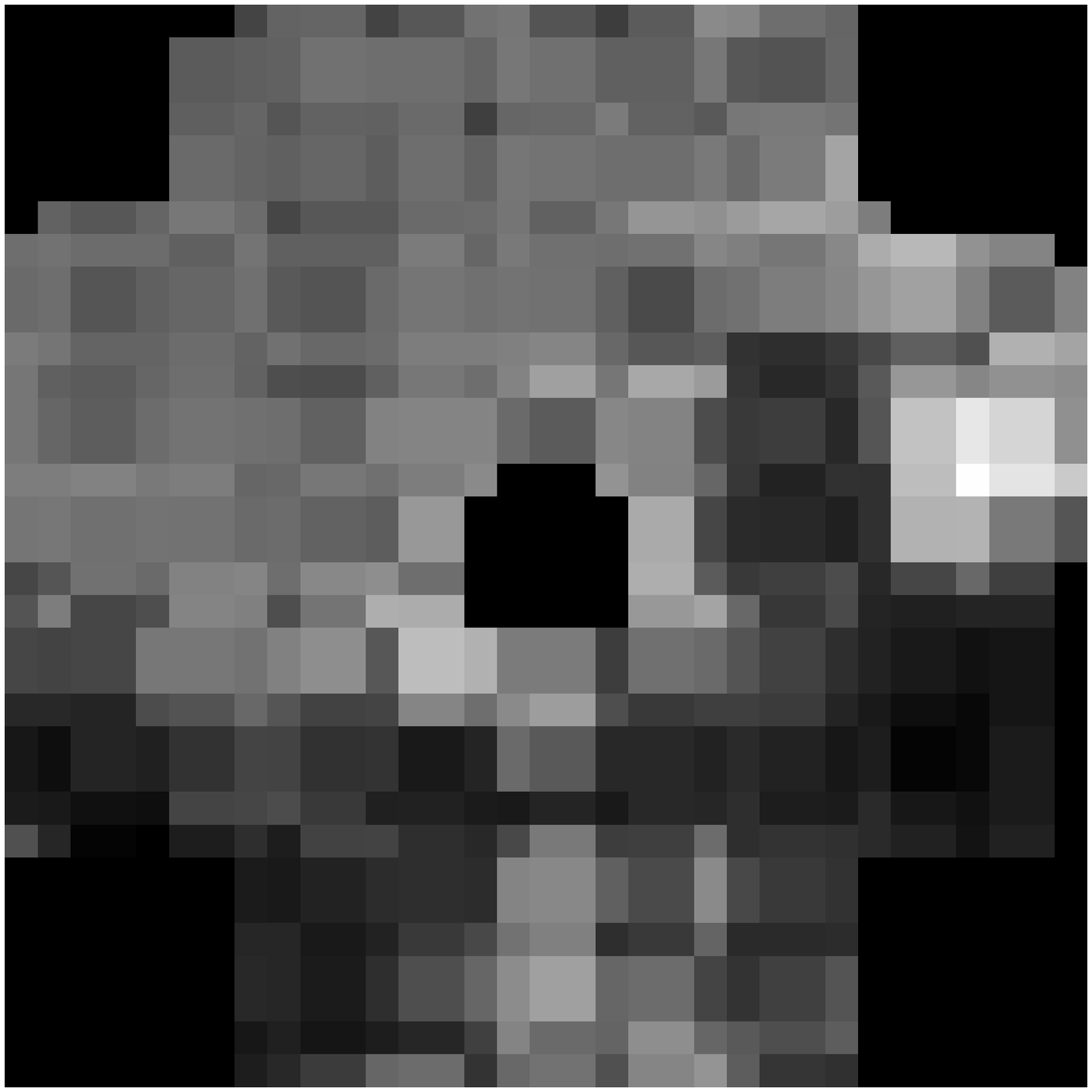}} at 25 50
\put {\includegraphics[height=8cm, angle=-90]{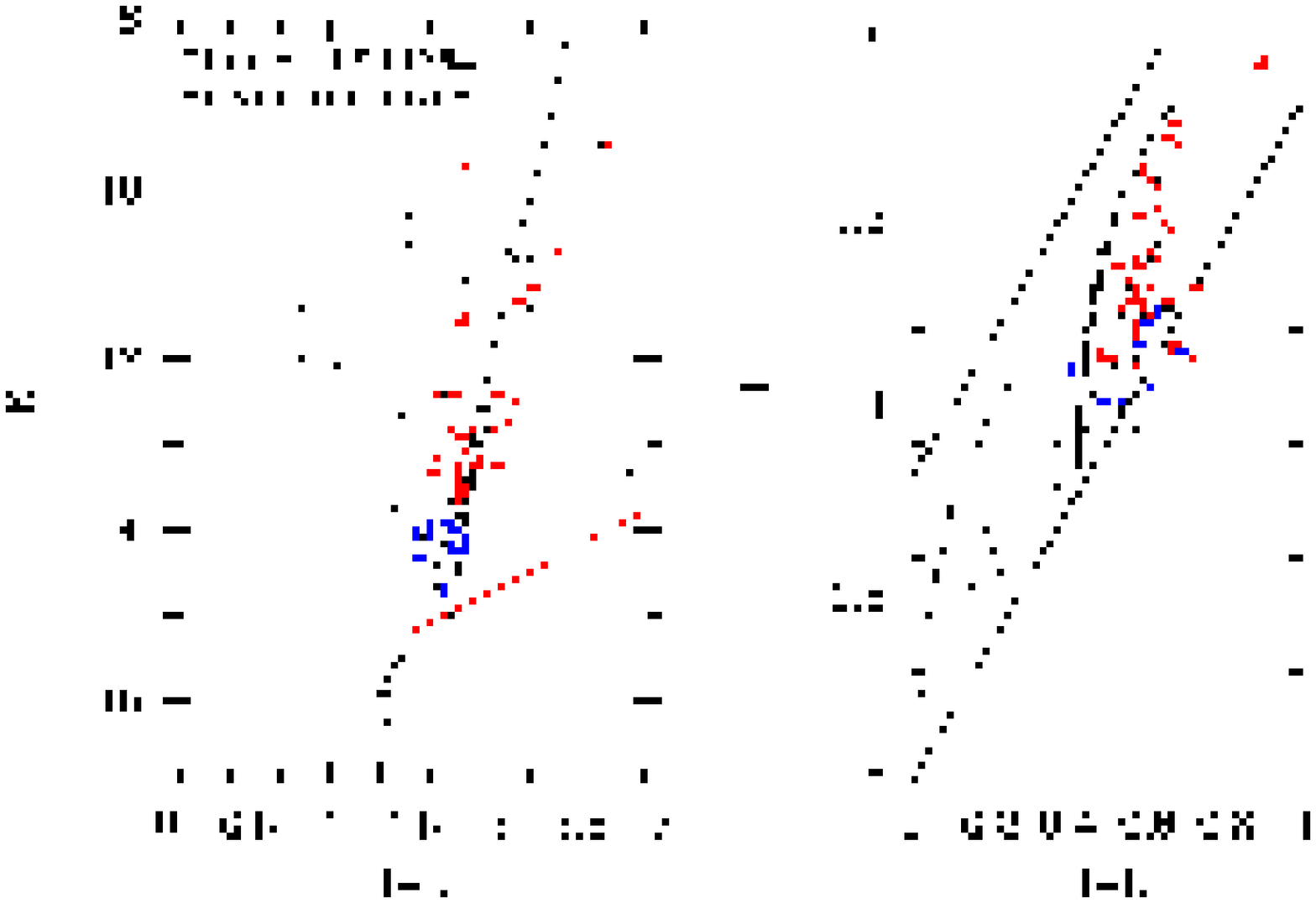}} at 0 0
\put {{\color{blue}\oval(30,30)}} at -25 58
\put {{\color{blue}\oval(30,30)}} at 20 58
\endpicture

\caption{\label{cl1754_jkk} As Fig.\,\ref{cl0002_jkk} but for the cluster
candidate FSR\,1754. The plotted isochrone has log(age)\,=\,9. See text for
other parameters.}  
\end{figure}

\subsection{FSR\,1754}

The image of this field shows no apparent overdensity of stars in the area of
the cluster candidate. This is confirmed by the SDM, which is virtually flat in
and around the object's position (see Fig.\,\ref{cl1754_jkk}). The southern and
western parts of the mosaic show a smaller number of stars. The REM indicates
significant differences in the average colour of the stars in this area, caused
by foreground clouds. We hence chose the eastern part of the image as the
control area.

Only about 10\,\% of the stars remain after the decontamination. In the CMD they
seem to show a populated giant branch (see Fig.\,\ref{cl1754_jkk}). However, the
slope would require metallicities in excess Z\,=\,0.03 to fit. Furthermore, the
positions of the stars in the CCD is not in agreement with the proposal of a
giant branch. Especially the fainter stars are too close to the bottom of the
reddening path. Furthermore, dust properties of $\beta <$\,1.4 would be required
to fit the stars with a single isochrone in CMD and CCD. The isochrone in
Fig.\,\ref{cl1754_jkk} is plotted to clarify this argument and has the following
parameters: Z\,=\,0.03, $\beta$\,=\,1.6, log(age)\,=\,9, d\,=\,10\,kpc, and
A$_K$\,=\,0.8\,mag. We conclude that FSR\,1754 is not a cluster. This partly
agrees with Bica et al. \cite{2008MNRAS.385..349B}, who classified this object
as an uncertain candidate. The two main sequences in their analysis of 2MASS
data are not evident in our analysis. In particular, the blue sequence
disappears in our decontaminated CMD, most probably caused by a better choice of
the control field (which is important given the large fluctuations visible in
the SDM and REM - see Fig.\,\ref{cl1754_jkk}).

\begin{figure}
\centering

\beginpicture
\setcoordinatesystem units <1mm,1mm> point at 0 0
\setplotarea x from 0 to 80 , y from 0 to 70
\put {\includegraphics[height=4.0cm]{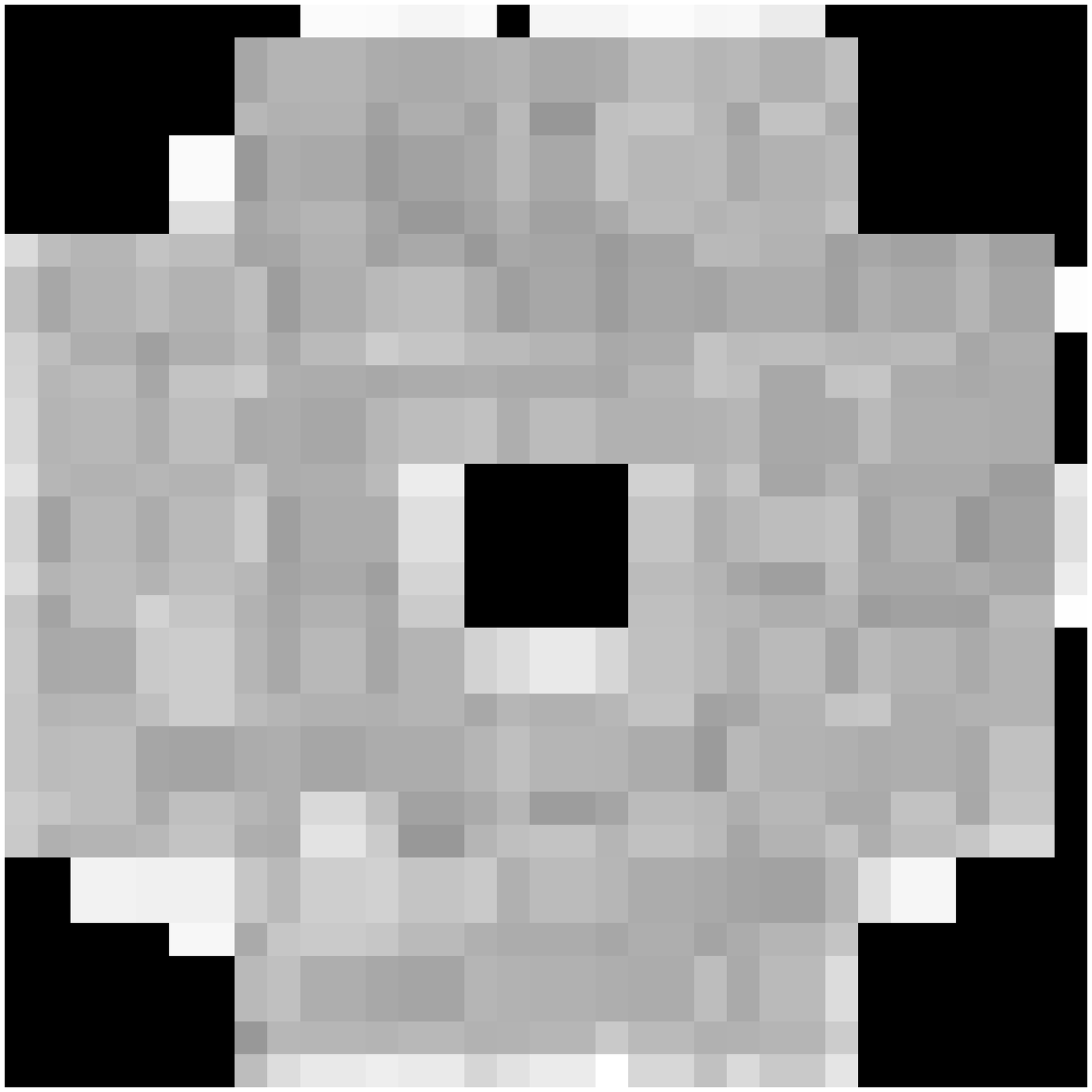}} at -20 50
\put {\includegraphics[height=4.0cm]{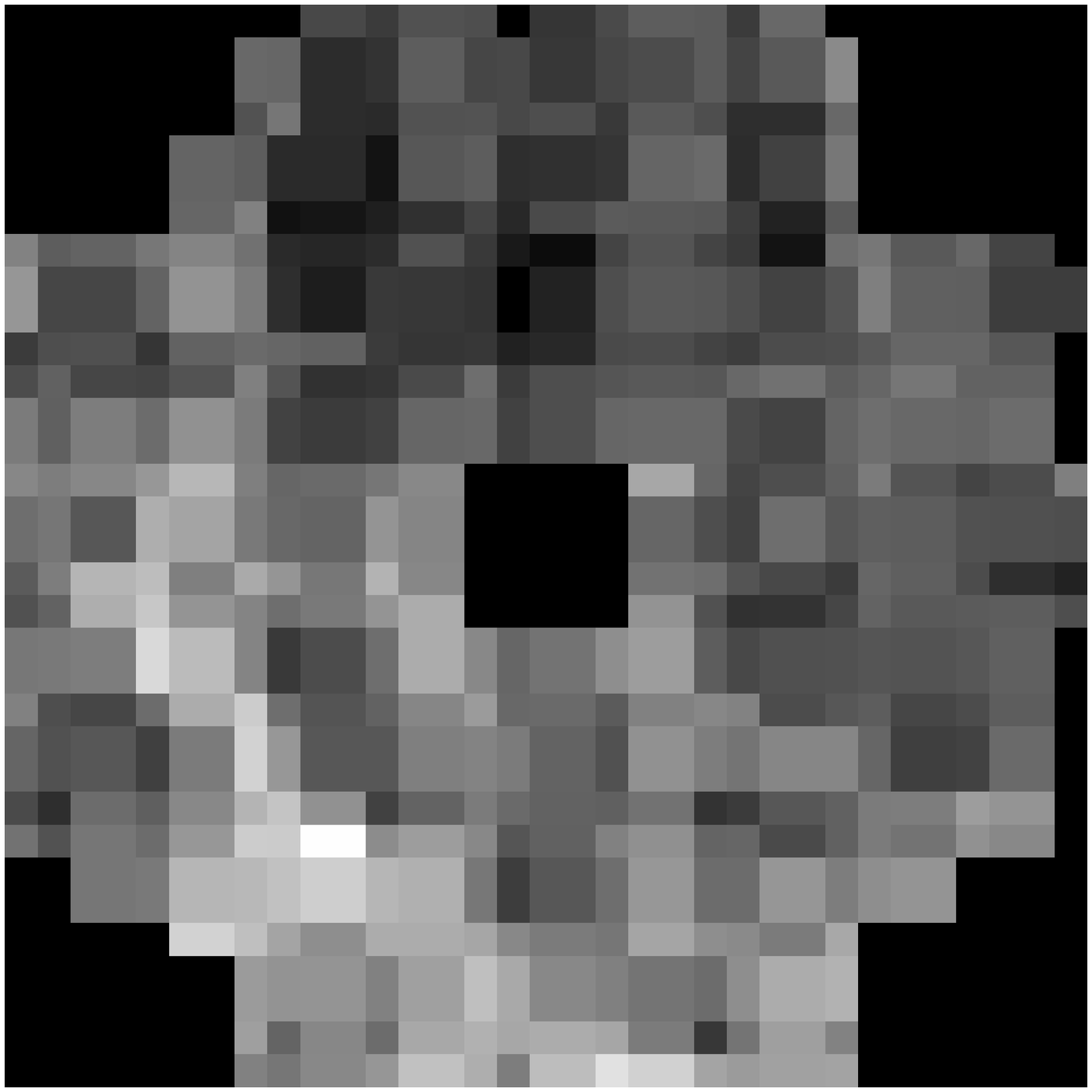}} at 25 50
\put {\includegraphics[height=8cm, angle=-90]{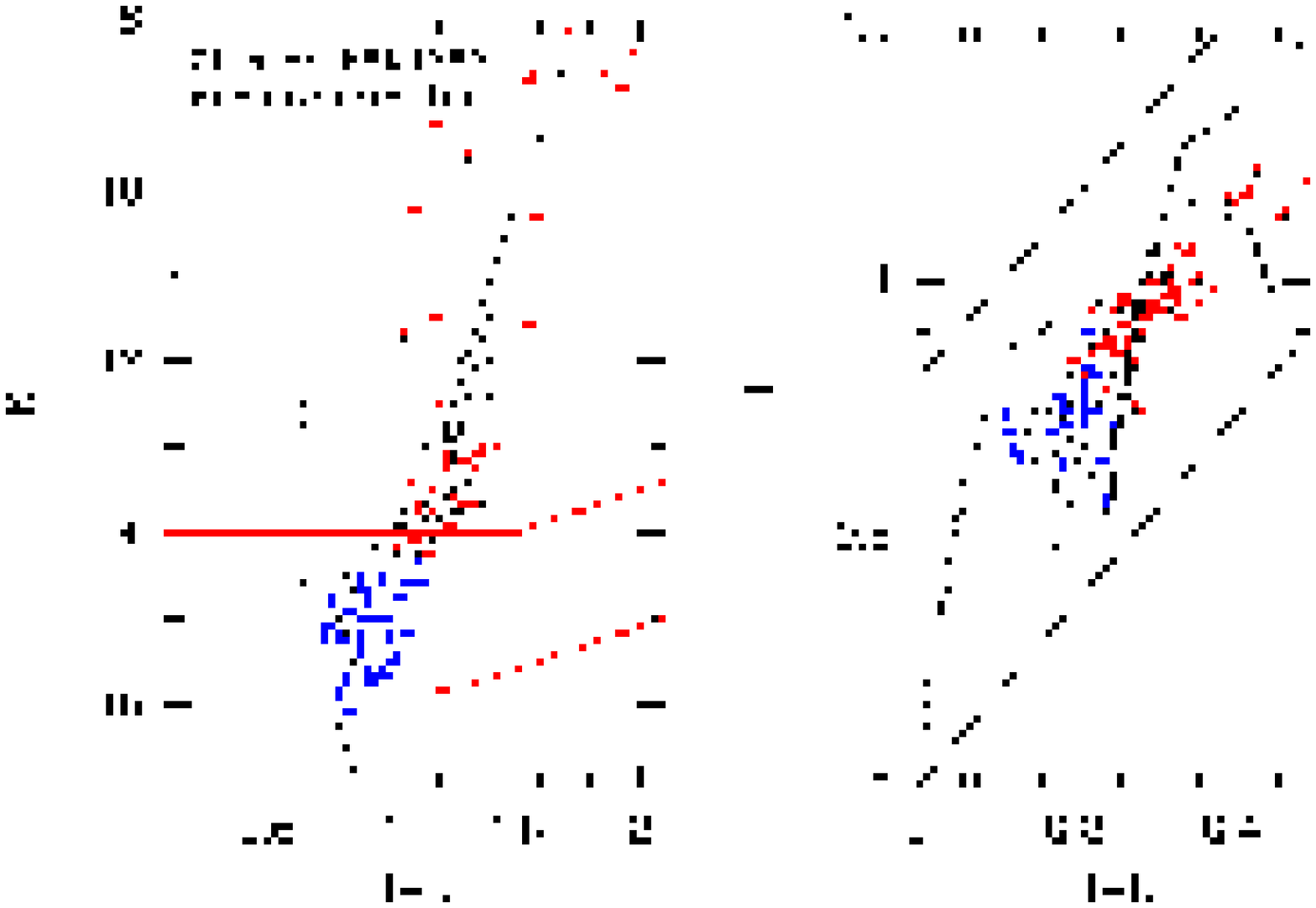}} at 0 0
\put {{\color{blue}\oval(30,30)}} at -25 58
\put {{\color{blue}\oval(30,30)}} at 20 58
\endpicture

\caption{\label{cl1767_jkk} As Fig.\,\ref{cl0002_jkk} but for the cluster
candidate FSR\,1767. The plotted isochrone has log(age)\,=\,9. See text for
other parameters. The upper completeness limit in the CMD indicates the 2MASS
limit. Clearly no isochrone can fit all the data.}   
\end{figure}

\begin{figure}
\centering
\includegraphics[height=8cm, angle=-90]{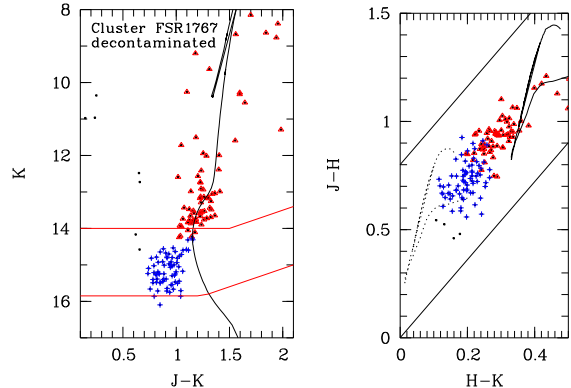}

\caption{\label{cl1767_jkk_wrongiso} As bottom panel of Fig.\,\ref{cl1767_jkk}
but using an isochrone with the cluster parameters for FSR\,1767 as suggested by
Bonatto et al. (2007); Age\,=\,10\,Gyr, Z\,=\,0.001, d\,=\,1.5kpc,
A$_K$\,=\,0.62\,mag. Clearly this isochrone does not fit the data.}   

\end{figure}

\subsection{FSR\,1767}

There is no apparent overdensity of stars in the image of the field around the
cluster candidate. The SDM shows, however, a small increase in stellar numbers.
In the REM we find that the cluster area suffers from less extinction than the
south-eastern part of the mosaic. We have therefore chosen the western part of
the image as control field.

The cluster was classified by Bonatto et al. \cite{2007MNRAS.381L..45B} as a
globular cluster. With its parameters, [Fe/H]\,=\,-1.2\,dex, $A_V$\,=\,6.2mag
and a distance of only 1.5\,kpc it would be the 2$^{nd}$ closest globular
cluster after FSR\,0584 (d\,=\,1.4\,kpc, Bica et al.
\cite{2007A&A...472..483B}), which is still under debate. The analysis of
FSR\,1767 in Bonatto et al.\cite{2007MNRAS.381L..45B} is based on 2MASS data
(the upper detection  limit in Fig.\,\ref{cl1767_jkk} and
\ref{cl1767_jkk_wrongiso}) and proper motions. Only a small number of cluster
red giants has been found, and the upper end of the main sequence has been
identified. With our much deeper images, we should be able to detect the main
sequence stars down to lower masses and thus verify the claim of Bonatto et al.
\cite{2007MNRAS.381L..45B} that this object is a globular cluster.

After decontamination only about 10\,\% of the stars remain in the cluster
candidate area. The stars identified in Bonatto et al.
\cite{2007MNRAS.381L..45B} as upper main sequence stars are also identifiable in
the CMD from our data (see Fig.\,\ref{cl1767_jkk}). However, overplotting an
isochrone with the given cluster parameters (and assuming an age of about
10\,Gyr) implies that the main sequence should continue towards fainter
magnitudes and slightly redder colours (see Fig.\,\ref{cl1767_jkk_wrongiso}).
Nothing like this is identifiable in our data. There is a second group of stars
above our detection limit (the lower line in Fig.\,\ref{cl1767_jkk}) remaining
in the CMD. We could interpret these as the top of the main sequence and the
original group of stars as giants. A fit in the CMD (see Fig.\,\ref{cl1767_jkk})
is then possible using an age of about 1\,Gyr, a reddening of $A_K$\,=\,0.5\,mag
and a distance of 6.5\,kpc. However, in the CCD we clearly see that the two
groups of stars cannot be fit by a single isochrone, since they possess
different extinction values. We  have to conclude that based on our deeper
observations, FSR\,1767 is not a globular cluster. It does not appear to be a
stellar cluster at all, but rather a locally decreased amount of extinction
mimicking a stellar overdensity. 

\begin{table*}
\centering
\renewcommand{\tabcolsep}{3pt}

\caption{\label{properties} Measured properties of the clusters investigated in
this paper. We list the FSR number, Right Ascension, Declination (J2000),
determined age, K-band extinction $A_K$, distance $d$, metallicity $Z$, radius
$r$, classification, and notes. Positions taken from: $^1$ Froebrich et al.
(2007b); $^2$ Froebrich et al. (2007a); $^3$ Bonatto \& Bica (2007b); $^4$
Kronberger et al. (2006); $^5$ measured in this paper. The classifications
stand for: NC - not a cluster; OC - open cluster; YOC - young open cluster (age
$<$\,100\,Myr); GC - globular cluster; ? - classification uncertain; {\bf $^{*}$ If the clusters is very old (12\,Gyr) the distance can be as
small as 5\,kpc; $^{**}$ Metallicity is assumed to be solar; $^{***}$ Uncertainty of the radii is 20\,\%; $^{****}$ Radius could not be
determined.}}

\begin{tabular}{lllrrlllll} 
Name & $\alpha$\,(2000) & $\delta$\,(2000) & Age\,[Gyr] & $A_K$\,[mag] & $d$\,[kpc] & $Z$ & $r$\,[pc]$^{***}$ & Class. & Notes \\
\hline
FSR\,0002 & 17:32:32$^1$   & $-$27:03:51$^1$ & -                         & -             & -                   & -                         & -         & NC     & \\
FSR\,0023 & 17:57:35$^1$   & $-$22:52:32$^1$ & -                         & -             & -                   & -                         & -         & NC     & \\
FSR\,0088 & 18:50:38$^1$   & $-$04:11:17$^1$ & 0.5$\pm$0.3               & 0.5$\pm$0.05  & 2.0$\pm$0.3         & 0.019$^{**}$              & $^{****}$ & OC     & \\
FSR\,0089 & 18:48:39$^3$   & $-$03:30:34$^3$ & 1.0$\pm$0.3               & 1.0$\pm$0.05  & 2.2$\pm$0.3         & 0.019$^{**}$              & $^{****}$ & OC     & confirms Bonatto \& Bica {\cite{2007A&A...473..445B}} \\
FSR\,0094 & 18:49:50$^1$   & $-$01:02:55$^1$ & -                         & -             & -                   & -                         & -         & NC     & \\
FSR\,1527 & 10:06:32$^1$   & $-$57:24:52$^1$ & -                         & -             & -                   & -                         & -         & NC     & \\
FSR\,1530 & 10:08:58.3$^5$ & $-$57:17:11$^5$ & $\le$0.004                & 0.9$\pm$0.05  & 2.5$^{+0.5}_{-1.0}$ & 0.019$^{**}$              & 0.15      & YOC    & \\
FSR\,1570 & 11:08:40.6$^4$ & $-$60:42:50$^4$ & 0.008$^{+0.008}_{-0.004}$ & 0.8$\pm$0.05  & 6.0$\pm$1.0         & 0.019$^{**}$              & 0.35      & YOC    & confirms Pasquali et al. {\cite{2006A&A...448..589P}} \\
FSR\,1659 & 13:38:01$^1$   & $-$62:27:55$^1$ & -                         & -             & -                   & -                         & -         & NC     & \\
FSR\,1712 & 15:54:46.3$^5$ & $-$52:31:47$^5$ & 0.8$^{+0.8}_{-0.4}$       & 1.4$\pm$0.1   & 1.8$\pm$0.3         & 0.019$^{**}$              & 0.20      & OC     & \\
FSR\,1716 & 16:10:29.0$^5$ & $-$53:44:48$^5$ & $>$2                      & 0.57$\pm$0.05 & 7.0$\pm$0.5$^{*}$   & 0.004$^{+0.004}_{-0.003}$ & 0.90      & OC/GC? & \\
FSR\,1735 & 16:52:10.6$^2$ & $-$47:03:29$^2$ & $>$8                      & 0.7$\pm$0.05  & 8.5$\pm$0.5         & 0.004$^{+0.004}_{-0.003}$ & 0.95      & GC?    & confirms Froebrich et al. {\cite{2007MNRAS.377L..54F}} \\
FSR\,1754 & 17:15:01$^1$   & $-$39:06:07$^1$ & -                         & -             & -                   & -                         & -         & NC     & partly confirms Bica et al. {\cite{2008MNRAS.385..349B}} \\
FSR\,1767 & 17:35:43$^1$   & $-$36:21:28$^1$ & -                         & -             & -                   & -                         & -         & NC     & conflicts with Bonatto et al. {\cite{2007MNRAS.381L..45B}} \\
\end{tabular}
\end{table*}

\section{Discussion}\label{discussion}

\subsection{Uncertainties}

 %
 %

\subsubsection{Photometry}

The nature of our observations, stellar clusters in crowded fields, implies that
the photometry suffers from relatively large errors. To ensure an as small as
possible influence of these uncertainties, we used only the most reliable
stellar magnitudes and colours for the analysis of the CMDs and CCDs. Only stars
with quality flags from the Source Extractor software (Bertin \& Arnouts
\cite{1996A&AS..117..393B}) better than 3 are used (except for FSR\,1735, were a
flag better than 4 was chosen due to the very large crowding and hence small
number of stars with a quality flag better than 3). Furthermore, the
completeness limit in the cluster fields will be at brighter magnitudes than in
the control fields. Thus, we only include stars in the analysis which are above
the completeness limit in the cluster areas. This is also the limit indicated in
all the CMDs. 

The observed scatter in the CMDs and also CCDs can thus be attributed to a
number of causes. i) uncertainty in the photometry due to crowding; ii) scatter
in the calibration due to the low spatial resolution of the 2MASS data; iii)
small scale variable extinction towards the stars (A$_V$ varies up to 0.5\,mag
from pixel to pixel in the fields close to some of the cluster candidates); iv)
unresolved binaries; Given all those sources of error, the observed scatter in
colour, for stars in a cluster with the same magnitude, of about 0.2\,mag to
0.4\,mag is understandable.

\subsubsection{Cluster Parameters}

In most cases the metallicity could not be determined. For all clusters with a
fitted age below a few Gyr we assume a solar metallicity. We hence labeled the
metallicities of these clusters with $^{**}$ in Table\,\ref{properties}. In the
cases of the two old clusters, FSR\,1716 and FSR\,1735, the metallicities that
fit the data range from Z\,=\,0.001 to Z\,=\,0.008 and the best fitting value of
Z\,=\,0.004 is listed in Table\,\ref{properties}. 

The determined values for extinction, distance and age are all dependent on each
other in a systematic way. By changing one parameter slightly we will still be
able to generate a good fit by adjusting the other two parameters. This is,
however, only possible within a certain range, outside which it becomes
impossible to fit CMDs and CCDs simultaneously. Typically the age of the
clusters can be constrained by better than a factor of two, the K-band
extinction to within 0.05\,mag, and the distance within 20\,\%. {\bf
Table\,\ref{properties} contains the uncertainties for the main parameters of
the individual clusters.}

The determined radii of the clusters vary dependent on which of the random
realisation of the decontamination processes we use. Generally, for the cluster
candidates with a high contrast between cluster area and field, the fitted radii
vary by about 20\,\%. For the cases with a low contrast, inconclusive results
are obtained, and the values in the Table\,\ref{properties} are marked by
$^{****}$. The star density in the cluster centre is also highly uncertain,
since it depends on: i) We only used stars with reliable photometry, hence we
will miss more stars towards the cluster centre; ii) The completeness limit in
the cluster centre differs from the control field, hence more stars are removed
during the decontamination procedure. We hence refrain from listing the central
star densities in Table\,\ref{properties}.

\begin{figure*}
\centering
\includegraphics[height=5.5cm, angle=-90, bb = 50 65 550 630]{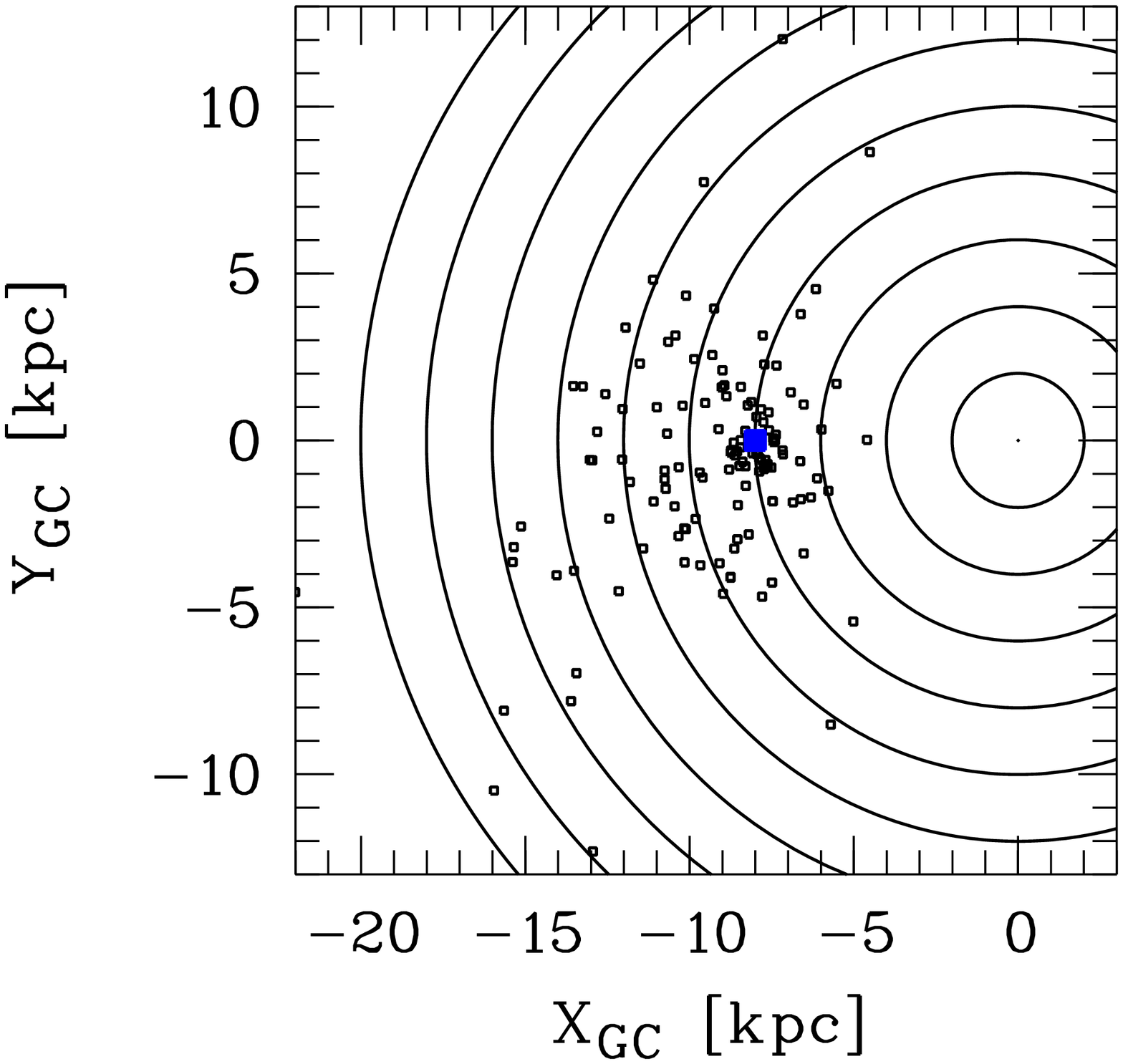} \hfill
\includegraphics[height=5.5cm, angle=-90, bb = 50 65 550 630]{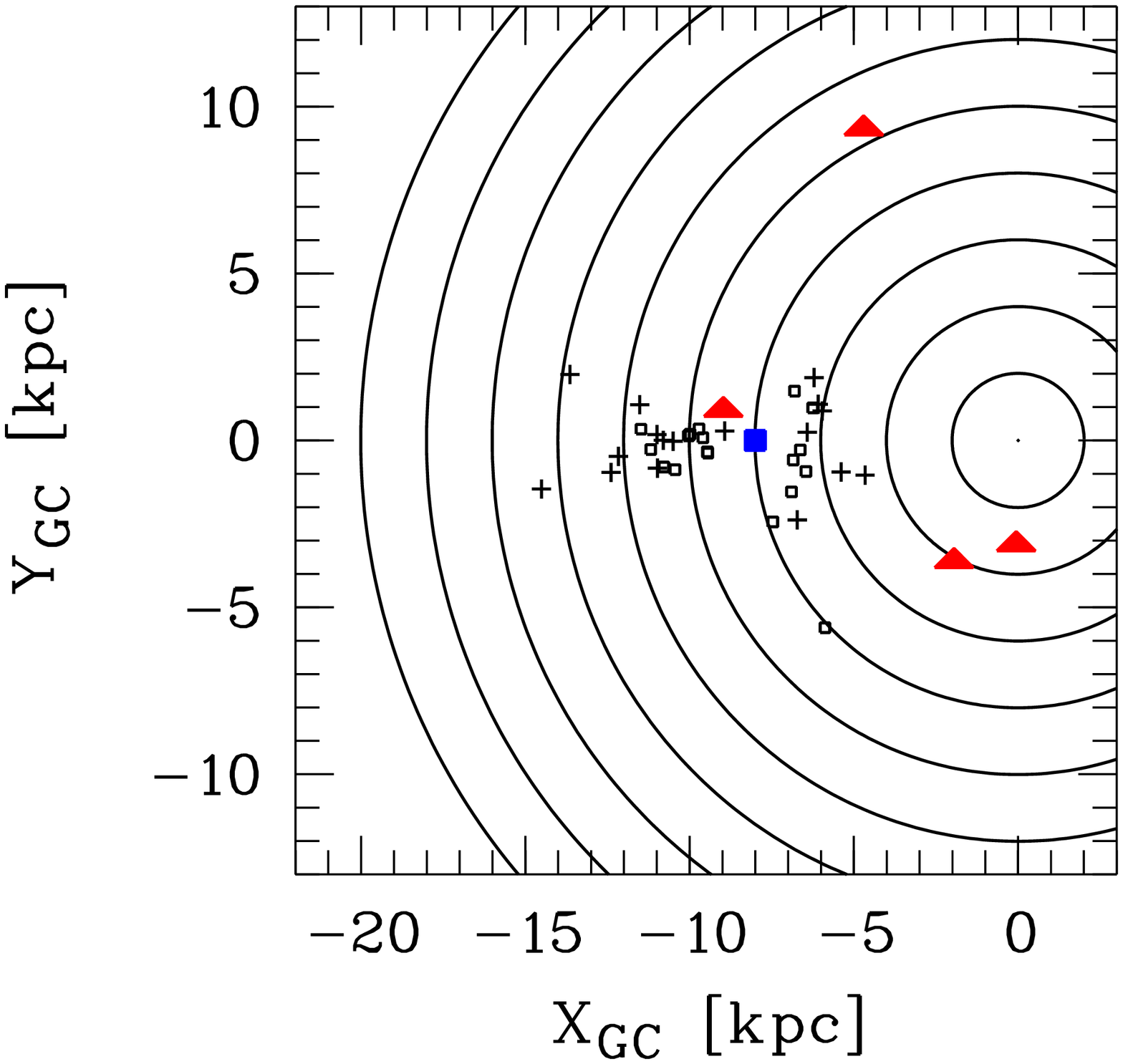} \hfill
\includegraphics[height=5.5cm, angle=-90, bb = 50 65 550 630]{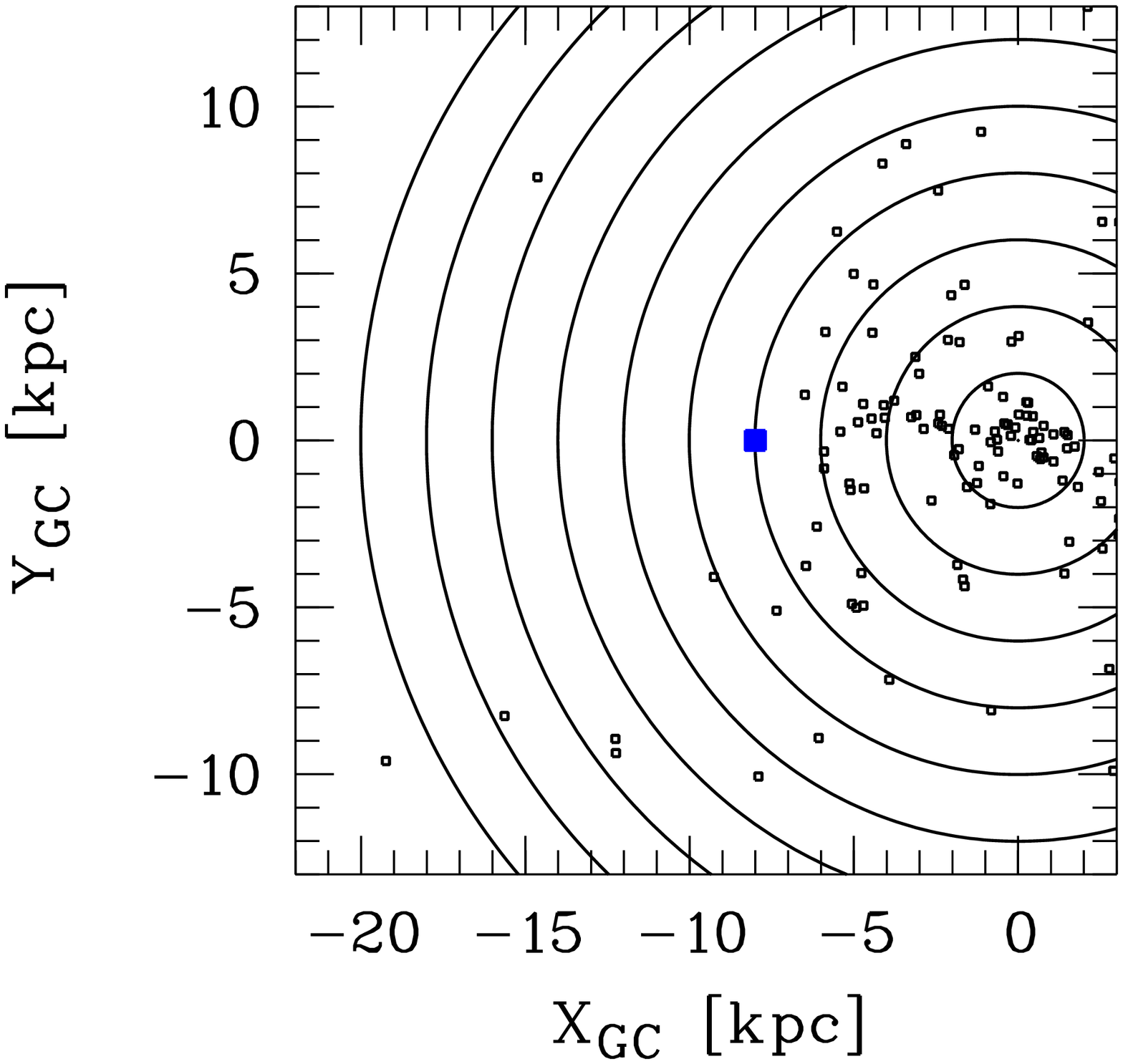} \\

\caption{\label{XY-dist} Distribution in the Galactic Plane of known old (age
$>$ 1\,Gyr) OCs ({\bf left}), classified FSR clusters ({\bf middle}), and known
GCs ({\bf right}). The blue square marks the position of the Sun. The red
triangles in the middle panel represent the FSR GC candidates, crosses the old
(age $\ge$ 1\,Gyr), and squares the young (age $<$ 1\,Gyr) OCs in the FSR
sample. We assumed a galactocentric distance of the Sun of 8\,kpc.}

\end{figure*}

\begin{figure*}
\centering
\includegraphics[height=5.5cm, angle=-90, bb = 50 65 550 630]{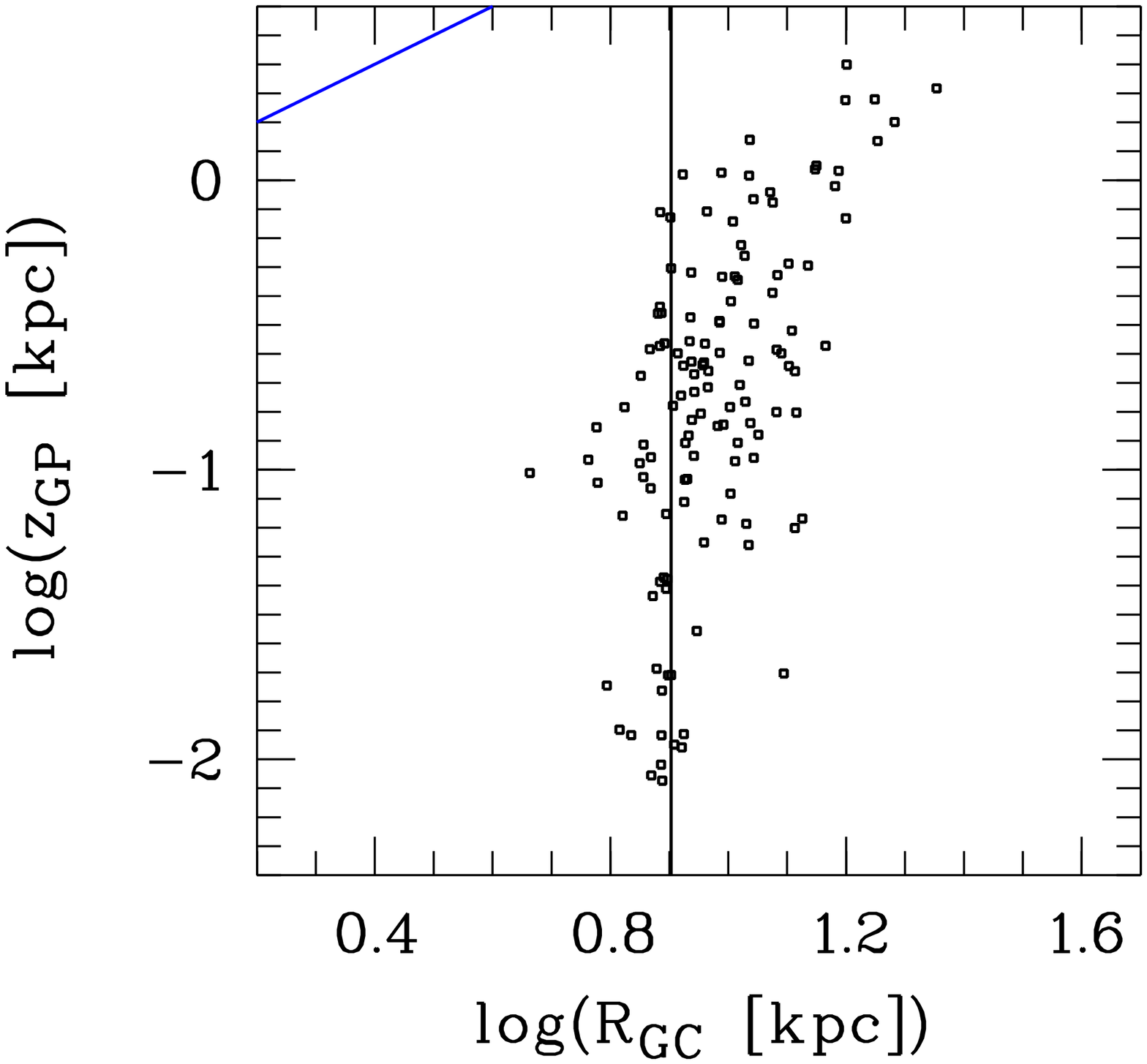} \hfill
\includegraphics[height=5.5cm, angle=-90, bb = 50 65 550 630]{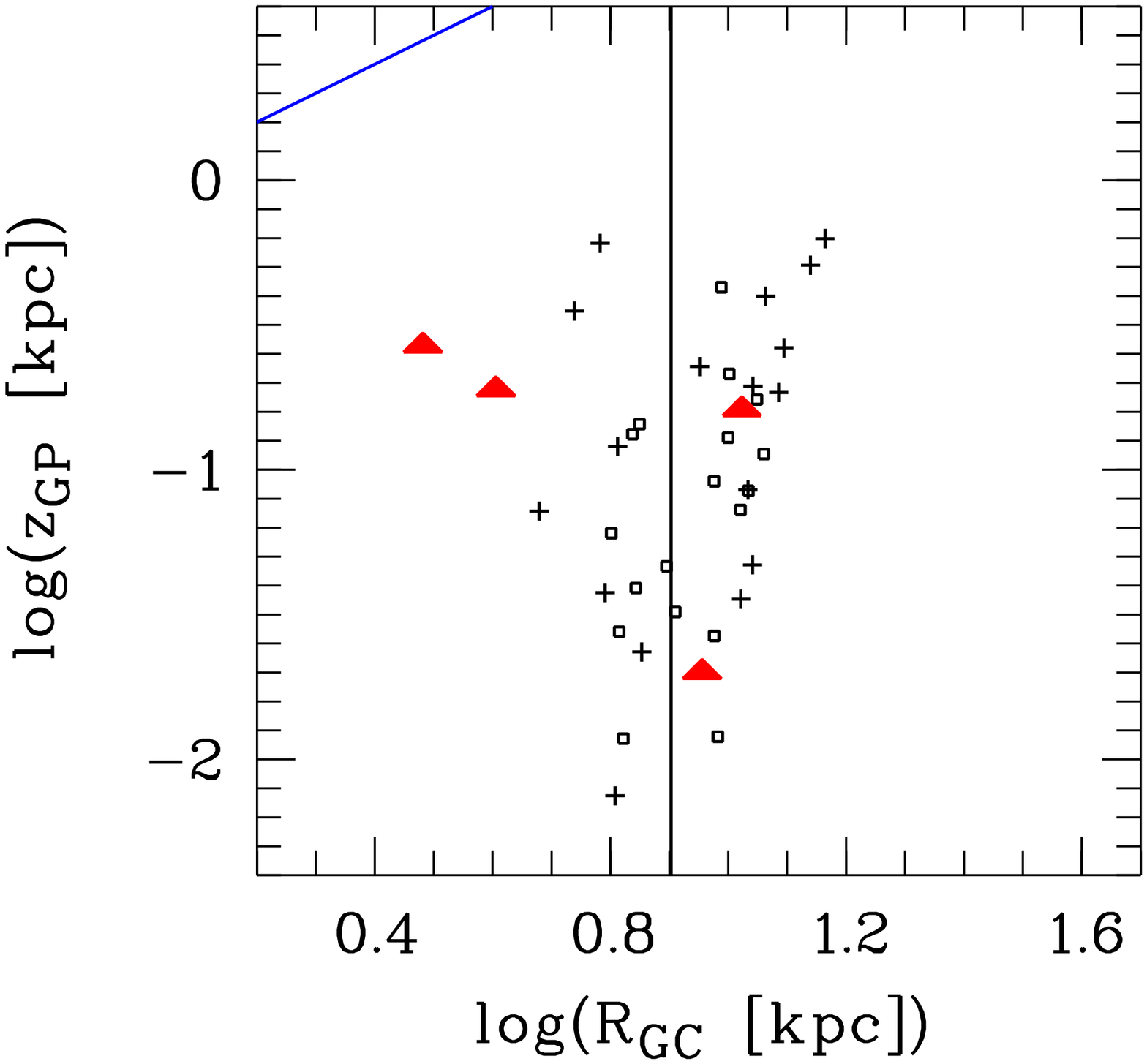} \hfill
\includegraphics[height=5.5cm, angle=-90, bb = 50 65 550 630]{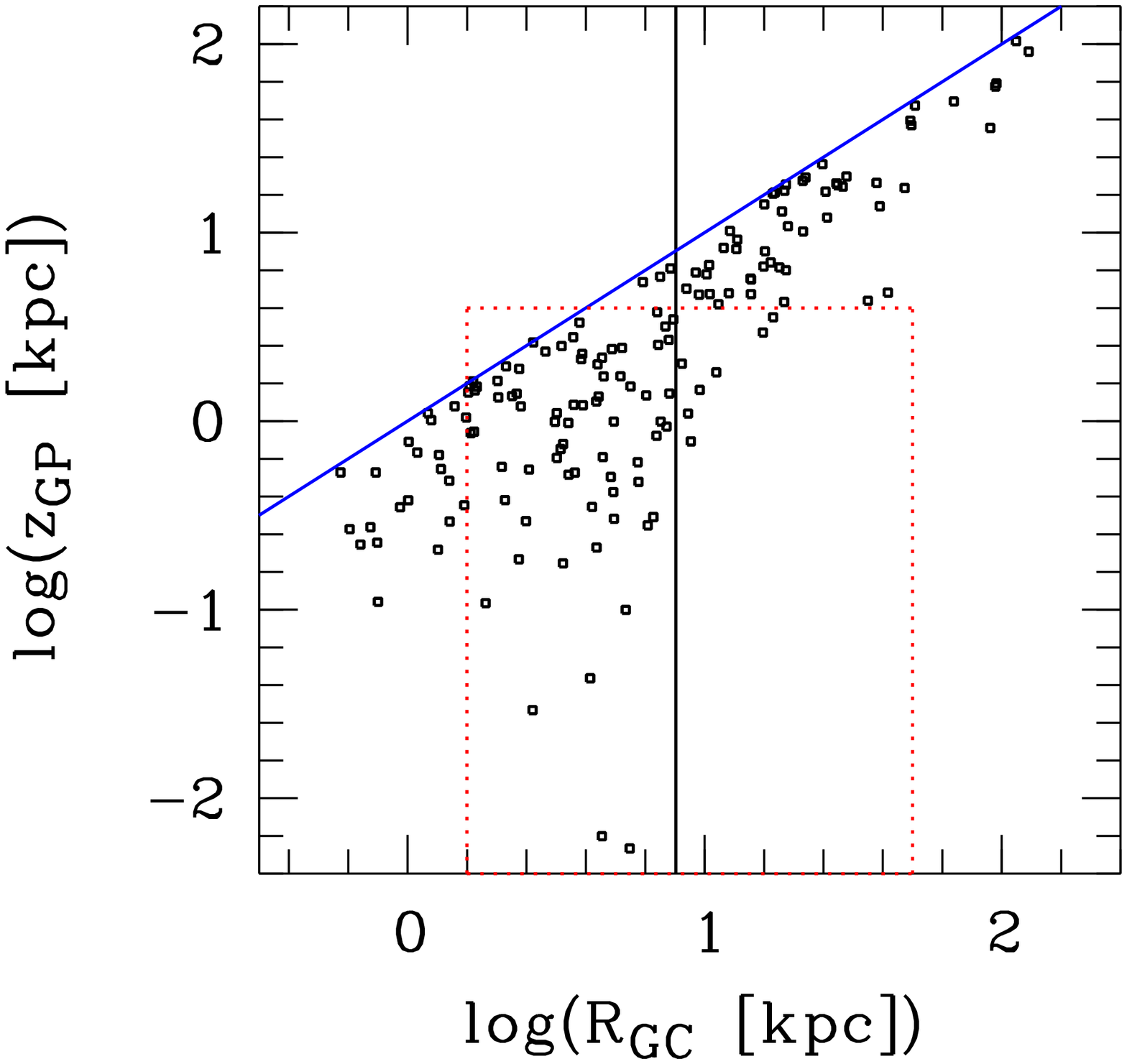} \\

\caption{\label{RZ-dist} Distance to the Galactic Centre ($R_{\rm GC}$) vs.
height above the Galactic Plane ($z_{\rm GP}$) of known old (age $>$ 1\,Gyr) OCs
({\bf left}), classified FSR clusters ({\bf middle}), and known GCs ({\bf
right}). The black vertical line marks the position of the Sun, the blue
diagonal line represents the maximum height above the plane for a given distance
from the centre. The red triangles in the middle panel represent the FSR GC
candidates, crosses the old (age $\ge$ 1\,Gyr), and squares the young (age $<$
1\,Gyr) OCs in the FSR sample. We assumed a galactocentric distance of the Sun
of 8\,kpc. The dotted box in the right panel marks the region shown in the left
and middle panel.}

\end{figure*}

\subsection{General FSR cluster properties}

A considerable number of clusters from the original FSR list have now been
analysed in more detail by a variety of authors. Here we provide a brief summary
of these investigations to date and a discussion of the properties of these new
star clusters.

We have summarised the properly classified FSR clusters in
Table\,\ref{fsrproperties}. The table lists the FSR catalogue number and other
identifications, Right Ascension and Declination (J2000), galactic coordinates,
determined distance, K-band extinction, age, and metallicity. Furthermore we
list the inferred position in the Galaxy, a classification, and the references
where the values are taken from. Table\,\ref{fsr_unknown_properties} lists the
FSR cluster candidates that have been investigated in detail but their nature
could not be established, or they are classified as not being a stellar cluster.
In these cases only the FSR number, other identifications, the coordinates,
classification, and references are listed.


So far 74 FSR cluster candidates have been investigated in more detail. For 38
of these clusters parameters could be determined. Another 3 are embedded
clusters with no parameters, 23 are not clusters, and 10 are uncertain cases. 
We cannot attempt to determine the contamination of the FSR sample from this
statistics, since the investigated cluster candidates are not selected in an
unbiased manner. There are, however, studies that have selected all FSR objects
within certain areas (Koposov et al. \cite{2008A&A...486..771K}, Bonatto \& Bica
\cite{2008A&A...485...81B}, Bica et al. \cite{2008MNRAS.385..349B}). These
studies determine contamination rates between 40 and 60\,\%, in agreement with
the original estimate of about 50\,\% in Froebrich et al.
\cite{2007MNRAS.374..399F}. 

Out of the now classified FSR clusters, 8 are young open clusters with ages
below 100\,Myr (this includes the 3 embedded clusters with no determined
parameters). A fraction of 50\,\% of the classified clusters have ages of more
than 1\,Gyr, and 20\,\% have ages above 2\,Gyr. This has increased the sample of
known old open clusters. In particular the FSR list revealed 7 old open clusters
inside the solar circle, significantly enhancing the known number of these
clusters. 

In Figs.\,\ref{XY-dist} and \ref{RZ-dist} we show the distribution of the
classified FSR clusters in the Galaxy, and how they compare to the sample of
known old (age\,$\ge$\,1\,Gyr) open clusters (taken from WEBDA\footnote{\tt
http://www.univie.ac.at/webda/}) and known globular clusters (taken from Harris
\cite{1996AJ....112.1487H}). We plot young and old open clusters and globular
cluster candidates among the FSR objects using different symbols. Despite the
unknown selection effects in the sample of properly classified FSR clusters, a
number of trends are evident: i) There is a large fraction of old open clusters
in the sample. This includes old clusters inside and outside the solar circle
and with distances of more than 1\,kpc from the Sun. The complete FSR list will
significantly increase the number of known old open clusters. ii) Two of the
globular cluster candidates (FSR\,1716, 1735) nicely follow the distribution of
the known globular clusters in the Milky Way. iii) The 10\,$\pm$\,3 'missing'
globulars near the Galactic Centre (Ivanov et al. \cite{2005A&A...442..195I})
are not contained in the FSR list. These clusters are either really not there,
hidden behind very dense clouds, or intrinsically faint. Given the fact that we
detected FSR\,1735, the number of intrinsically bright globular clusters near
the Galactic Centre that have been missed by the FSR survey is very small.

\section{Conclusions}

We have properly (re)-classified 14 star cluster candidates from the list of
Froebrich et al. \cite{2007MNRAS.374..399F}. We utilised new deep NIR photometry
obtained at the NTT. Star density maps, extinction maps, colour-magnitude
diagrams, and colour-colour diagrams, facilitated by radial star density
profiles and luminosity functions were used for the analysis of the individual
objects. Seven FSR cluster candidates are found to be real stellar clusters,
seven candidates are only star density enhancements.

Out of the seven classified star clusters, we have identified two young clusters
with massive stars (FSR\,1530, FSR\,1570 - investigated also by Pasquali et al.
\cite{2006A&A...448..589P}); three intermediate aged open clusters (FSR\,0088,
FSR\,0089 - confirms Bonatto \& Bica \cite{2007A&A...473..445B}, FSR\,1712);
and two globular cluster candidates (FSR\,1716, FSR\,1735 confirms Froebrich et
al. \cite{2007MNRAS.377L..54F}) with well populated giant branches.

We show that 2MASS data alone can sometimes be misleading or insufficient when
classifying (globular) cluster candidates. Our analysis of deep high spatial
resolution photometry is superior in some cases, especially for cluster
candidates in crowded fields near the Galactic Plane. One particular example is
the cluster candidate FSR\,1767. It has been classified as a globular cluster by
Bonatto et al. \cite{2007MNRAS.381L..45B}. Our new data shows that their
interpretation of the insufficient 2MASS data is incorrect, and we conclude that
FSR\,1767 is not a globular cluster. Most probably this object is just an
apparent stellar overdensity caused by a locally decreased amount of extinction,
mimicking a stellar cluster.

So far a total of 74 cluster candidates from the FSR list (containing 1021
candidate clusters) has been analysed in detail. The statistics points to a
contamination of the FSR sample with stellar overdensities of 40\,\% to 60\,\%,
in agreement with the original estimate of Froebrich et al.
\cite{2007MNRAS.374..399F}. A large fraction of the so far investigated FSR
clusters have turned out to be old stellar clusters with ages above 1\,Gyr. This
significantly increases the number of known old clusters, in particular inside
the solar circle. Mining the FSR catalogue for such type of clusters has the
potential to provide new insights into long-term cluster evolution and the
associated physical processes.

 %
 %

\section*{acknowledgments}

This publication makes use of data products from the Two Micron All Sky Survey,
which is a joint project of the University of Massachusetts and the Infrared
Processing and Analysis Center/California Institute of Technology, funded by the
National Aeronautics and Space Administration and the National Science
Foundation. This research has made use of the SIMBAD database, operated at CDS,
Strasbourg, France. This research has made use of the WEBDA database, operated
at the Institute for Astronomy of the University of Vienna. These observations
have been funded by the Optical Infrared Coordination network (OPTICON), a major
international collaboration supported by the Research Infrastructures Programme
of the European Commission's Sixth Framework Programme.

\clearpage

\begin{appendix}

\section{K-band Luminosity functions}

\begin{figure}
\centering
\includegraphics[height=8cm, angle=-90]{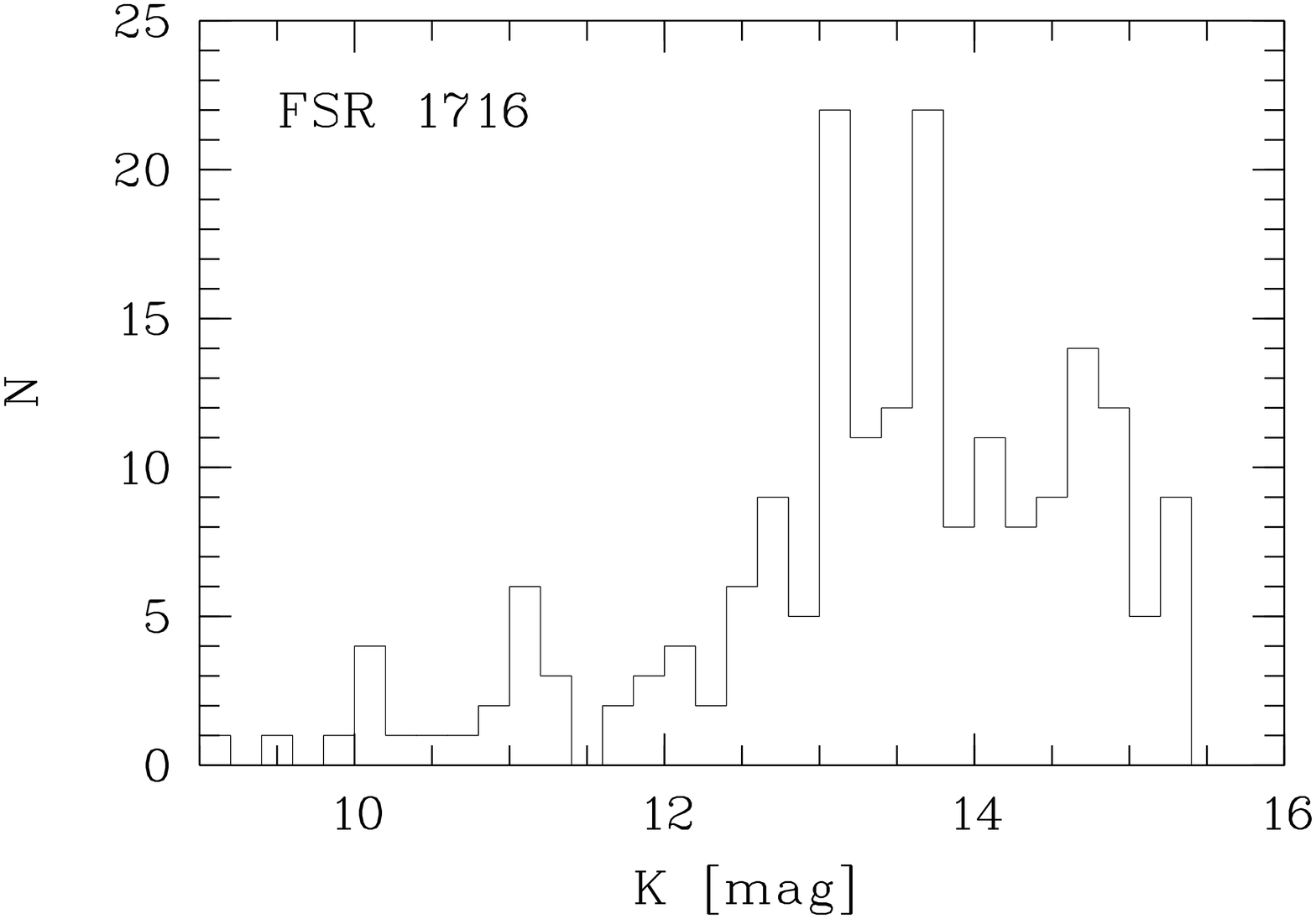} 
\caption{\label{klum_1716} One realisation of the K-band luminosity function of
the decontaminated cluster area of FSR\,1716.}
\end{figure}

\begin{figure}
\centering
\includegraphics[height=8cm, angle=-90]{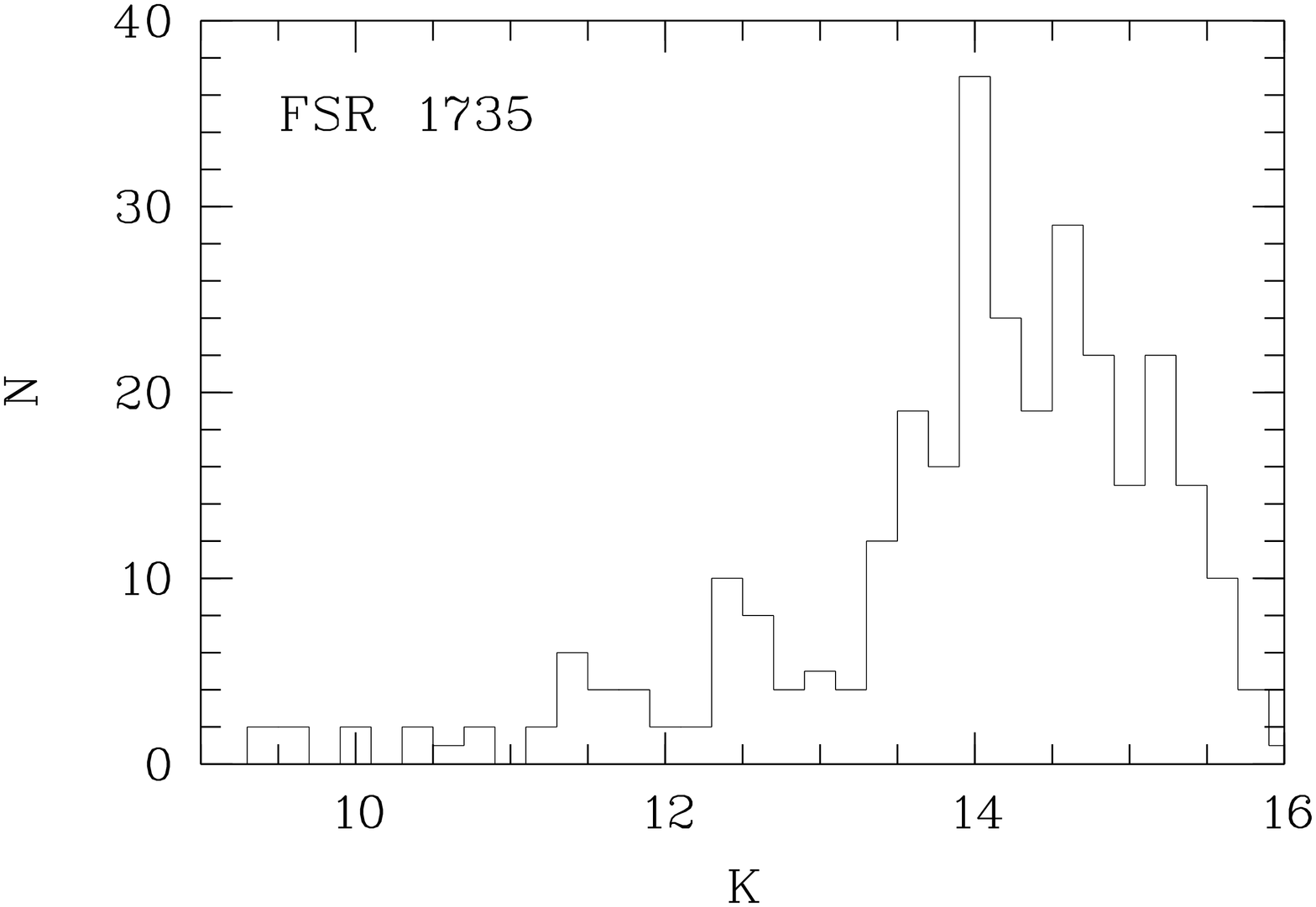} 
\caption{\label{klum_1735} K-band luminosity function of the decontaminated
cluster area of FSR\,1735.}
\end{figure}

\clearpage

\section{K-band radial star density profiles}

\begin{figure}
\centering
\includegraphics[height=8cm, angle=-90]{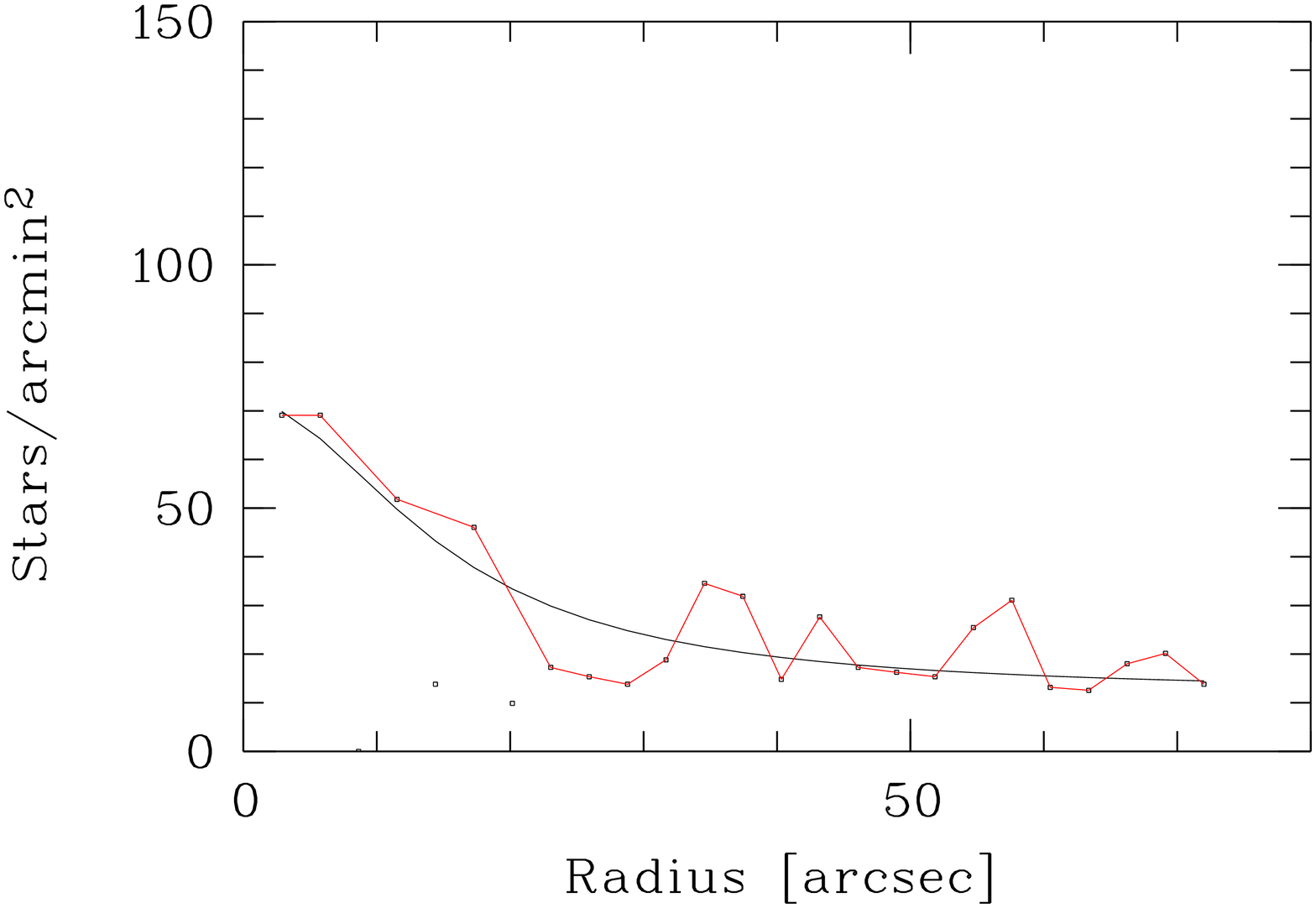} 
\caption{\label{krad_0088} One realisation of the K-band RDP (dots, red line) of
the decontaminated region of FSR\,0088. The solid black line represents the best
fit using a core radius of 15". The dots are RDP points more than 2$\sigma$ away
from the general profile, and are hence not used in the fit.}
\end{figure}

\begin{figure}
\centering
\includegraphics[height=8cm, angle=-90]{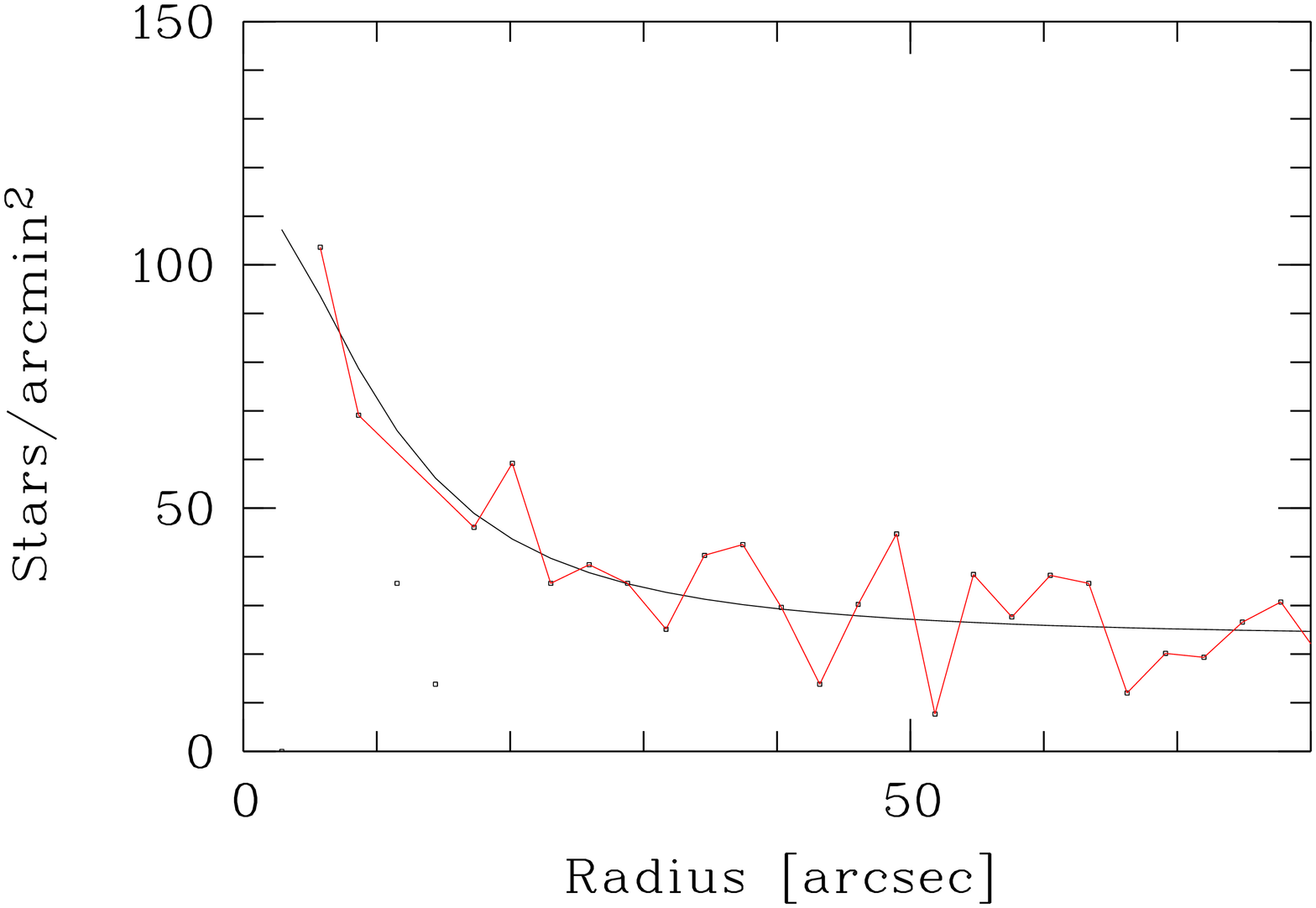} 
\caption{\label{krad_0089} As Fig.\,\ref{krad_0088} but for FSR\,0089. The solid
black line represents the best fit using a core radius of 11".}
\end{figure}

\begin{figure}
\centering
\includegraphics[height=8cm, angle=-90]{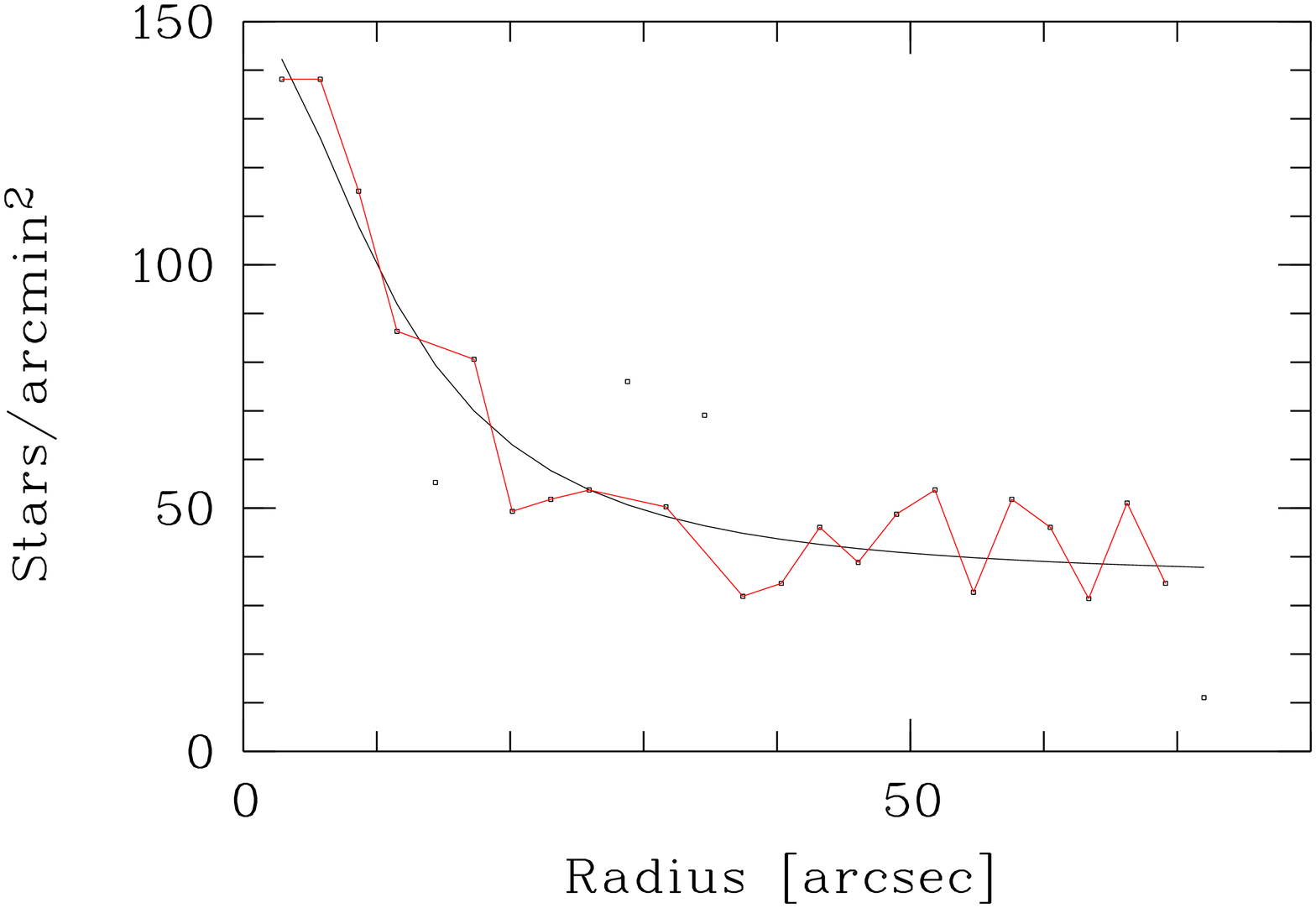} 
\caption{\label{krad_1530} As Fig.\,\ref{krad_0088} but for FSR\,1530. The solid
black line represents the best fit using a core radius of 11.5".}
\end{figure}

\begin{figure}
\centering
\includegraphics[height=8cm, angle=-90]{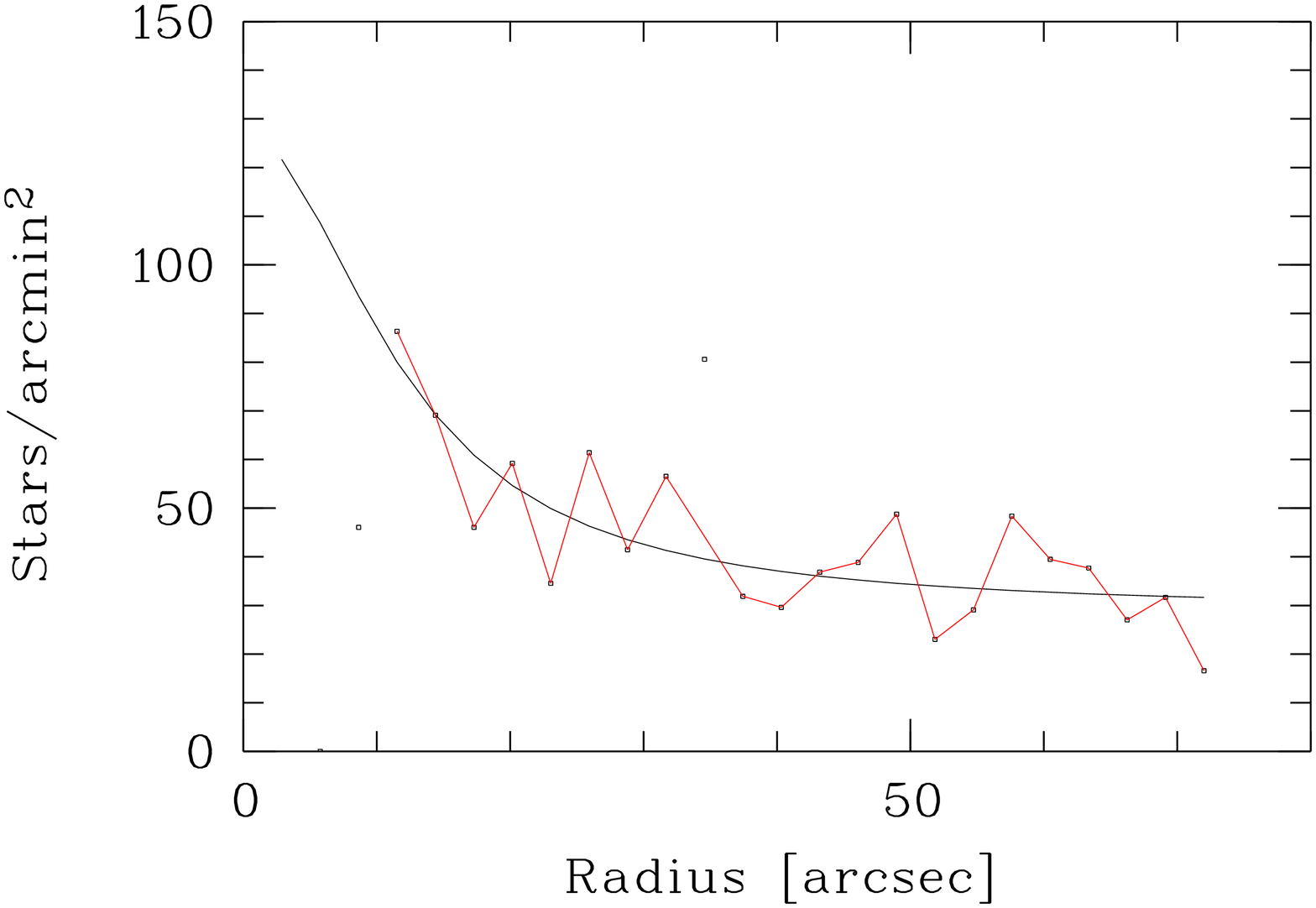} 
\caption{\label{krad_1570} As Fig.\,\ref{krad_0088} but for FSR\,1570. The solid
black line represents the best fit using a core radius of 12".}
\end{figure}

\begin{figure}
\centering
\includegraphics[height=8cm, angle=-90]{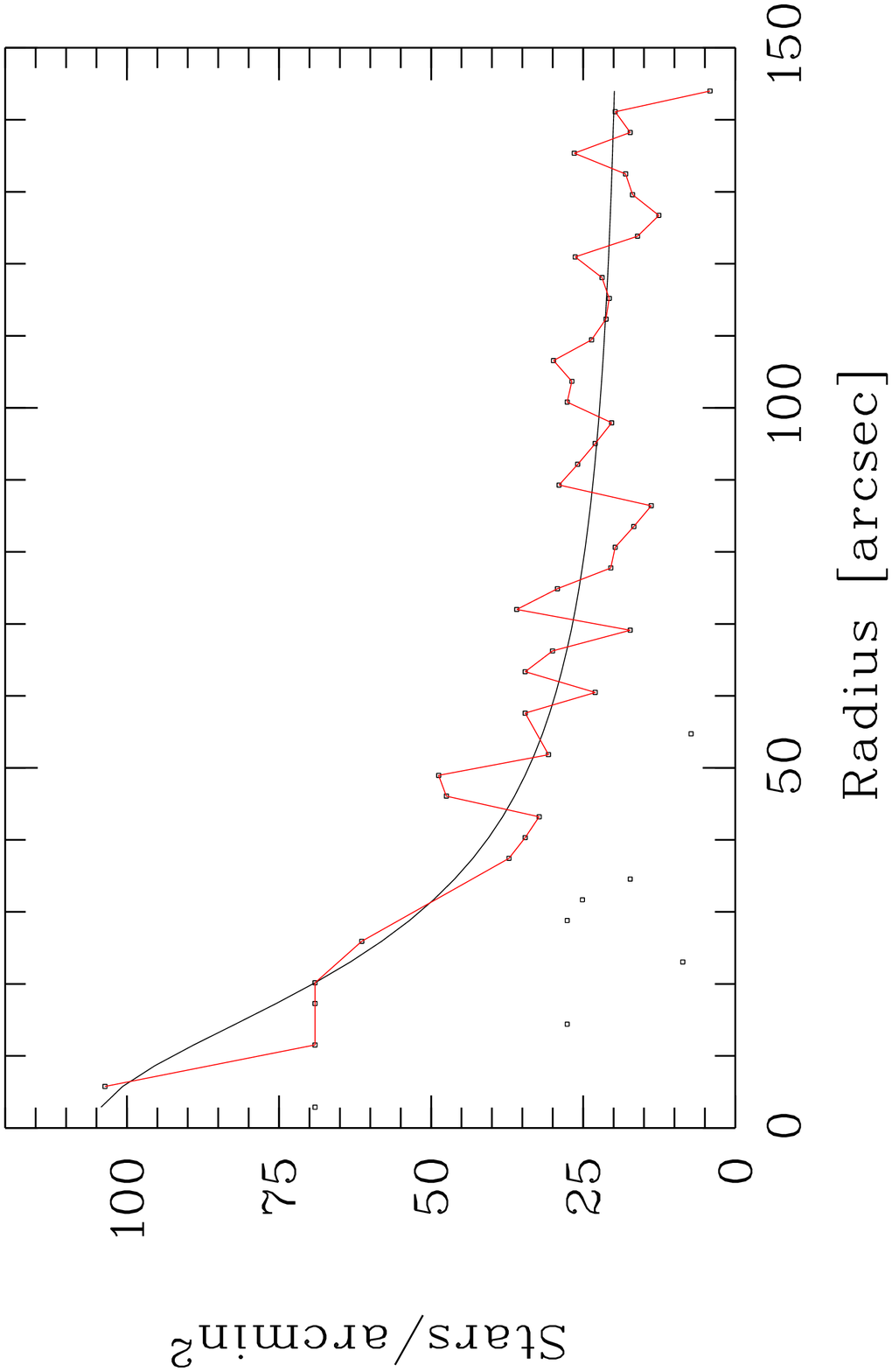} 
\caption{\label{krad_1712} As Fig.\,\ref{krad_0088} but for FSR\,1712. The solid
black line represents the best fit using a core radius of 24".}
\end{figure}

\begin{figure}
\centering
\includegraphics[height=8cm, angle=-90]{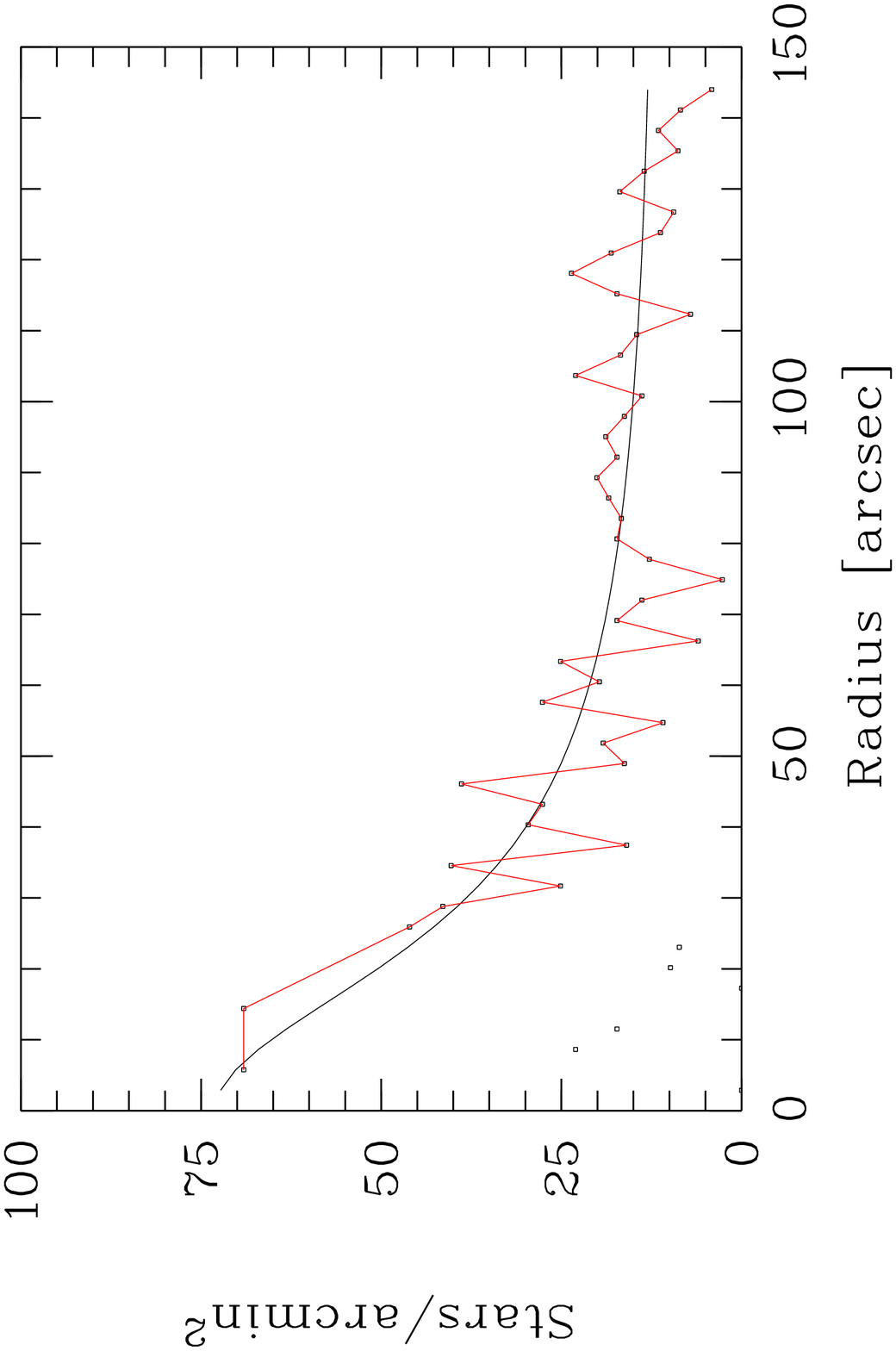} 
\caption{\label{krad_1716} As Fig.\,\ref{krad_0088} but for FSR\,1716. The solid
black line represents the best fit using a core radius of 26.5".}
\end{figure}

\begin{figure}
\centering
\includegraphics[height=8cm, angle=-90]{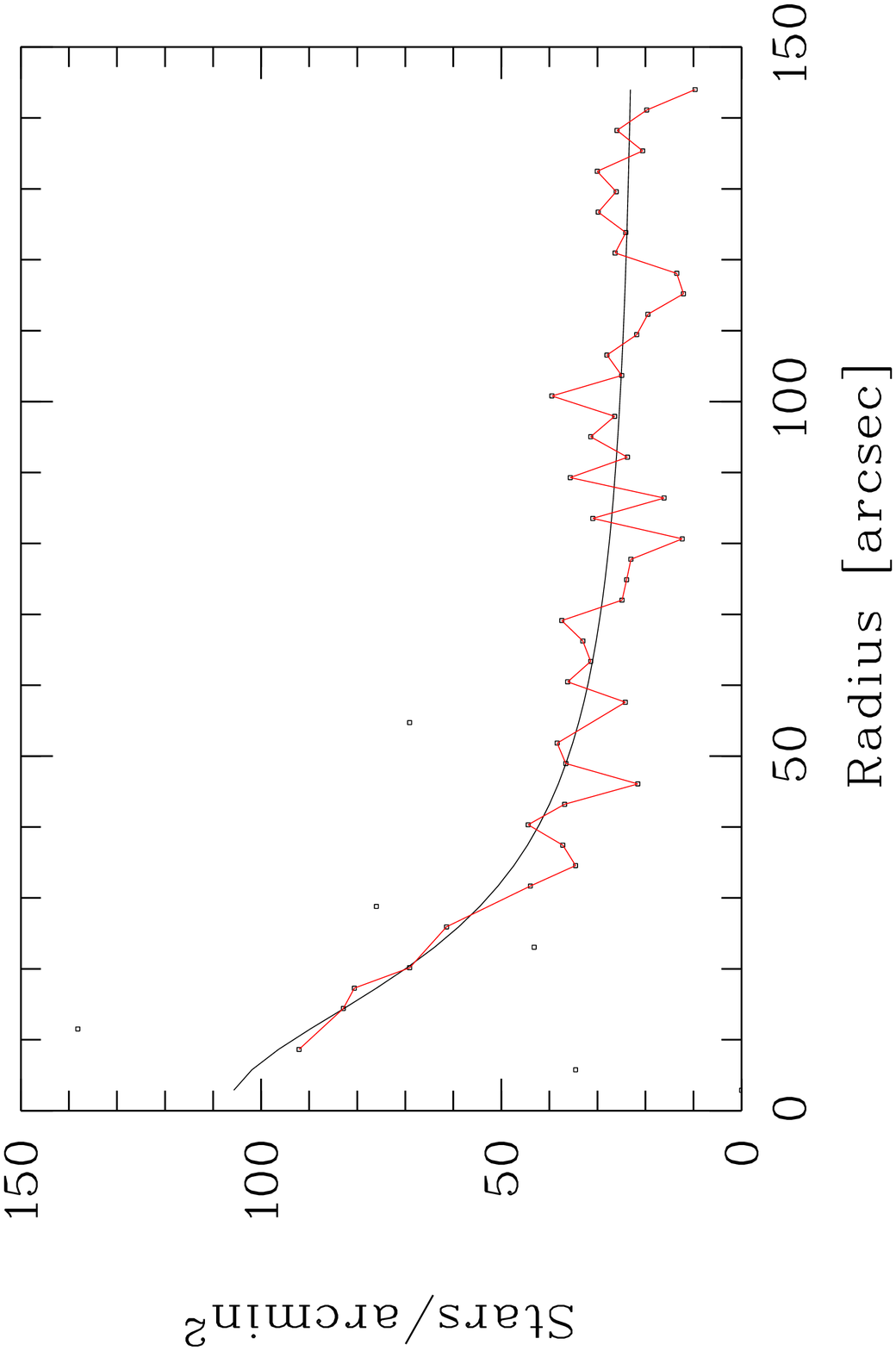} 
\caption{\label{krad_1735} As Fig.\,\ref{krad_0088} but for FSR\,1735. The solid
black line represents the best fit using a core radius of 23".}
\end{figure}

\clearpage

\section{K-band images of the cluster candidates}

\begin{figure}
\centering
\includegraphics[height=7.5cm]{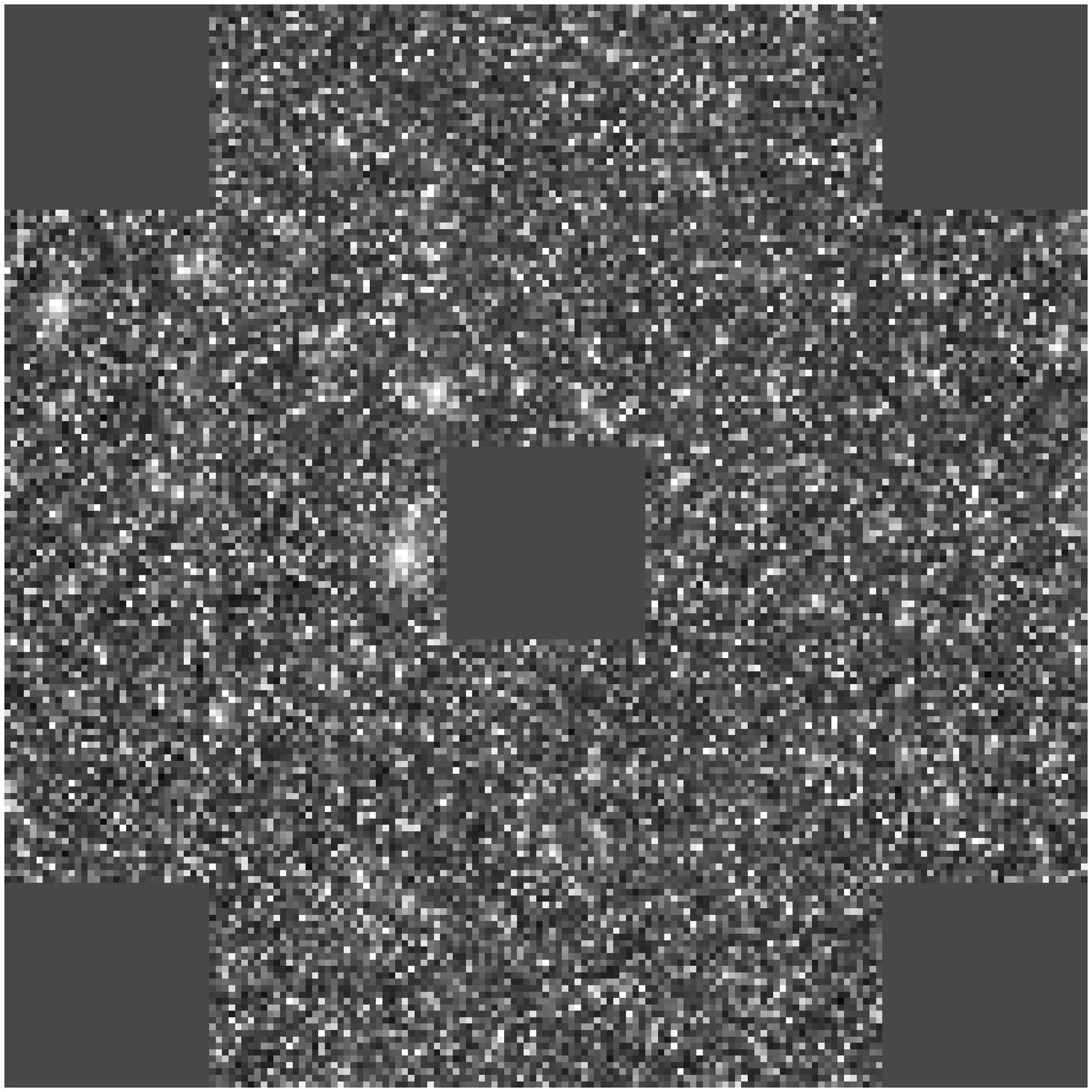} 
\caption{\label{cl0002_k} Grey scale representation of the K-band mosaic of
cluster candidate FSR\,0002. The image size is about 11.7'\,x\,11.7', north is
up and east to the left. The cluster candidate is positioned in the upper left
part of the mosaic.}
\end{figure}

\begin{figure}
\centering
\includegraphics[height=7.5cm]{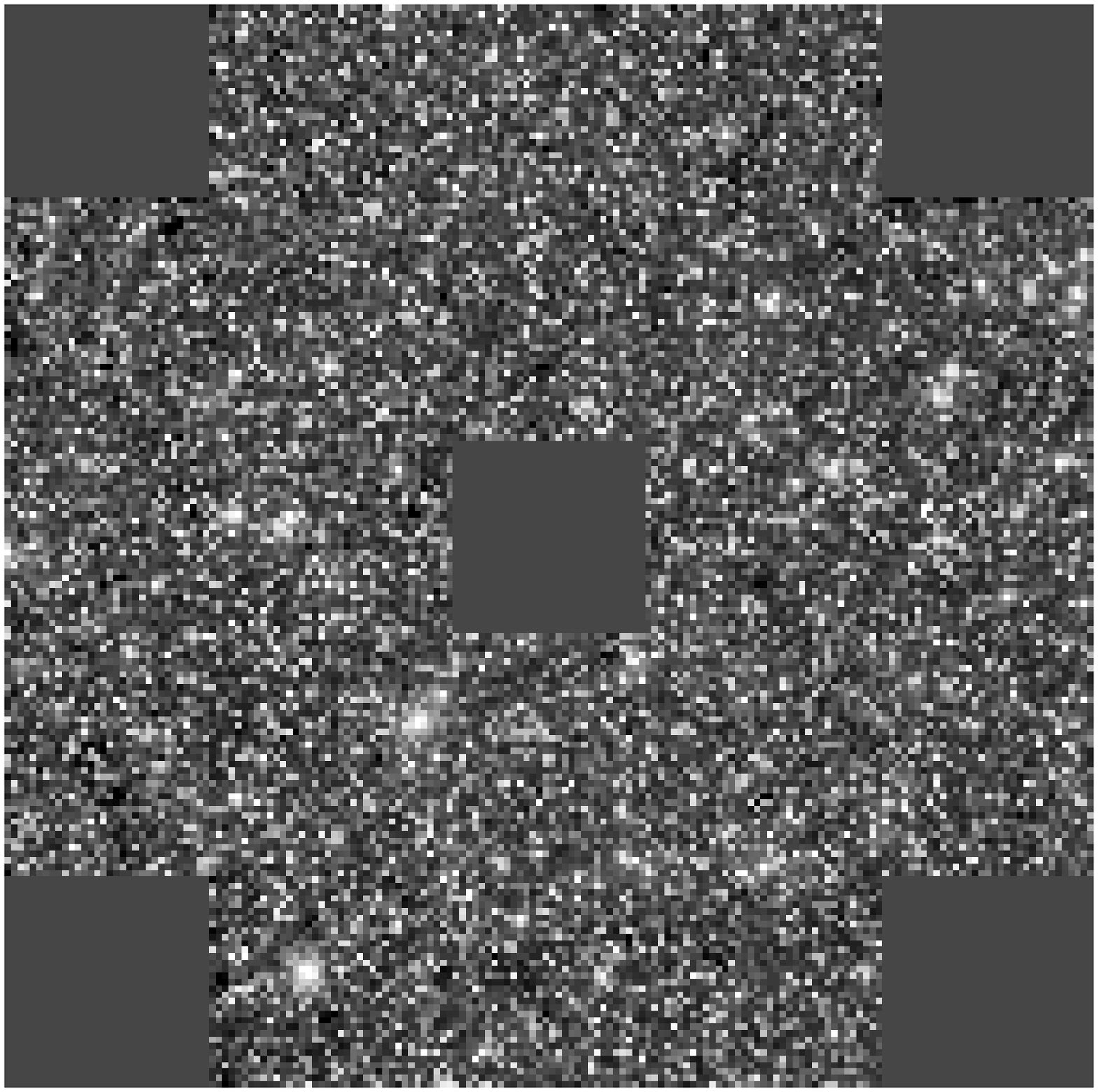} 
\caption{\label{cl0023_k} As Fig.\,\ref{cl0002_k} but for the cluster candidate
FSR\,0023.}
\end{figure}

\begin{figure}
\centering
\includegraphics[height=7.5cm]{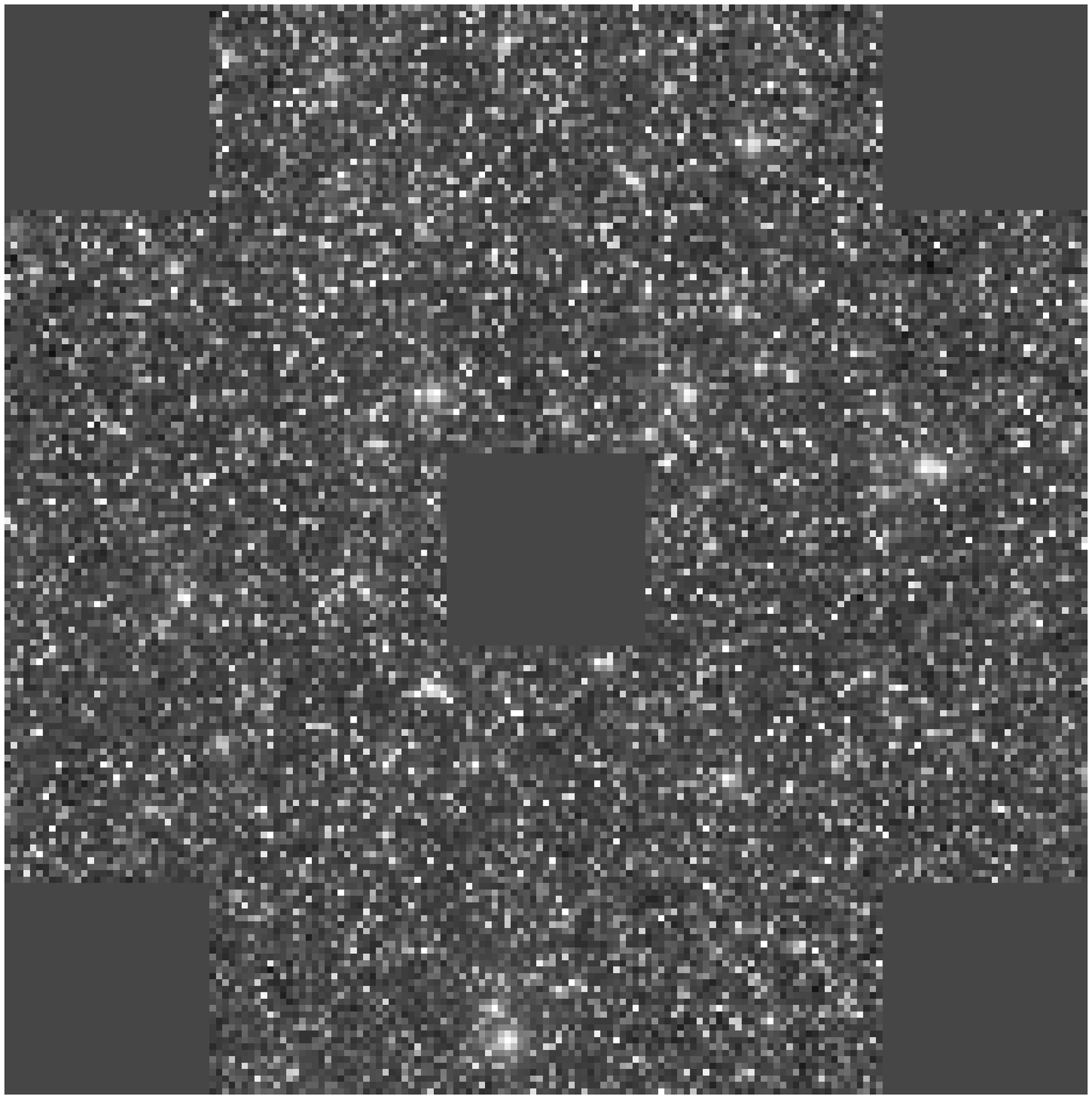} 
\caption{\label{cl0088_k} As Fig.\,\ref{cl0002_k} but for the cluster candidate
FSR\,0088.}
\end{figure}

\begin{figure}
\centering
\includegraphics[height=7.5cm]{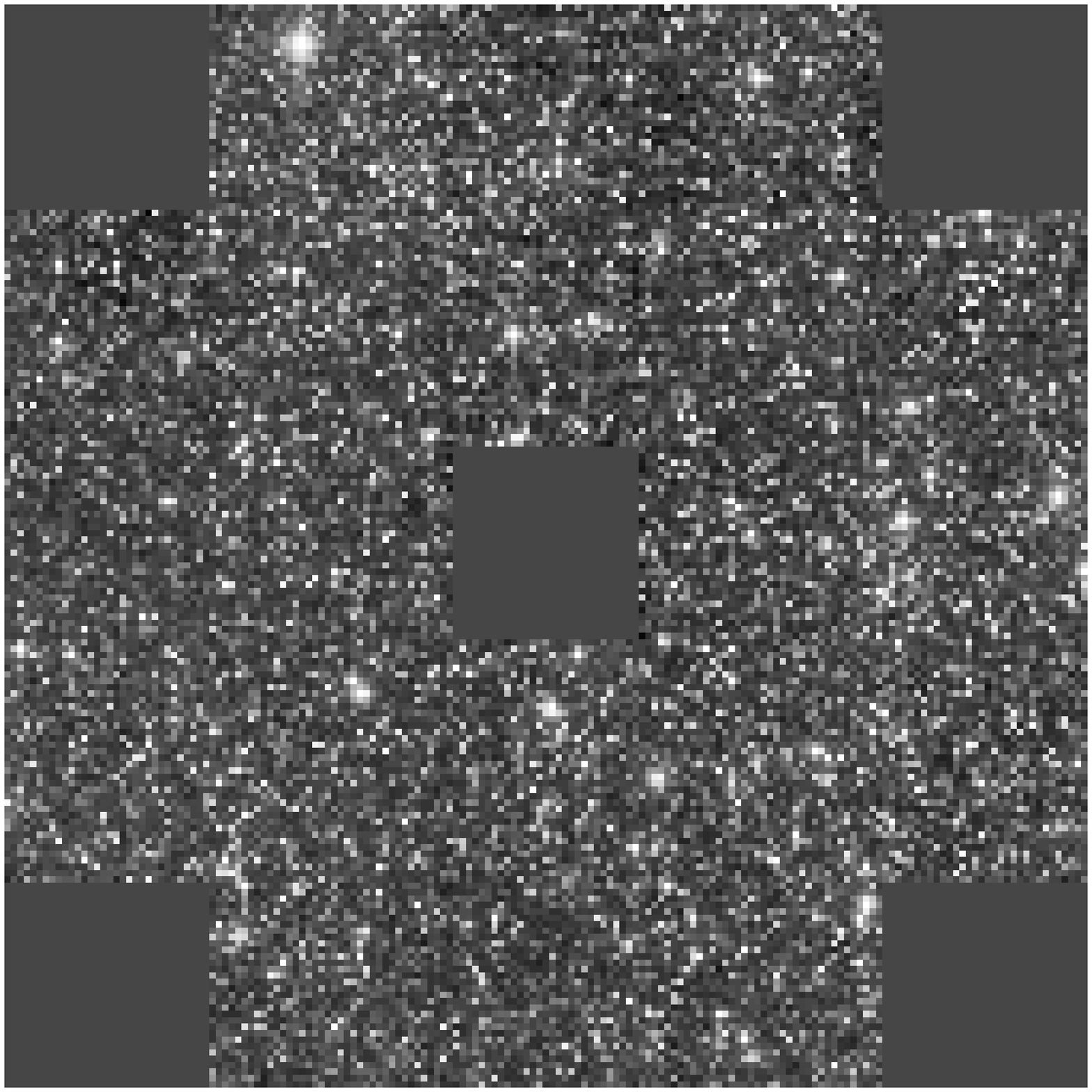} 
\caption{\label{cl0089_k} As Fig.\,\ref{cl0002_k} but for the cluster candidate
FSR\,0089.}
\end{figure}

\begin{figure}
\centering
\includegraphics[height=7.5cm]{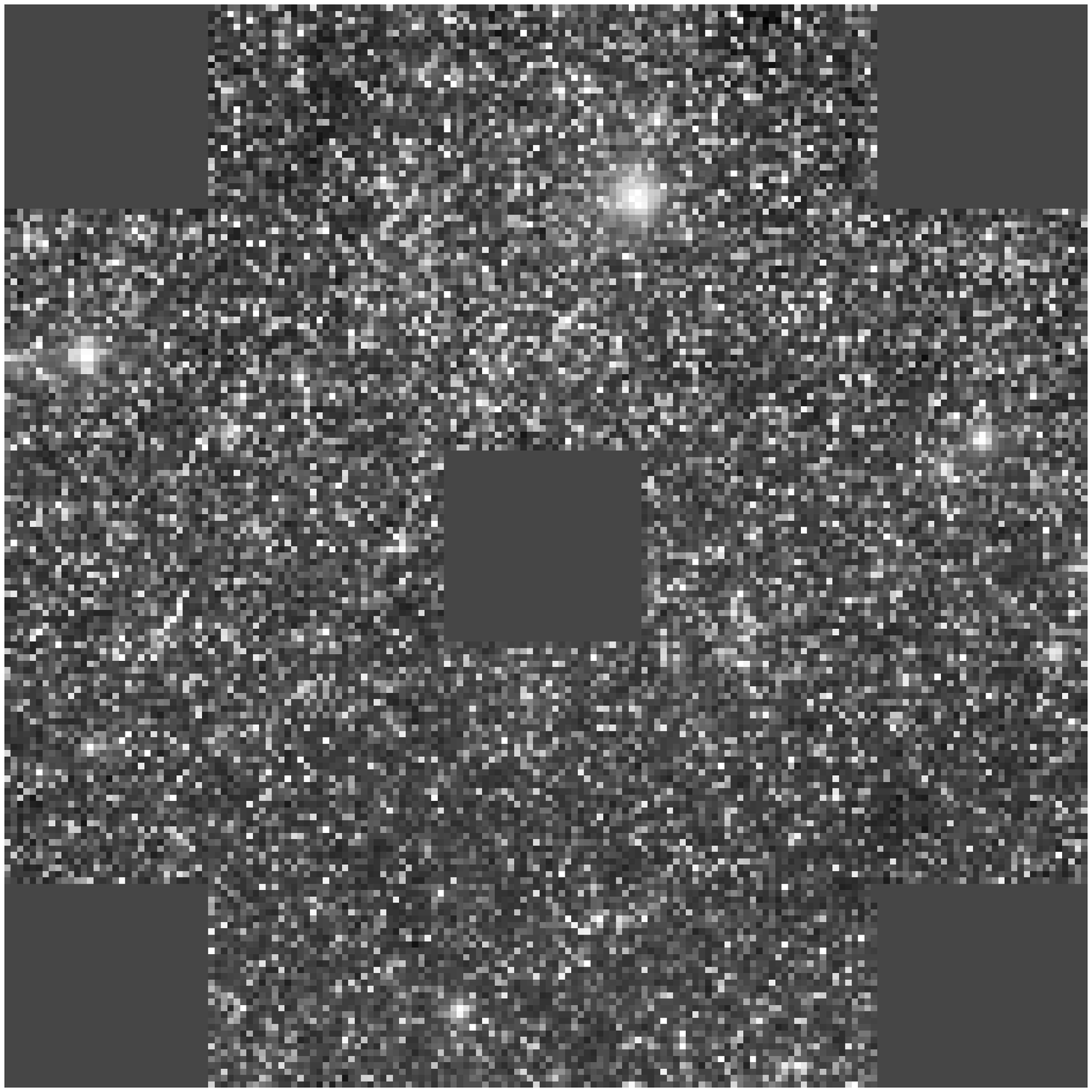} 
\caption{\label{cl0094_k} As Fig.\,\ref{cl0002_k} but for the cluster candidate
FSR\,0094.}
\end{figure}

\begin{figure}
\centering
\includegraphics[height=7.5cm]{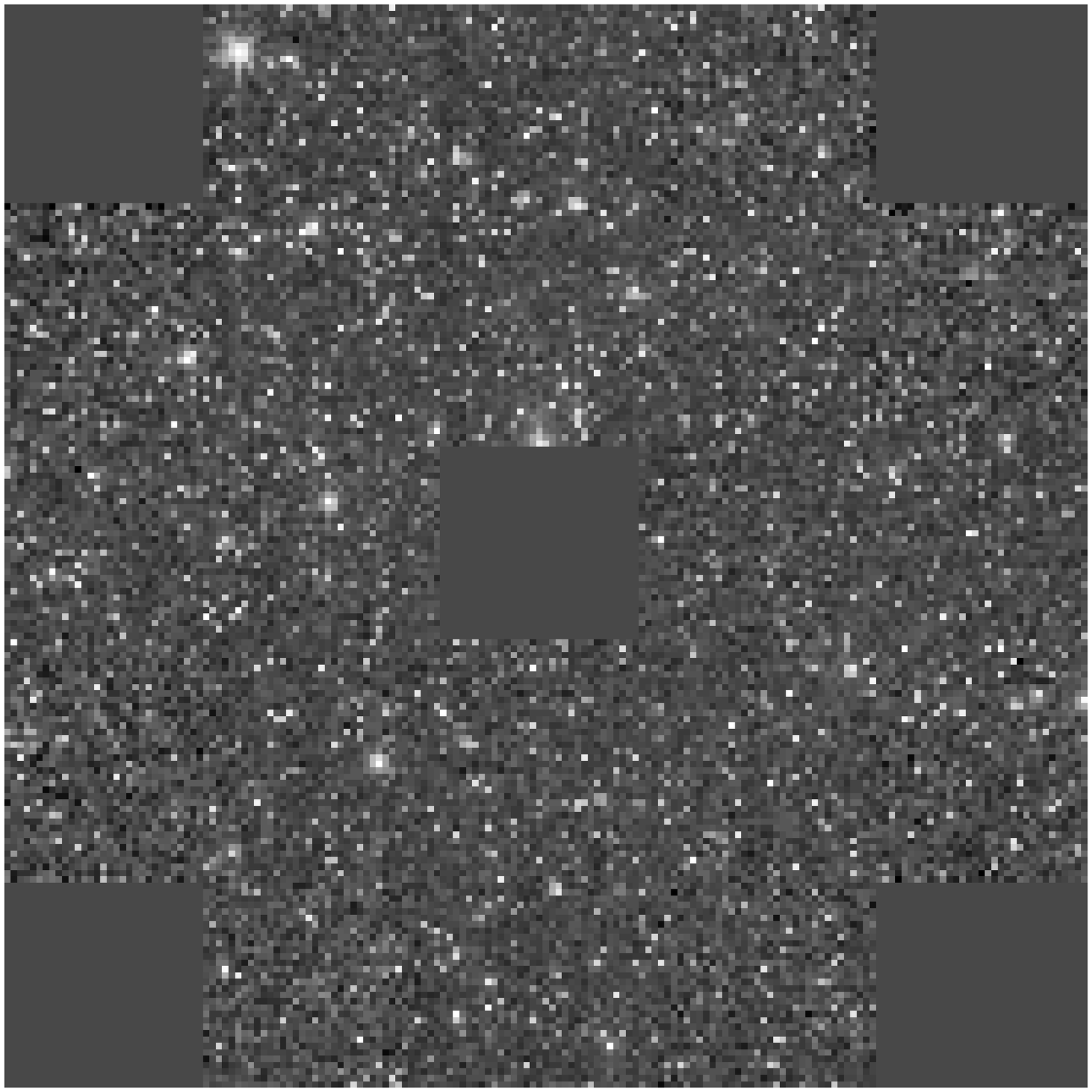} 
\caption{\label{cl1527_k} As Fig.\,\ref{cl0002_k} but for the cluster candidate
FSR\,1527.}
\end{figure}

\begin{figure}
\centering
\includegraphics[height=7.5cm]{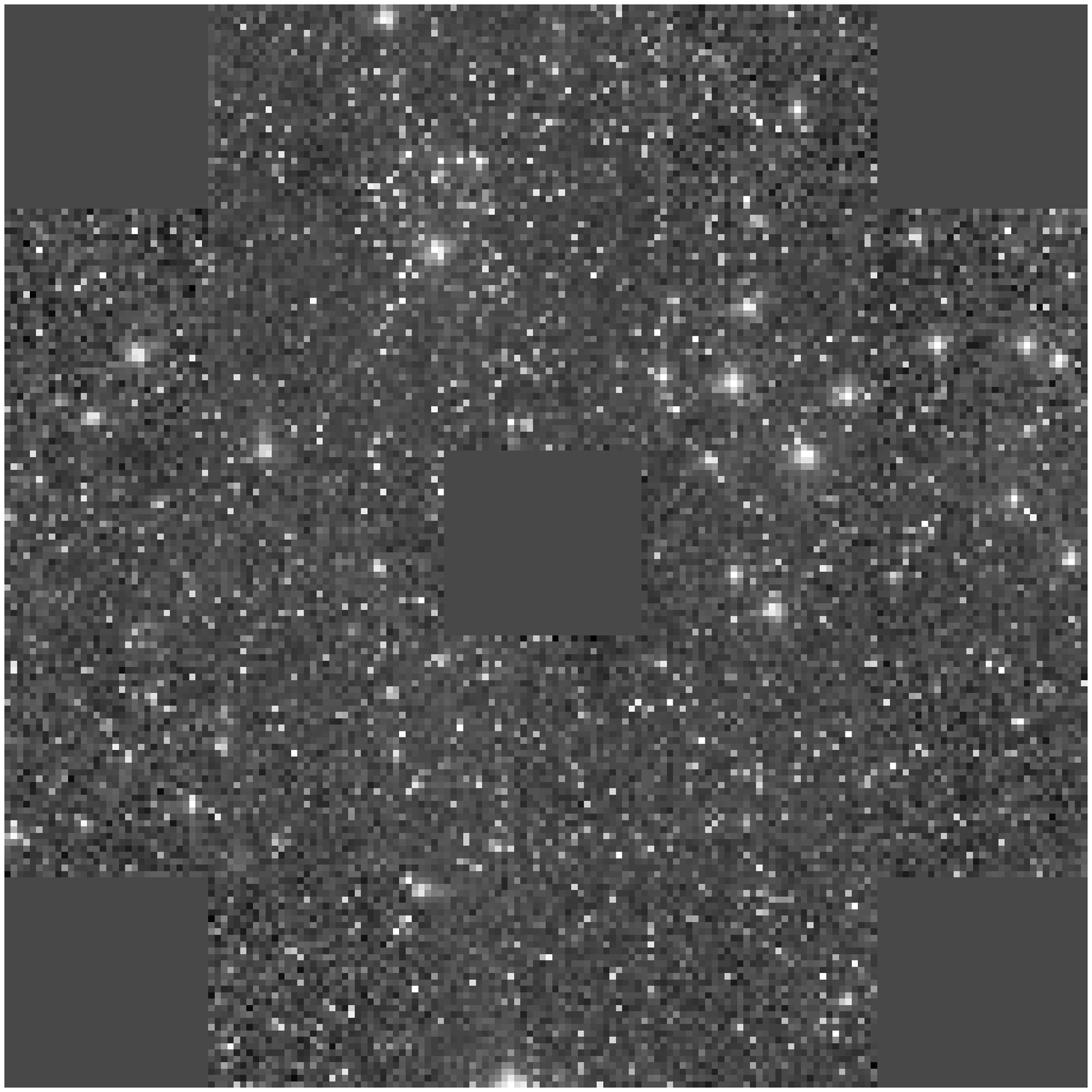} 
\caption{\label{cl1530_k} As Fig.\,\ref{cl0002_k} but for the cluster candidate
FSR\,1530.}
\end{figure}

\begin{figure}
\centering
\includegraphics[height=7.5cm]{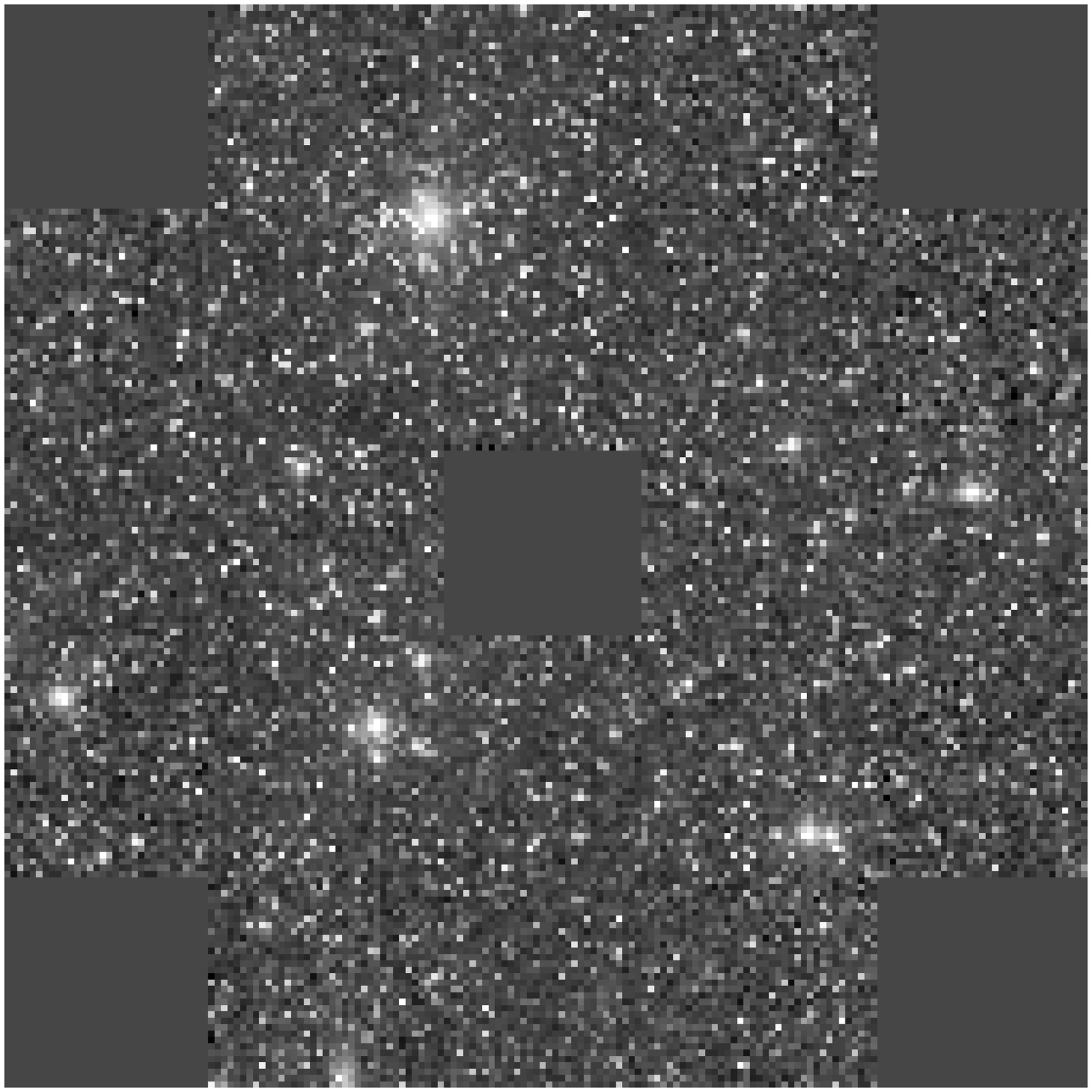} 
\caption{\label{cl1570_k} As Fig.\,\ref{cl0002_k} but for the cluster candidate
FSR\,1570.}
\end{figure}

\begin{figure}
\centering
\includegraphics[height=7.5cm]{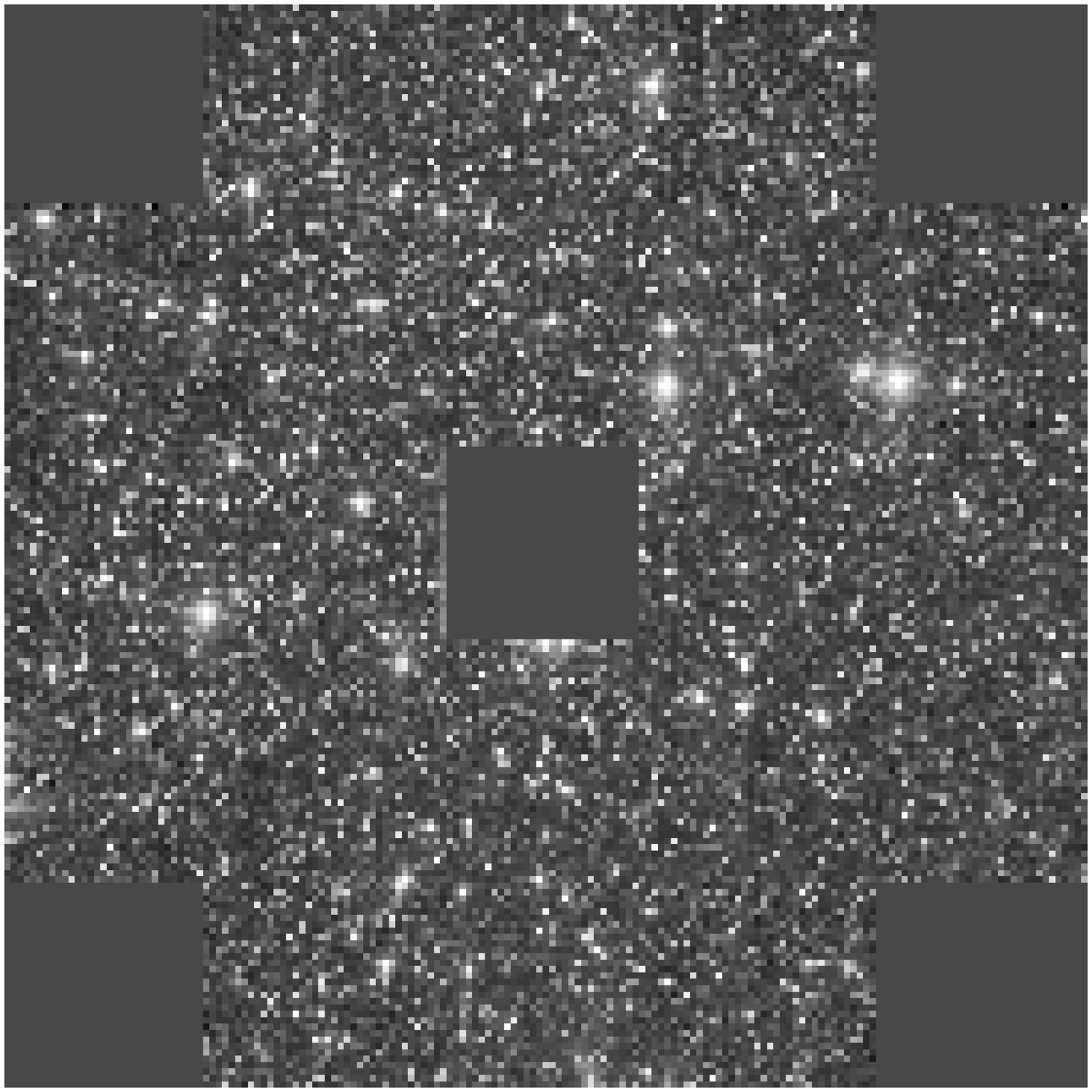} 
\caption{\label{cl1659_k} As Fig.\,\ref{cl0002_k} but for the cluster candidate
FSR\,1659.}
\end{figure}

\begin{figure}
\centering
\includegraphics[height=7.5cm]{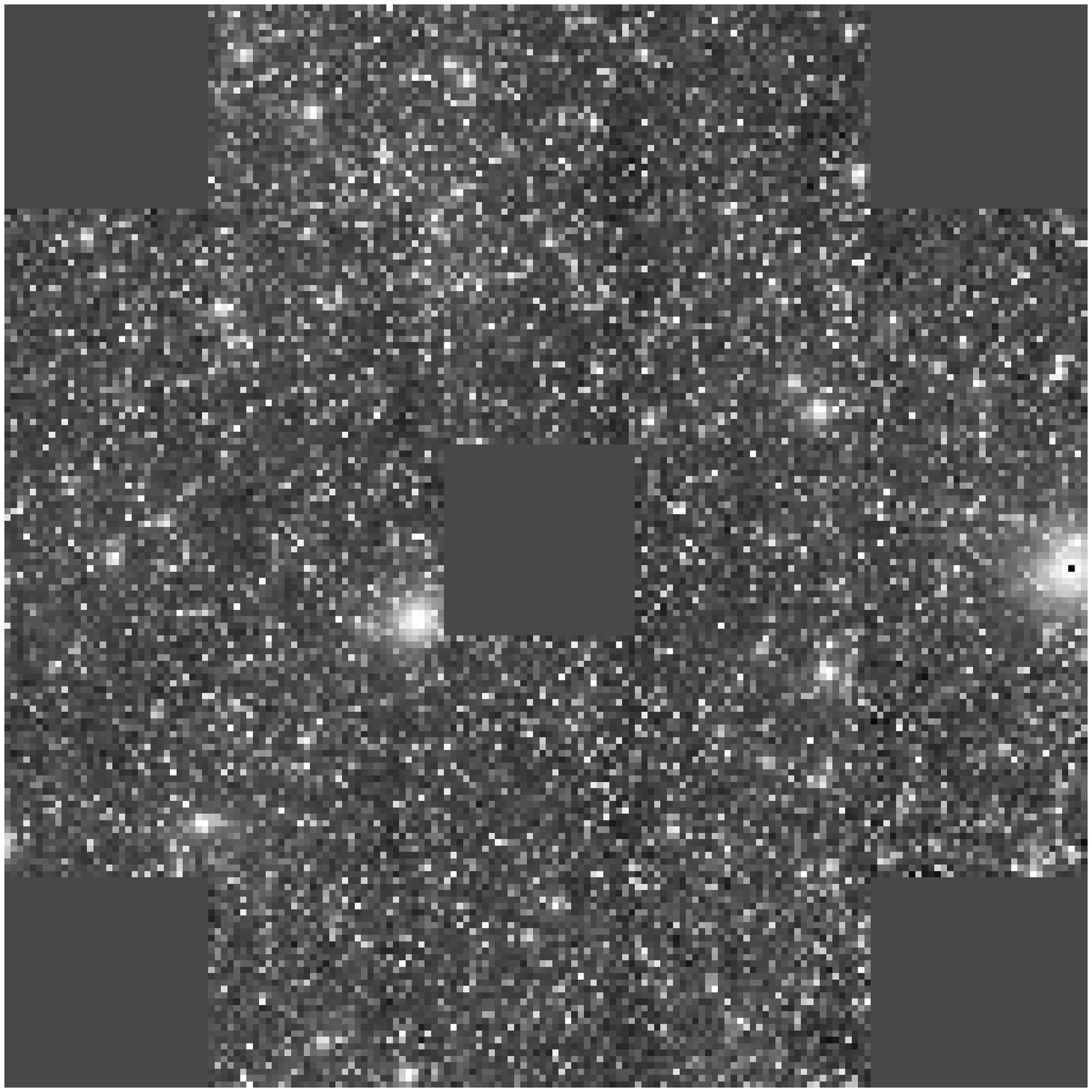} 
\caption{\label{cl1712_k} As Fig.\,\ref{cl0002_k} but for the cluster candidate
FSR\,1712.}
\end{figure}

\begin{figure}
\centering
\includegraphics[height=7.5cm]{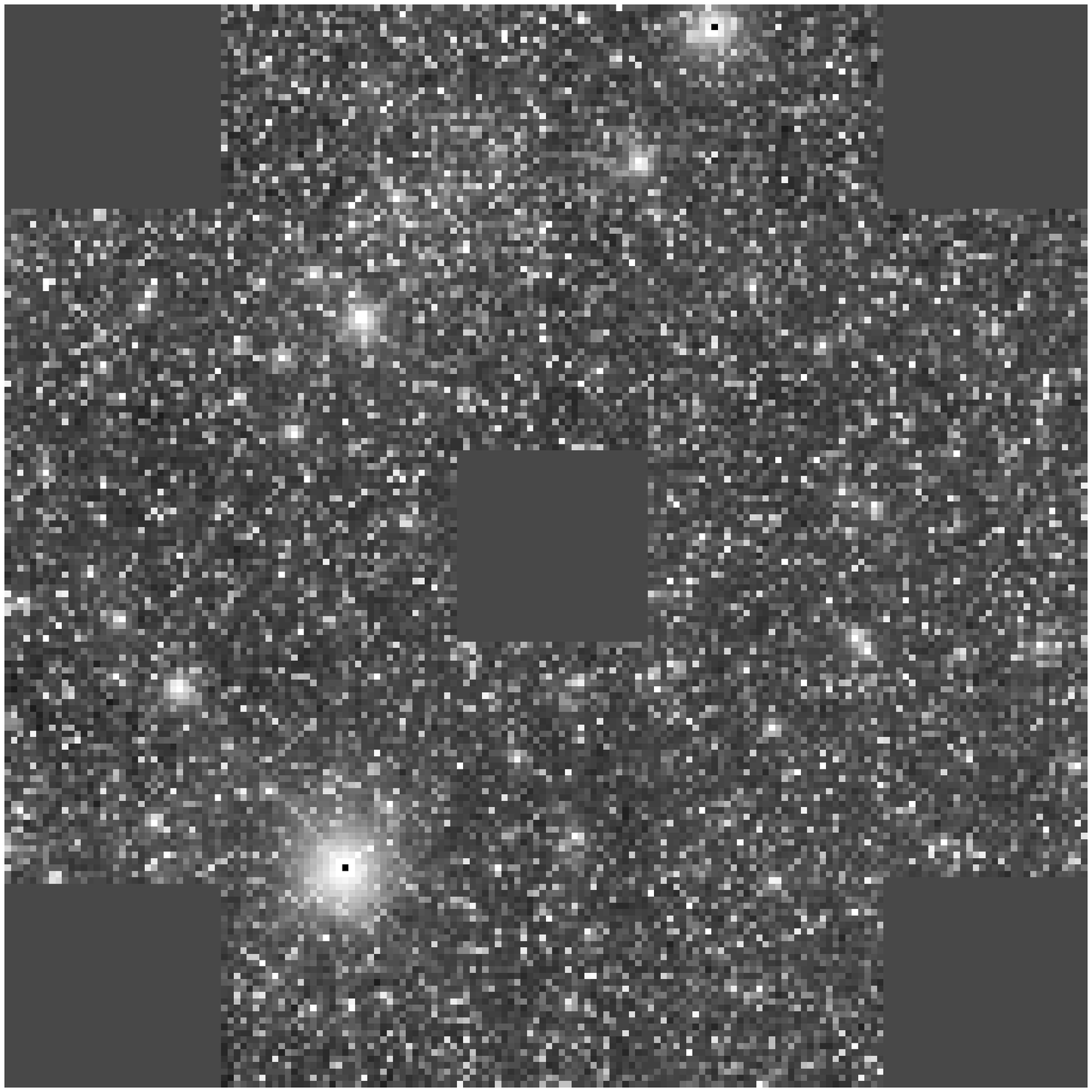} 
\caption{\label{cl1716_k} As Fig.\,\ref{cl0002_k} but for the cluster candidate
FSR\,1716.}
\end{figure}

\begin{figure}
\centering
\includegraphics[height=7.5cm]{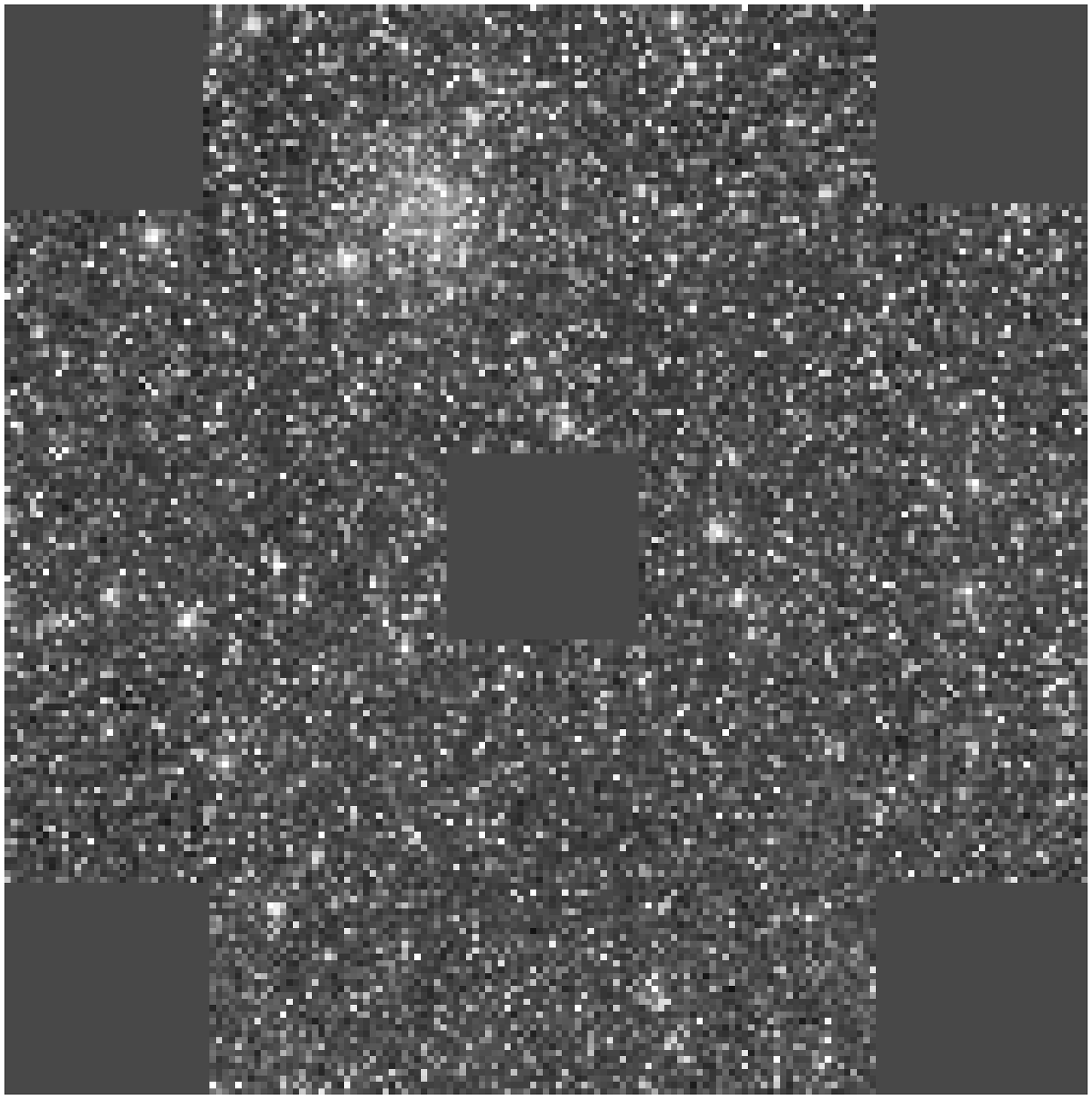} 
\caption{\label{cl1735_k} As Fig.\,\ref{cl0002_k} but for the cluster candidate
FSR\,1735.}
\end{figure}

\begin{figure}
\centering
\includegraphics[height=7.5cm]{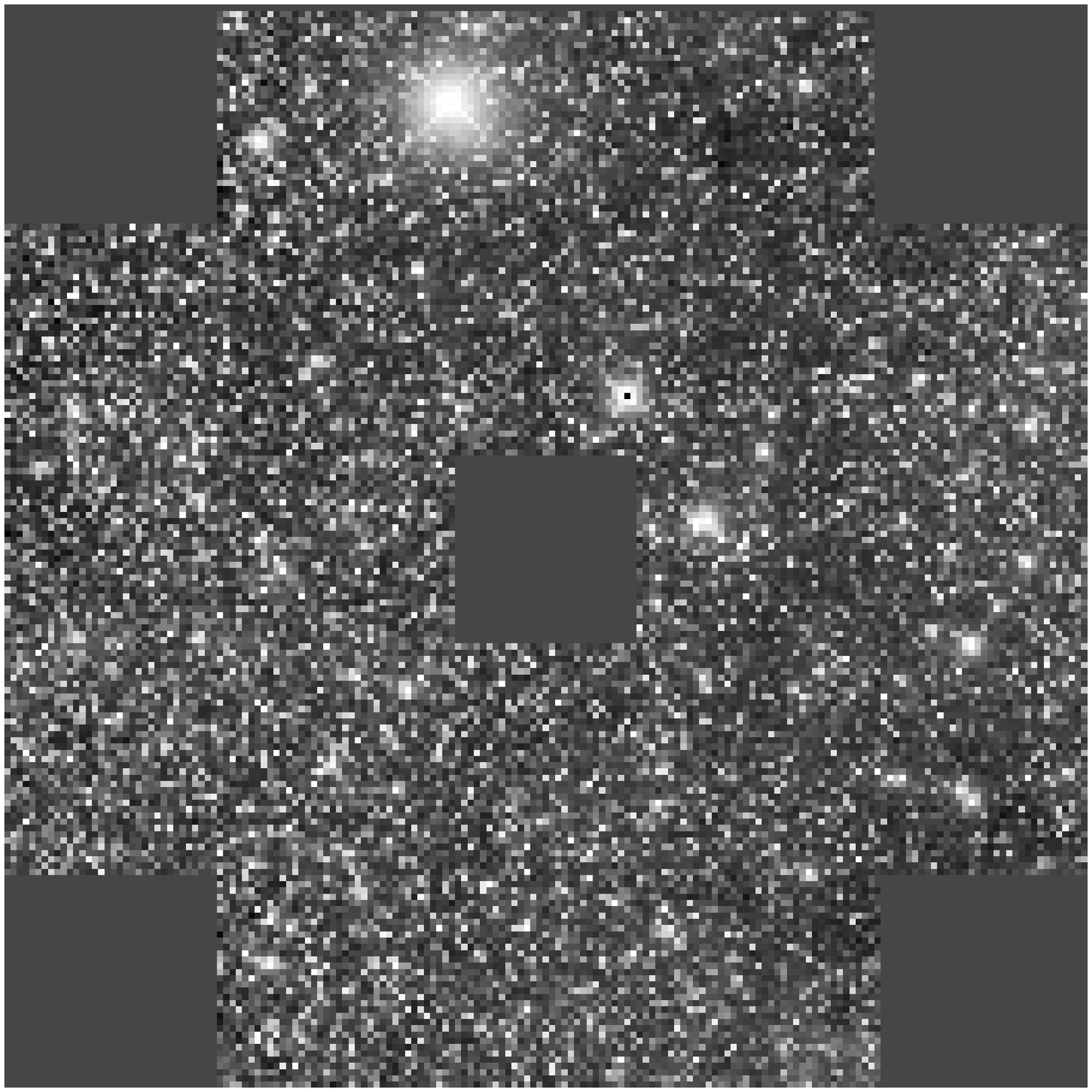} 
\caption{\label{cl1754_k} As Fig.\,\ref{cl0002_k} but for the cluster candidate
FSR\,1754.}
\end{figure}

\begin{figure}
\centering
\includegraphics[height=7.5cm]{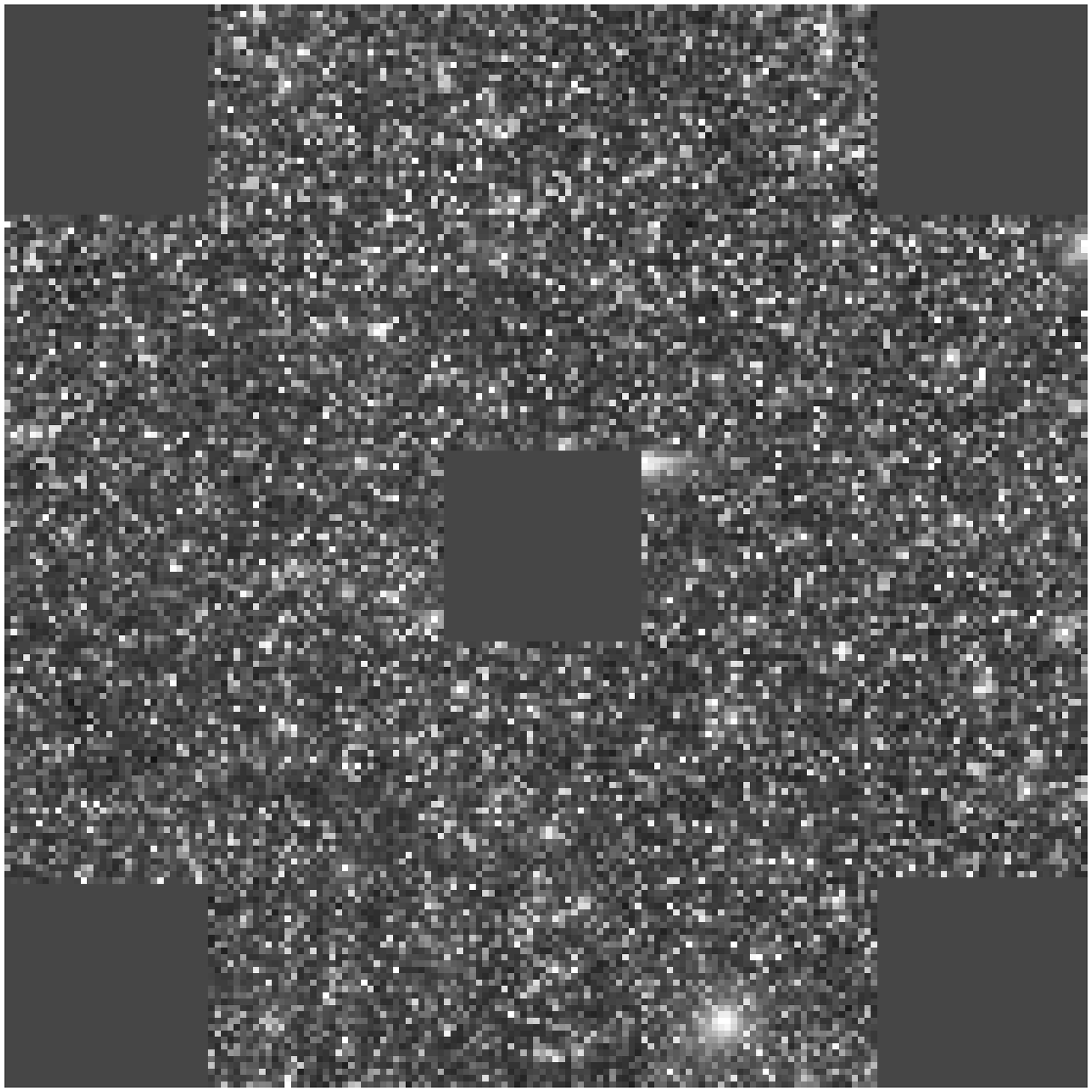} 
\caption{\label{cl1767_k} As Fig.\,\ref{cl0002_k} but for the cluster candidate
FSR\,1767.}
\end{figure}

\clearpage
\onecolumn

\section{Parameters of properly classified FSR clusters}

\begin{table}
\renewcommand{\tabcolsep}{3pt}

\caption{\label{fsrproperties} Summary of the properties of all FSR clusters
analysed in detail so far and that are clearly star clusters. The table lists:
FSR number and other identifications, Right Ascension, Declination (J2000),
Galactic Coordinates ($l$,$b$), Distance $d$, Extinction $A_K$, Age, Metallicity
$Z$, Galactocentric Distance $R_{{\rm GC}}$, Distance above the  Galactic Plane
$z_{{\rm GP}}$, Classification, and References. The classifications stand for:
OC - open cluster; YOC - young open cluster (age $<$\,100\,Myr); GC - globular
cluster; ? - classification, or parameter uncertain. We assumed a galactocentric
distance of the Sun of 8\,kpc.} 

\begin{tabular}{p{2cm}llrrrlllrrlp{3cm}} 
Name(s) & $\alpha$\,(2000) & $\delta$\,(2000) & $l$ & $b$ & $d$ & $A_K$ & Age & $Z$ & $R_{{\rm GC}}$ & $z_{{\rm GP}}$ & Class. & Reference \\
  &   &   & [deg] & [deg] & [kpc] & [mag] & [Gyr] &   & [kpc] & [pc] &   &   \\
\hline
FSR\,0031 									& 18:06:29.0 & $-$21:22:33 &  8.9062 &$-$0.2680 &  1.6 & 0.5 &   1.1    & 0.019?& 6.4 & -7.5 & OC     & Bonatto \& Bica {\cite{2007A&A...473..445B}} \\
FSR\,0070 									& 19:30:02.0 & $-$15:10:01 & 23.4404 &$-$15.2646&  2.3 & 0.1 &   5.0    & 0.019?& 6.1 & -610 & OC     & Bica et al. {\cite{2008MNRAS.385..349B}} \\
FSR\,0088 									& 18:50:38.0 & $-$04:11:17 & 29.1119 &$-$1.7298 &  2.0 & 0.5 &   0.5    & 0.019?& 6.3 & -60  & OC     & this paper \\
FSR\,0089 									& 18:48:39.0 & $-$03:30:34 & 29.4908 &$-$0.9803 &  2.2 & 1.0 &   1.0    & 0.019?& 6.2 & -38  & OC     & Bonatto \& Bica {\cite{2007A&A...473..445B}} \newline this paper \\
FSR\,0124 									& 19:06:52.0 & $+$13:15:21 & 46.4754 &$+$2.6526 &  2.6 & 0.4 &   1.0    & 0.019?& 6.5 &  120 & OC     & Bica et al. {\cite{2008MNRAS.385..349B}} \\
FSR\,0133 									& 19:29:48.5 & $+$15:33:36 & 51.1120 &$-$1.1780 &  1.9 & 0.7 &   0.6    & 0.019?& 7.0 & -39  & OC     & Bica et al. {\cite{2008MNRAS.385..349B}} \\
FSR\,0190 									& 20:05:31.3 & $+$33:34:09 & 70.7303 &$+$0.9498 & 10.0 & 0.8 &   $>$7   & 0.002 &10.5 &  170 & OC/GC? & Froebrich et al. {\cite{2008MNRAS.383L..45F}} \\
FSR\,0584 									& 02:27:15.0 & $+$61:37:28 &134.0579 &$+$0.8399 &  1.4 & 1.0 &   10.0?  &0.0002?& 9.0 &  21  & GC?    & Bica et al. {\cite{2007A&A...472..483B}} \\
FSR\,0705 \newline NGC\,1798 							& 05:11:43.0 & $+$47:41:42 &160.7071 &$+$4.8607 &  6.0 & 0.1 &   1.5    & 0.019?&13.8 &  510 & OC     & Bonatto \& Bica {\cite{2008A&A...485...81B}} \\
FSR\,0729 \newline NGC\,1883 							& 05:25:55.0 & $+$46:29:46 &163.0794 &$+$6.1646 &  3.7 & 0.05&   1.0    & 0.019?&11.6 &  400 & OC     & Bonatto \& Bica {\cite{2008A&A...485...81B}} \\
FSR\,0730 \newline NGC\,2126 \newline Melotte\,39 \newline Collinder\,78 	& 06:02:34.6 & $+$49:51:36 &163.2463 &$+$13.1319&  1.0 & 0.05&   1.2    & 0.019?& 8.9 &  230 & OC     & Bonatto \& Bica {\cite{2008A&A...485...81B}} \\
FSR\,0756 									& 04:24:13.4 & $+$29:42:14 &168.6494 &$-$13.7190&  1.8 & 0.3 &   0.3    & 0.019?& 9.7 & -430 & OC     & Bonatto \& Bica {\cite{2008A&A...485...81B}} \\
FSR\,0793 \newline Berkeley\,69 						& 05:24:21.6 & $+$32:36:03 &174.4474 &$-$1.8561 &  3.5 & 0.15&   0.9    & 0.019?&11.5 & -110 & OC     & Bonatto \& Bica {\cite{2008A&A...485...81B}} \\
FSR\,0795 \newline Koposov\,10 							& 05:47:28.6 & $+$35:25:56 &174.6448 &$+$3.7033 &  2.0 & 0.35&  $<$0.4  & 0.019?&10.0 &  130 & OC     & Koposov et al. {\cite{2008A&A...486..771K}} \\
FSR\,0802 \newline Koposov\,12 							& 06:00:56.2 & $+$35:16:36 &176.1598 &$+$6.0004 &  2.05& 0.13&   0.8    & 0.019?&10.0 &  210 & OC     & Koposov et al. {\cite{2008A&A...486..771K}} \\
FSR\,0810 \newline Berkeley\,71 						& 05:40:57.0 & $+$32:16:16 &176.6344 &$+$0.8968 &  3.0 & 0.25&   1.0    & 0.019?&11.0 &  47  & OC     & Bonatto \& Bica {\cite{2008A&A...485...81B}} \\
FSR\,0814 \newline Koposov\,36 							& 05:36:46.1 & $+$31:11:46 &177.0691 &$-$0.4288 &  1.6 & 0.3 & $<$0.22  & 0.019?& 9.6 & -12  & YOC    & Bonatto \& Bica {\cite{2008A&A...485...81B}} \newline Koposov et al. {\cite{2008A&A...486..771K}} \\
FSR\,0828 \newline Koposov\,43 							& 05:52:14.6 & $+$29:55:09 &179.9001 &$+$1.7425 &  2.8 & 0.17&   2.0    & 0.019?&10.9 &  85  & OC     & Koposov et al. {\cite{2008A&A...486..771K}} \\
FSR\,0834 \newline Czernik\,23 							& 05:50:07.0 & $+$28:53:28 &180.5474 &$+$0.8191 &  2.5 & 0.0 &   4.5    & 0.019?&10.5 &  36  & OC     & Bonatto \& Bica {\cite{2008A&A...485...81B}} \\
FSR\,0856 \newline Koposov\,53 							& 06:08:56.2 & $+$26:15:49 &184.9029 &$+$3.1308 &  3.2 & 0.15& $<$0.32  & 0.019?&11.2 &  170 & OC     & Koposov et al. {\cite{2008A&A...486..771K}} \\
FSR\,0869 \newline Koposov\,63 							& 06:10:01.9 & $+$24:32:55 &186.5267 &$+$2.5208 &  4.2 & 0.15&   1.4    & 0.019?&12.2 &  180 & OC     & Bonatto \& Bica {\cite{2008A&A...485...81B}} \newline Koposov et al. {\cite{2008A&A...486..771K}} \\
FSR\,0911 \newline Cl\,1 in Bochum\,1	                         		& 06:25:01.0 & $+$19:50:55 &192.3100 &$+$3.3600 &  4.5 & 0.16&   7.0    & 0.019?&12.4 & -580 & OC     & Bonatto \& Bica {\cite{2008A&A...485...81B}} \newline Bica et al. {\cite{2008arXiv0807.4077B}} \\
FSR\,0917 \newline Berkeley\,23 						& 06:33:16.2 & $+$20:31:08 &192.6094 &$+$5.3837 &  6.7 & 0.03&   1.2    & 0.019?&14.6 &  630 & OC     & Bonatto \& Bica {\cite{2008A&A...485...81B}} \\
FSR\,0923 									& 06:10:36.0 & $+$16:58:16 &193.2296 &$-$1.0187 &  1.5 & 0.45&   0.5    & 0.019?& 9.5 & -27  & OC     & Bonatto \& Bica {\cite{2008A&A...485...81B}} \\
FSR\,0932 									& 06:04:26.4 & $+$14:33:20 &194.6240 &$-$3.4866 &  1.5 & 0.3 &   0.15   & 0.019?& 9.5 & -91  & OC     & Bonatto \& Bica {\cite{2008A&A...485...81B}} \\
FSR\,0942 									& 06:05:58.0 & $+$13:40:06 &195.5810 &$-$3.5950 &  3.1 & 0.2 &   1.0    & 0.019?&11.0 & -190 & OC     & Bonatto \& Bica {\cite{2008A&A...485...81B}} \\
FSR\,0948 									& 06:25:52.8 & $+$15:50:15 &195.9645 &$+$1.6710 &  2.9 & 0.15&   0.03   & 0.019?&10.8 &  85  & YOC    & Bonatto \& Bica {\cite{2008A&A...485...81B}} \\
FSR\,0974 									& 06:32:41.3 & $+$12:31:55 &199.6598 &$+$1.6015 &  2.6 & 0.2 &   0.4    & 0.019?&10.5 &  73  & OC     & Bonatto \& Bica {\cite{2008A&A...485...81B}} \\
FSR\,1530 									& 10:08:58.3 & $-$57:17:11 &282.3294 &$-$1.0628 &  2.5 & 0.9 &$\le$0.004& 0.019?& 7.9 & -46  & YOC    & this paper \\
FSR\,1570 									& 11:08:40.6 & $-$60:42:50 &290.6888 &$-$0.3083 &  6.0 & 0.8 &   0.008  & 0.019?& 8.1 & -32  & YOC    & Pasquali et al. {\cite{2006A&A...448..589P}} \newline this paper \\
FSR\,1603 									& 12:09:45.0 & $-$62:59:17 &298.2191 &$-$0.4984 &  2.7 & 0.2 &   1.0    & 0.019?& 7.1 & -23  & OC     & Bica \& Bonatto {\cite{2008MNRAS.384.1733B}} \\
FSR\,1644 \newline Harvard\,8 \newline Cr\,268 					& 13:18:02.9 & $-$67:04:34 &305.5257 &$-$4.3385 &  1.9 & 0.1 &   0.6    & 0.019?& 7.1 & -140 & OC     & Bica et al. {\cite{2008MNRAS.385..349B}} \\
FSR\,1712 									& 15:54:46.3 & $-$52:31:47 &328.8084 &$+$0.8786 &  1.8 & 1.4 &   0.8    & 0.019?& 6.5 &  28  & OC     & this paper \\
FSR\,1716 									& 16:10:29.0 & $-$53:44:48 &329.7779 &$-$1.5893 &  7.0 & 0.57&   $>$2   & 0.004 & 4.0 & -190 & OC/GC? & this paper \\
FSR\,1723 \newline ESO\,275SC1 							& 15:55:05.0 & $-$46:00:51 &333.0269 &$+$5.8517 &  1.3 & 0.01&   0.8    & 0.019?& 6.9 &  130 & OC     & Bica et al. {\cite{2008MNRAS.385..349B}} \\
FSR\,1735 									& 16:52:10.6 & $-$47:03:29 &339.1877 &$-$1.8533 &  8.5 & 0.7 &   $>$8   & 0.004 & 3.0 & -270 & GC?    & Froebrich et al. {\cite{2007MNRAS.377L..54F}} \newline this paper \\
FSR\,1737 									& 16:18:21.0 & $-$40:14:35 &340.0953 &$+$7.2499 &  2.8 & 0.2 &   $>$5   & 0.019?& 5.5 &  350 & OC     & Bica et al. {\cite{2008MNRAS.385..349B}} \\
FSR\,1744 									& 16:51:36.0 & $-$42:24:55 &342.7060 &$+$1.1783 &  3.5 & 0.85&   1.0    & 0.019?& 4.8 &  72  & OC     & Bonatto \& Bica {\cite{2007A&A...473..445B}} \\
FSR\,1755 									& 17:12:20.0 & $-$38:27:44 &348.2458 &$+$0.4825 &  1.4 & 0.45&$<$0.005  & 0.019?& 6.6 &  12  & YOC    & Bica \& Bonatto {\cite{2008MNRAS.384.1733B}} \\     
\end{tabular}
\end{table}

\section{List of other investigated FSR candidates}

\begin{table}

\caption{\label{fsr_unknown_properties} Summary of the properties of all FSR
clusters analysed in detail so far and were the classification is unclear, no
parameters are derived or which are clearly not star clusters. The table lists:
FSR number and other identifications, Right Ascension, Declination (J2000),
Galactic Coordinates (l,b), Classification, and References. The classifications
stand for: NC - not a cluster; OC - open cluster; EC - embedded cluster; ? -
classification uncertain.} 

\begin{tabular}{p{1.5cm}llrrlp{3cm}} 
\footnotesize
Name & $\alpha$\,(2000) & $\delta$\,(2000) & l\,[deg] & b\,[deg] & Class. & Reference \\
\hline
FSR\,0002 			& 17:32:32.0 & $-$27:03:51 &  0.0462 &$+$3.4416 & NC     & this paper \\
FSR\,0010 			& 16:40:49.0 & $-$16:01:09 &  2.1485 &$+$19.6176& OC?    & Bica et al. {\cite{2008MNRAS.385..349B}} \\
FSR\,0023 			& 17:57:35.0 & $-$22:52:32 &  6.5839 &$+$0.7827 & NC     & this paper \\
FSR\,0041 			& 17:03:30.0 & $-$08:51:13 & 11.7384 &$+$19.1982& NC     & Bica et al. {\cite{2008MNRAS.385..349B}} \\
FSR\,0091 			& 17:38:21.0 & $+$05:43:14 & 29.6887 &$+$18.8502& NC     & Bica et al. {\cite{2008MNRAS.385..349B}} \\
FSR\,0094 			& 18:49:50.0 & $-$01:02:55 & 31.8158 &$-$0.1213 & NC     & this paper \\
FSR\,0098 			& 18:47:36.0 & $+$00:35:46 & 33.0251 &$+$1.1256 & OC?    & Bica et al. {\cite{2008MNRAS.385..349B}} \\
FSR\,0114 			& 20:09:09.0 & $-$02:13:03 & 40.0167 &$-$18.2675& NC     & Bica et al. {\cite{2008MNRAS.385..349B}} \\
FSR\,0119 			& 18:23:05.0 & $+$15:49:12 & 44.0994 &$+$13.2956& NC     & Bica et al. {\cite{2008MNRAS.385..349B}} \\
FSR\,0128 			& 20:31:10.1 & $+$04:45:07 & 49.2922 &$-$19.6403& NC     & Bica et al. {\cite{2008MNRAS.385..349B}} \\
FSR\,0744 			& 04:59:30.0 & $+$38:00:42 &167.0846 &$-$2.7641 & NC     & Bonatto \& Bica {\cite{2008A&A...485...81B}} \\
FSR\,0773 			& 04:29:37.0 & $+$26:00:14 &172.3124 &$-$15.3117&   ?    & Bonatto \& Bica {\cite{2008A&A...485...81B}} \\
FSR\,0776 			& 06:07:24.0 & $+$39:49:34 &172.7400 &$+$9.2946 & NC     & Bonatto \& Bica {\cite{2008A&A...485...81B}} \\
FSR\,0784 \newline Koposov\,7 	& 05:40:44.1 & $+$35:55:25 &173.5096 &$+$2.7915 & EC     & Koposov et al. {\cite{2008A&A...486..771K}} \\
FSR\,0801 			& 04:47:04.8 & $+$24:54:00 &175.7930 &$-$12.9947& NC     & Bonatto \& Bica {\cite{2008A&A...485...81B}} \\
FSR\,0839 \newline Koposov\,41 	& 06:03:58.0 & $+$30:15:41 &180.8678 &$+$4.1118 & EC     & Koposov et al. {\cite{2008A&A...486..771K}} \\
FSR\,0841 			& 05:06:13.4 & $+$21:33:27 &181.2319 &$-$11.4928& NC     & Bonatto \& Bica {\cite{2008A&A...485...81B}} \\
FSR\,0849 \newline Koposov\,58 	& 05:51:11.0 & $+$25:46:41 &183.3426 &$-$0.5718 & EC     & Koposov et al. {\cite{2008A&A...486..771K}} \\
FSR\,0851 			& 05:14:44.9 & $+$19:47:31 &183.8703 &$-$10.8657&   ?    & Bonatto \& Bica {\cite{2008A&A...485...81B}} \\
FSR\,0855 			& 05:42:21.6 & $+$22:49:48 &184.8249 &$-$3.8180 &   ?    & Bonatto \& Bica {\cite{2008A&A...485...81B}} \\
FSR\,0882 			& 05:27:51.1 & $+$16:53:49 &188.0656 &$-$9.8578 &   ?    & Bonatto \& Bica {\cite{2008A&A...485...81B}} \\
FSR\,0884 			& 05:32:21.0 & $+$17:11:02 &188.4011 &$-$8.7954 &   ?    & Bonatto \& Bica {\cite{2008A&A...485...81B}} \\
FSR\,0894 			& 06:04:05.0 & $+$20:16:51 &189.5866 &$-$0.7570 & NC     & Bonatto \& Bica {\cite{2008A&A...485...81B}} \\
FSR\,0927 			& 06:24:10.0 & $+$18:01:30 &193.8370 &$+$2.3286 & NC     & Bonatto \& Bica {\cite{2008A&A...485...81B}} \\
FSR\,0956 			& 06:12:25.0 & $+$13:00:26 &196.9185 &$-$2.5390 & NC     & Bonatto \& Bica {\cite{2008A&A...485...81B}} \\
FSR\,1527 			& 10:06:32.0 & $-$57:24:52 &282.1369 &$-$1.3586 & NC     & this paper \\
FSR\,1635 			& 12:54:57.0 & $-$43:29:24 &303.6073 &$+$19.3771& NC     & Bica et al. {\cite{2008MNRAS.385..349B}} \\
FSR\,1647 			& 13:45:48.0 & $-$73:57:29 &306.7313 &$-$11.4887& NC     & Bica et al. {\cite{2008MNRAS.385..349B}} \\
FSR\,1659 			& 13:38:01.0 & $-$62:27:55 &308.2860 &$-$0.0787 & NC     & this paper \\
FSR\,1685 			& 14:57:14.9 & $-$64:57:22 &315.7477 &$-$5.2629 & NC     & Bica et al. {\cite{2008MNRAS.385..349B}} \\
FSR\,1695 			& 14:33:38.0 & $-$49:10:09 &319.5904 &$+$10.3687& NC     & Bica et al. {\cite{2008MNRAS.385..349B}} \\
FSR\,1740 			& 17:49:17.8 & $-$51:31:55 &340.7279 &$-$12.0481& OC?    & Bica et al. {\cite{2008MNRAS.385..349B}} \\
FSR\,1754 			& 17:15:03.4 & $-$39:05:46 &348.0432 &$-$0.3191 & NC     & Bica et al. {\cite{2008MNRAS.385..349B}} \newline this paper \\
FSR\,1767 			& 17:35:43.0 & $-$36:21:28 &352.6010 &$-$2.1662 & NC     & Bonatto et al. {\cite{2007MNRAS.381L..45B}} \newline this paper \\
FSR\,1769 			& 17:04:41.3 & $-$31:00:43 &353.3068 &$+$6.1797 & OC?    & Bica et al. {\cite{2008MNRAS.385..349B}} \\
\end{tabular}
\end{table}

\end{appendix}

\label{lastpage}

\end{document}